\documentclass[orivec,a4paper,10pt]{llncs}
\usepackage{amsmath}
\usepackage{amsfonts}
\usepackage{amssymb}
\usepackage{latexsym}
\usepackage{stmaryrd}

\usepackage{url}

\usepackage{graphicx}
\usepackage{wrapfig}
\usepackage{pgf}
\usepackage{tikz}
\usetikzlibrary{automata,arrows,intersections,positioning,calc,shapes}
\tikzstyle{state}=[circle,draw=black,inner sep=0pt,minimum size=10pt]
\tikzstyle{class green}=[fill=green, regular polygon, regular polygon sides=3]
\tikzstyle{class blue}=[fill=blue, regular polygon, regular polygon sides=4]
\tikzstyle{class red}=[fill=red, regular polygon, regular polygon sides=5]

\tikzstyle{vertex}=[inner sep=0pt,minimum size=10pt]

\tikzstyle{point}=[circle,draw=black,fill=black,inner sep=0pt,minimum size=0.75pt]

\usepackage[noend]{algorithmic}
\usepackage{algorithm}

\usepackage{paralist}

\usepackage{cite}

\usepackage{times}

\allowdisplaybreaks

\newenvironment{myproof}{\begin{proof}}{\qed\end{proof}}

\newtheorem{result}{Result}

\newcommand{\bigO}[1]{\mathcal{O}(#1)}

\renewcommand{\epsilon}{\varepsilon}

\newcommand{\nat}{\mathbb{N}}

\newcommand{\sd}{\mu} % states distribution
\newcommand{\gd}{\rho} % generic distribution
\newcommand{\dirac}[1]{\delta_{#1}}

\newcommand{\probeval}[2]{#1(#2)}

\newcommand{\Disc}[1]{\mathrm{Disc}(#1)}
\newcommand{\SubDisc}[1]{\mathrm{SubDisc}(#1)}
\newcommand{\Supp}[1]{\mathrm{Supp}(#1)}

\newcommand{\family}[2]{{{\{#1\}}_{#2}}}

\newcommand{\aut}[1][A]{\mathcal{#1}}
\newcommand{\stateSet}{S}
\newcommand{\actionSet}{\Sigma}
\newcommand{\internalActionSet}{H}
\newcommand{\externalActionSet}{E}
\newcommand{\startState}[1][s]{\bar{#1}}
\newcommand{\transitionRelation}{\mathit{D}}

\newcommand{\dotteddiamond}{\setbox0=\vbox{\hbox{$\diamond$}}{\ooalign{\hfil\box0\hfil\cr\hfil$\mkern-0.5mu \cdot$\hfil\crcr}}}

\newcommand{\NetworkTBetaMuTransRel}[5]{G(#1,#2,#3,#4,#5)}%{G^{#1,#2,#3}_{#4,#5}}
\newcommand{\LPproblemTBetaMuTransRel}[5]{{\hyperWeakCombinedAllowedTransition{#1}{#2}{\dotteddiamond\liftrel[{#5}]#3}{#4}}}%{P^{#1,#2,#3}_{#4,#5}}
\newcommand{\LPproblemMatchingRelGdaBetaaTranaGdbBetabTranb}[7]{{#2 \stackrel{#3}{\Longrightarrow}^{\raisebox{-3pt}{$\scriptstyle #4$}}_{\combined} \dotteddiamond\liftrel[{#1}]\dotteddiamond{}\hspace{2pt} {}^{\raisebox{-3pt}{$\scriptstyle #7$}}_{\kern4pt \combined}{\stackrel{\mkern.5mu #6}{\Longleftarrow}}#5}}%{P(#1,#2,#3,#4,#5,#6,#7)}

\newcommand{\transitionsFromState}[1]{\transitionRelation(#1)}
\newcommand{\transitionsWithLabel}[1]{\transitionRelation(#1)}
\newcommand{\allowedTransitions}{A}

\newcommand{\hidden}{\tau}

\newcommand{\tr}{\mathit{tr}}
\newcommand{\source}[1]{\mathit{src}(#1)}
\newcommand{\action}[1]{\mathit{act}(#1)}
\newcommand{\target}[1]{\mathit{trg}(#1)}

\newcommand{\prefix}{\leqslant}
\newcommand{\frags}[1]{\mathit{frags}(#1)}
\newcommand{\finiteFrags}[1]{\mathit{frags}^{*}(#1)}

\newcommand{\length}[1]{\lvert #1 \rvert}
\newcommand{\cone}[1]{C_{#1}}
\newcommand{\trace}[1]{\mathit{trace}(#1)}
\newcommand{\last}[1]{\mathit{last}(#1)}
\newcommand{\first}[1]{\mathit{first}(#1)}
\newcommand{\emptytrace}{\epsilon}

\newcommand{\combined}{\mathrm{C}}
\newcommand{\strongTransition}[3]{{#1 \stackrel{#2}{\longrightarrow} #3}}

\newcommand{\weakCombinedTransition}[3]{{#1 \stackrel{#2}{\Longrightarrow}_{\combined} #3}}

\newcommand{\hyperWeakCombinedTransition}[3]{{#1 \stackrel{#2}{\Longrightarrow}_{\combined} #3}}

\newcommand{\weakCombinedAllowedTransition}[4]{{#1 \stackrel{#2}{\Longrightarrow}^{\raisebox{-3pt}{$\scriptstyle #4$}}_{\combined} #3}}

\newcommand{\hyperWeakCombinedAllowedTransition}[4]{{#1 \stackrel{#2}{\Longrightarrow}^{\raisebox{-3pt}{$\scriptstyle #4$}}_{\combined} #3}}

\newcommand{\rel}[1][\relsymbol]{\mathrel{\mathcal{#1}}}
\newcommand{\idrelsymbol}{I}
\newcommand{\idrel}{\rel[\idrelsymbol]}
\newcommand{\liftrel}[1][\rel]{\mathrel{\mathcal{L}(#1)}}

\newcommand{\sched}{\sigma}
\newcommand{\schedeval}[2]{#1(#2)}

\newcommand{\equivclass}[0]{\mathcal{C}}
\newcommand{\relclass}[2]{[#1]_{#2}}
\newcommand{\quotienting}[2]{#1/#2}
\newcommand{\partitionset}[2][\relsymbol]{\quotienting{#2}{\mathcal{#1}}}

\newcommand{\weakBisim}{\approx}

\newcommand{\incomingflow}{\vec{f}}
\newcommand{\netsource}{\vartriangle}
\newcommand{\netsink}{\blacktriangledown}

\newcommand{\acronym}[1]{\ensuremath{\textsl{#1}}}
\newcommand{\PA}{\acronym{PA}}

\newcommand{\IMC}{\acronym{IMC}}
\newcommand{\LCMC}{\acronym{LCMC}}
\newcommand{\MDP}{\acronym{MDP}}

\newcommand{\proc}[1]{\textnormal{\scshape#1}}
\newcommand{\progHeader}[1]{#1}

\newcommand{\partitioning}{\rel[W]}

\newcommand{\stategreen}{
   \begin{tikzpicture}
     \node[state, class green, minimum size=9pt]    {};
   \end{tikzpicture}
}

\newcommand{\stategreenInSubscript}{
   \begin{tikzpicture}
     \node[state, class green, minimum size=5pt]    {};
   \end{tikzpicture}
}

\newcommand{\stateblue}{
   \begin{tikzpicture}
     \node[state, class blue, minimum size=9pt]    {};
   \end{tikzpicture}
}

\newcommand{\stateblueInSubscript}{
   \begin{tikzpicture}
     \node[state, class blue, minimum size=5pt]    {};
   \end{tikzpicture}
}

\newcommand{\statered}{
   \begin{tikzpicture}
     \node[state, class red, minimum size=9pt]    {};
   \end{tikzpicture}
}  

\newcommand{\stateredInSubscript}{
   \begin{tikzpicture}
     \node[state, class red, minimum size=5pt]    {};
   \end{tikzpicture}
}

\advance\hoffset-1in
\advance\voffset-1in
\advance\hoffset2.95mm

\title{Deciding Probabilistic Automata Weak Bisimulation in Polynomial Time}
\author{Holger Hermanns \and Andrea Turrini}
\institute{Saarland University -- Computer Science, Saarbr\"{u}cken, Germany}

\begin{document}

\maketitle
%\sloppy
\begin{abstract}
\noindent
Deciding in an efficient way weak probabilistic bisimulation in the
context of probabilistic automata is an open problem for about a
decade.  In this work we close this problem by proposing a procedure
that checks in polynomial time the existence of a weak combined
transition satisfying the step condition of the bisimulation. This
enables us to arrive at a polynomial time algorithm for deciding weak
probabilistic bisimulation.  We also present several extensions to
interesting related problems setting the ground for the development of
more effective and compositional analysis algorithms for probabilistic
systems.
\end{abstract}

\section{Introduction}
\label{sec:introduction}

\emph{Probabilistic automata} (\PA{}) constitute a mathematical framework for
the specification of probabilistic concurrent
systems~\cite{Seg95,CSV07}.  Probabilistic automata extend classical
concurrency models in a simple yet conservative fashion.  In
probabilistic automata, there is no global notion of time, and
probabilistic experiments can be performed inside a transition.  This
embodies a clear separation between probability and nondeterminism,
and is represented by transitions of the form
$\strongTransition{s}{a}{\sd}$, where $s$ is a state, $a$ is an action
label, and $\sd$ is a probability distribution on states. Labeled
transition systems are instances of this model family, obtained by
restricting to Dirac distributions (assigning full probability to
single states). Thus, foundational concepts and results of standard
concurrency theory are retained in full and extend smoothly to the
model of probabilistic automata.  The \PA{} model is akin to Markov
decision processes (\MDP{})~\cite{Der70}, and its foundational beauty can
be paired with powerful model checking techniques, as implemented for
instance in the PRISM tool~\cite{prism}. Variations of this model are
Labeled Concurrent Markov Chains~(\LCMC{}) and alternating
Models~\cite{Var85,Han91,PLS00}.  We refer the interested reader to
\cite{Seg06} for a survey on \PA{} and other models.

If facing a concrete probabilistic system, we can conceive several
different \PA{} models to reflect its behavior. For instance, we can
use different state names, encode diverse information in the states,
represent internal computations with different action labels, and so
on.  \emph{Bisimulation relations} constitute a powerful tool allowing
us to check whether two models describe essentially the same
system. They are then called bisimilar.  The bisimilarity of two
systems can be viewed in terms of a game played between a challenger
and a defender. In each step of the infinite bisimulation game, the
challenger chooses one automaton, makes a step, and the defender
matches it with a step of the other automaton.  Depending on how we
want to treat internal computations, this leads to \emph{strong} and
\emph{weak} bisimulations: the former requires that each single step
of the challenger automaton is matched by an equally labeled single
step of the defender automaton, the latter allows the matching up to
internal computation steps. On the other hand, depending on how
nondeterminism is resolved, probabilistic bisimulation can be varied
by allowing the defender to match the challenger's step by a convex
combination of enabled probabilistic transitions.  This results in a
spectrum of four bisimulations: strong \cite{Seg95,Han91,Var85},
strong probabilistic \cite{Seg95}, weak \cite{PLS00,Seg95}, and weak
probabilistic \cite{Seg95} bisimulation.

Besides comparing automata, bisimulation relations allow us to reduce
the size of an automaton without changing its properties (i.e., with
respect to logic formulae satisfied by it). This is particularly
useful to alleviate the state explosion problem notoriously
encountered in model checking. 

Polynomial decision algorithms for strong (probabilistic) bisimulation
\cite{CS02} and weak bisimulation \cite{PLS00} are known. However,
\PA{} weak bisimulation lacks in transitivity and this severely limits
its usefulness. On the other hand weak probabilistic bisimulation is
indeed transitive, while the only known algorithm for such
bisimulation is exponential \cite{CS02} in the size of
the probabilistic automaton. % \PA{} model \cite{Seg95}.

In this context, it is worth to note that \LCMC{} weak bisimulation
\cite{PLS00} and \PA{} weak probabilistic bisimulation \cite{Seg95}
coincide \cite{ST05} when \LCMC{} is seen as a \PA{} with restrictions on
the structure of the automaton and that restricted versions of \PA{}
weak probabilistic bisimulations, such as normed \cite{BS00} and delay
\cite{Sto02} bisimulation, can be decided in polynomial time.
Following \cite{ST05}, an \LCMC{} is just a \PA{} where each state with
outgoing transitions enables either labeled transitions each one
leading to a single state, or a single transition leading to a
probability distribution over states and this constraint on the
structure of the automaton is enough to reduce the complexity of the
decision procedure at the expense of the loss of using combined
transitions and nondeterminism to simplify the automaton.

Lately, the model of \PA{} has been enhanced with memoryless
continuous time, integrated into the model of Markov
automata~\cite{EHZ10a,EHZ10b,DH11}. This extension is also rooted in
interactive Markov chains (\IMC{})~\cite{Her02}, another model with a
well-understood compositional theory. \IMC{}s are applied in a large
spectrum of practical applications, ranging from networked hardware on
chips~\cite{CHLS09} to water treatment facilities~\cite{HKRRS10} and
ultra-modern satellite designs~\cite{EKNPY12}.  The standard
analysis trajectory for \IMC{} revolves around compositional applications
of weak bisimulation minimization, a strategy that has been proven
very effective~\cite{HK00,BHHJPPRWB09,CHLS09}, and is based on a
polynomial time weak bisimulation decision
algorithm~\cite{Her02,WHHSB06}. Owed to the unavailability of effective
algorithms for \PA{} weak probabilistic bisimulations, this
compositional minimization strategy has thus far not been applied in
the \PA{} (or \MDP{}) setting. We aim at making this possible, and
furthermore, we intend to repeat and extend the successful
applications of \IMC{} in the extended Markov automata setting. For this,
a polynomial time decision procedures for weak probabilistic
bisimulation on \PA{} is the essential building block.

In this paper we show that \PA{} weak probabilistic bisimulation can
be decided in polynomial time, thus just as all other bisimulations on
\PA{}.  To arrive there, we provide a decision procedure that follows
the standard partition refinement approach \cite{CS02,KS90,PT87} and
that is based on a Linear Programming (LP) problem. The crucial step
is that we manage to generate and decide an LP problem that proves or
disproves the existence of a weak step in time polynomial in the size
of an automaton which in turn encodes a weak transition linear in its
size.  This enables us to decide in polynomial time whether the
defender has a matching weak transition step - opposed to the
exponential time required thus far~\cite{CS02} for this. Apart from this
result, which closes successfully the open problem of~\cite{CS02}, we
show how our LP approach can be extended to hyper-transitions (weak
transitions leaving a probability distribution instead of a single
state) and to the novel concepts of allowed weak/hyper-transitions
(weak/hyper-transitions involving only a restricted set of
transitions) and of equivalence matching (given two states, check
whether each one enables a weak transition matchable by the other).
Hyper-transitions naturally occur in weak probabilistic bisimulation
on Markov automata, and in the bisimulation formulation of
probabilistic forward simulation~\cite{EHZ10b,Seg95}.

\noindent \textbf{Organization of the paper.} 
After the preliminaries in Section~\ref{sec:Preliminaries}, we present in Section~\ref{sec:weakTransitionAsLPP} the polynomial LP problem that models weak transitions together with several extensions that can be computed in polynomial time as well. 
Then, in Section~\ref{sec:decision}, we recast the algorithm proposed in \cite{CS02} that decides whether two probabilistic automata are weak probabilistic bisimilar and we show that the decision procedure is polynomial.
We conclude the paper in Section~\ref{sec:conclusion} with some remarks, 
followed by appendixes containing all detailed proofs.

%%% Local Variables: 
%%% mode: latex
%%% TeX-master: "only-\PA{}-decisionProcedure"
%%% End: 

\section{Mathematical Preliminaries}
\label{sec:Preliminaries}

For a generic set $X$, denote by $\Disc{X}$ the set of discrete probability distributions over $X$, and by $\SubDisc{X}$ the set of discrete sub-probability distributions over $X$. 
Given $\gd \in \SubDisc{X}$, we denote by $\Supp{\gd}$ the set $\{x \in X \mid \probeval{\gd}{x} >0\}$, by $\probeval{\gd}{\bot}$ the value $1 - \probeval{\gd}{X}$ where $\bot \notin X$, and by $\dirac{x}$ the \emph{Dirac} distribution such that $\probeval{\gd}{x} = 1$ for $x \in X \cup \{\bot\}$.
For a sub-probability distribution $\gd$, we also write $\gd = \{p_x x \mid x \in X, p_x = \probeval{\gd}{x} \}$.
The lifting $\liftrel$ \cite{LS89}  of a relation $\rel \subseteq X \times Y$ is defined as follows: for $\gd_{X} \in \Disc{X}$ and $\gd_{Y} \in \Disc{Y}$, $\gd_{X} \liftrel \gd_{Y}$ holds if there exists a \emph{weighting function} $w \colon X \times Y \to [0,1]$ such that
\begin{inparaenum}[\scriptsize$(1)$]
% \begin{enumerate}
\item $w(x,y) > 0$ implies $x \rel y$,
\item $\sum_{y \in Y} w(x,y) = \gd_{X}(x)$, and
\item $\sum_{x \in X} w(x,y) = \gd_{Y}(y)$.
% \end{enumerate}
\end{inparaenum}
When $\rel$ is an equivalence relation on a set $X$, $\gd_{1} \liftrel \gd_{2}$ holds if for each $\equivclass \in \partitionset{X}$, $\probeval{\gd_{1}}{\equivclass} = \probeval{\gd_{2}}{\equivclass}$.

A Probabilistic Automaton (\PA{}) $\aut$ is a tuple $(\stateSet, \startState, \actionSet, \transitionRelation)$, where $\stateSet$ is a set of \emph{states}, $\startState \in \stateSet$ is the \emph{start state}, $\actionSet$ is the set of \emph{actions}, and $\transitionRelation \subseteq \stateSet \times \actionSet \times \Disc{\stateSet}$ is a \emph{probabilistic transition relation}.
The set $\actionSet$ is parted in two sets $\internalActionSet$ and $\externalActionSet$ of internal (hidden) and external actions, respectively;
we let $s$,$t$,$u$,$v$, and their variants with indices range over $\stateSet$, $a$, $b$ range over actions, and $\hidden$ range over hidden actions.
In this work we consider only finite \PA{}s, i.e., automata such that $\stateSet$ and $\transitionRelation$ are finite.

A transition $\tr = (s, a, \sd) \in \transitionRelation$, also denoted by $\strongTransition{s}{a}{\sd}$, is said to \emph{leave} from state $s$, to be \emph{labeled} by $a$, and to \emph{lead} to $\sd$, also denoted by $\sd_{\tr}$.
We denote by $\source{\tr}$ the \emph{source} state $s$, by $\action{\tr}$ the \emph{action} $a$, and by $\target{\tr}$ the \emph{target} distribution $\sd$.
We also say that $s$ enables action $a$, that action $a$ is enabled from $s$, and that $(s,a,\sd)$ is enabled from $s$.
Finally, we denote by $\transitionsFromState{s}$ the set of transitions enabled from $s$, i.e., $\transitionsFromState{s} = \{ \tr \in \transitionRelation \mid \source{\tr} = s \}$, and similarly by $\transitionsWithLabel{a}$ the set of transitions with action $a$, i.e., $\transitionsWithLabel{a} = \{ \tr \in \transitionRelation \mid \action{\tr} = a \}$.

An \emph{execution fragment} of a \PA{} $\aut$ is a finite or infinite sequence of alternating states and actions $\alpha = s_{0} a_{1} s_{1} a_{2} s_{2} \dots$ starting from a state $s_{0}$, also denoted by $\first{\alpha}$, and, if the sequence is finite, ending with a state, such that for each $i > 0$ there exists a transition $(s_{i-1}, a_{i}, \sd_{i}) \in \transitionRelation$ such that $\probeval{\sd_{i}}{s_{i}}> 0$.
If the sequence $\alpha$ is finite, then denote by $\last{\alpha}$ the last state of $\alpha$.
The \emph{length} of $\alpha$, denoted by $\length{\alpha}$, is the number of occurrences of actions in $\alpha$.
If $\alpha$ is infinite, then $\length{\alpha} = \infty$.
Denote by $\frags{\aut}$ the set of execution fragments of $\aut$ and by $\finiteFrags{\aut}$ the set of finite execution fragments of $\aut$.
An execution fragment $\alpha$ is a \emph{prefix} of an execution fragment $\alpha'$, denoted by $\alpha \prefix \alpha'$, if the sequence $\alpha$ is a prefix of the sequence $\alpha'$.
The \emph{trace} of $\alpha$, denoted by $\trace{\alpha}$, is the sub-sequence of external actions of $\alpha$.
For instance, for $a \in \externalActionSet$, $\trace{s_{0}a s_{1}} = \trace{s_{0} \hidden s_{1} \hidden \dots \hidden s_{n-1} a s_{n}} = a$, also denoted by $\trace{a}$, and $\trace{s_{0}} = \trace{s_{0} \hidden s_{1} \hidden \dots \hidden s_{n}} = \emptytrace$, the empty sequence, also denoted by $\trace{\hidden}$.

A \emph{scheduler} for a \PA{} $\aut$ is a function $\sched \colon \finiteFrags{\aut} \to \SubDisc{\transitionRelation}$ such that for each finite execution fragment $\alpha$, $\schedeval{\sched}{\alpha} \in \SubDisc{\transitionsFromState{\last{\alpha}}}$.
A scheduler is \emph{determinate} \cite{CS02} if for each pair of execution fragments $\alpha$, $\alpha'$, if $\trace{\alpha} = \trace{\alpha'}$ and $\last{\alpha} = \last{\alpha'}$, then $\schedeval{\sched}{\alpha} = \schedeval{\sched}{\alpha'}$.
Given a scheduler $\sched$ and a finite execution fragment $\alpha$,
the distribution $\schedeval{\sched}{\alpha}$ describes how
transitions are chosen to move on from $\last{\alpha}$.
% The resulting combined transition is the combined transition according to $\schedeval{\sched}{\alpha}$.
% We denote by $\sd_{\schedeval{\sched}{\alpha}}$ the corresponding target distribution.
% 
A scheduler $\sched$ and a state $s$ induce a probability distribution $\sd_{\sched,s}$ over execution fragments as follows.
The basic measurable events are the cones of finite execution fragments, where the cone of a finite execution fragment $\alpha$, denoted by $\cone{\alpha}$, is the set $\{ \alpha' \in \finiteFrags{\aut} \mid \alpha \prefix \alpha' \}$.
The probability $\sd_{\sched,s}$ of a cone $\cone{\alpha}$ is defined recursively as follows:
\[
 \probeval{\sd_{\sched,s}}{\cone{\alpha}} =
 \begin{cases}
    0 & \text{if $\alpha = t$ for a state $t \neq s$,} \\
    1 & \text{if $\alpha = s$,} \\
    \probeval{\sd_{\sched,s}}{\cone{\alpha'}} \cdot \sum_{\tr \in \transitionsWithLabel{a}} \probeval{\schedeval{\sched}{\alpha'}}{\tr} \cdot  \probeval{\sd_{\tr}}{t}
        & \text{if $\alpha = \alpha'a t$.}
 \end{cases}
\]
Standard measure theoretical arguments ensure that $\sd_{\sched,s}$ extends uniquely to the $\sigma$-field generated by cones.
We call the measure $\sd_{\sched,s}$ a \emph{probabilistic execution fragment} of $\aut$ and we say that it is generated by $\sched$ from $s$.
% If $s = \startState$, we call the measure $\sd_{\sched,\startState}$ a \emph{probabilistic execution} $\sd_{\sched}$ of $\aut$ and we say that it is generated by $\sched$ from $\startState$.
Given a finite execution fragment $\alpha$, we define $\probeval{\sd_{\sched,s}}{\alpha}$ as $\probeval{\sd_{\sched,s}}{\alpha} = \probeval{\sd_{\sched,s}}{\cone{\alpha}} \cdot \probeval{\schedeval{\sched}{\alpha}}{\bot}$, where $\probeval{\schedeval{\sched}{\alpha}}{\bot}$ is the probability of chosing no transitions, i.e., of terminating the computation after $\alpha$ has occurred.

We say that there is a \emph{weak combined transition} from $s \in \stateSet$ to $\sd \in \Disc{\stateSet}$ labeled by $a \in \actionSet$ that is induced by $\sched$, denoted by $\weakCombinedTransition{s}{a}{\sd}$, if there exists a scheduler $\sched$ such that the following holds for the induced probabilistic execution fragment $\sd_{\sched,s}$:
\begin{inparaenum}[\scriptsize$(1)$]
% \begin{enumerate}
\item $\probeval{\sd_{\sched,s}}{\finiteFrags{\aut}} = 1$;
\item for each $\alpha \in \finiteFrags{\aut}$, if $\probeval{\sd_{\sched,s}}{\alpha} > 0$ then $\trace{\alpha} = \trace{a}$;
\item for each state $t$, $\probeval{\sd_{\sched,s}}{\{ \alpha \in \finiteFrags{\aut} \mid \last{\alpha} = t \}} = \probeval{\sd}{t}$.
% \end{enumerate}
\end{inparaenum}
See \cite{Seg06} for more details on weak combined transitions.

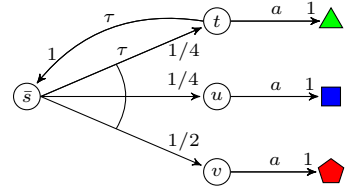
\begin{wrapfigure}[7]{r}{6cm}
\vspace{-3mm}

\begin{tikzpicture}[->,>=stealth',shorten >=1pt,auto]
\scriptsize
\path[use as bounding box] (-0.5,-1.5) rectangle (4.5,0.5);
\tikzstyle{every state}=[fill=none,draw=black,text=black,shape=circle]

\node[state] (s0) at (0,0) {$\startState$};
\node[state] (s1) at (2.5,1) {$t$};
\node[state] (s2) at (2.5,0) {$u$};
\node[state] (s3) at (2.5,-1) {$v$};
\node[state,class green] (green) at (4,1) {};
\node[state,class blue] (blue) at (4,0) {};
\node[state,class red] (red) at (4,-1) {};

\draw[name path=t1label] (s0.east) to node [above] {$\hidden$} (s1);
\draw[name path=t1] (s0.east) to node [below, very near end] {$1/4$} (s1);
\draw[name path=t2] (s0.east) to node [above, very near end] {$1/4$} (s2);
\draw (s1.west) to [bend right] node [above, very near end] {$1$} (s0);
\draw (s1.west) to [bend right] node [above] {$\hidden$} (s0);
\draw[name path=t3] (s0.east) to node [above, very near end] {$1/2$} (s3);
\draw (s1.east) to node [above, very near end] {$1$} (green);
\draw (s1.east) to node [above] {$a$} (green);
\draw (s2.east) to node [above, very near end] {$1$} (blue);
\draw (s2.east) to node [above] {$a$} (blue);
\draw (s3.east) to node [above, very near end] {$1$} (red);
\draw (s3.east) to node [above] {$a$} (red);

\path[name path=c] (s0) circle (1.25);
\draw[name intersections={of=t3 and c, by=i3},
      name intersections={of=t1 and c, by=i1},
      -,shorten >=0pt]
  (i3) to[bend right] (i1); % well, it isn't an arc
\end{tikzpicture}
\vspace{-5mm}
\caption{The probabilistic automaton $\aut[E]$}
\label{fig:weakTransExample}
\end{wrapfigure}
% AT: dirty trick to avoid problems with wrapfigure...
\vskip 1\baselineskip
\noindent \textit{Example~1.\hspace{1ex}}
Consider the automaton $\aut[E]$ depicted in Figure~\ref{fig:weakTransExample} and denote by $\tr$ the only transition enabled by $\startState$; $\aut[E]$ enables the weak combined transition $\weakCombinedTransition{\startState}{a}{\sd}$ where $\sd = \{\frac{1}{16}\stategreen, \frac{5}{16}\stateblue, \frac{10}{16}\statered\}$ via the scheduler $\sched$ defined as follows: $\schedeval{\sched}{\startState} = \schedeval{\sched}{\startState \hidden t \hidden \startState} = \dirac{\tr}$, $\schedeval{\sched}{\startState \hidden t} = \dirac{\strongTransition{t}{\hidden}{\dirac{\startState}}}$, $\schedeval{\sched}{\startState \hidden u} = \schedeval{\sched}{\startState \hidden t \hidden \startState \hidden u} = \dirac{\strongTransition{u}{a}{\dirac{\stateblueInSubscript}}}$,
$\schedeval{\sched}{\startState \hidden v} = \schedeval{\sched}{\startState \hidden t \hidden \startState \hidden v} = \dirac{\strongTransition{v}{a}{\dirac{\stateredInSubscript}}}$,
$\schedeval{\sched}{\startState \hidden t \hidden \startState \hidden t} = \dirac{\strongTransition{t}{a}{\dirac{\stategreenInSubscript}}}$, and 
$\schedeval{\sched}{\alpha} = \dirac{\bot}$ for each other finite execution fragment $\alpha$.
For instance, state $\stateblue$ is reached with probability $\probeval{\sd_{\sched',\startState}}{\{ \alpha \in \finiteFrags{\aut[E]} \mid \last{\alpha} = \stateblue \}} = \probeval{\sd_{\sched',\startState}}{\{\startState \hidden u a \stateblue, \startState \hidden t \hidden \startState \hidden u a \stateblue\}} = 1 \cdot \frac{1}{4} \cdot 1 \cdot 1 \cdot 1 + 1 \cdot \frac{1}{4} \cdot 1 \cdot 1 \cdot \frac{1}{4} \cdot 1 \cdot 1 \cdot 1 = \frac{5}{16} = \probeval{\sd}{\stateblue}$, as required.
\vskip 1\baselineskip

We say that there is a \emph{hyper-transition} from $\gd \in \Disc{\stateSet}$ to $\sd \in \Disc{\stateSet}$ labeled by $a \in \actionSet$, denoted by $\hyperWeakCombinedTransition{\gd}{a}{\sd}$, if there exists a family of weak combined transitions $\family{\weakCombinedTransition{s}{a}{\sd_{s}}}{s \in \Supp{\gd}}$ such that $\sd = \sum_{s \in \Supp{\gd}} \probeval{\gd}{s} \cdot \sd_{s}$, i.e., for each $t \in \stateSet$, $\probeval{\sd}{t} = \sum_{s \in \Supp{\gd}} \probeval{\gd}{s} \cdot \probeval{\sd_{s}}{t}$.

\begin{definition}
Let $\aut_{1}$, $\aut_{2}$ be two probabilistic automata.  An
equivalence relation $\rel$ on the disjoint union $\stateSet_{1}
\uplus \stateSet_{2}$ is a \emph{weak probabilistic bisimulation} if,
for each pair of states $s,t \in \stateSet_{1} \uplus \stateSet_{2}$
such that $s \rel t$, if $\strongTransition{s}{a}{\sd_{s}}$ for some
probability distribution $\sd_{s}$, then there exists a probability
distribution $\sd_{t}$ such that
$\weakCombinedTransition{t}{a}{\sd_{t}}$ and $\sd_{s} \liftrel
\sd_{t}$.

Two probabilistic automata $\aut_{1}$ and $\aut_{2}$ are weakly
probabilistic bisimilar if there exists a weak probabilistic
bisimulation $\rel$ on $\stateSet_{1} \uplus \stateSet_{2}$ such that
$\startState_{1} \rel \startState_{2}$.  We denote the coarsest weak
probabilistic bisimulation by $\weakBisim$, and call it weak
probabilistic bisimilarity.
\end{definition}

This is the central definition around which the paper revolves. Weak
probabilistic bisimilarity is an equivalence relation preserved by
standard process algebraic composition operators on \PA{} \cite{PS04}.  The
definition of bisimulation can be reformulated as follows, by simple manipulation of quantifiers:

\begin{definition}
Given two
\PA{}s $\aut_{1}$, $\aut_{2}$, an equivalence relation $\rel$ on
$\stateSet_{1} \uplus \stateSet_{2}$ is a \emph{weak probabilistic
  bisimulation} if, for each transition $(s,a,\sd_{s}) \in
\transitionRelation_{1} \uplus \transitionRelation_{2}$ and each state
$t$ such that $s \rel t$, there exists $\sd_{t}$ such that
$\weakCombinedTransition{t}{a}{\sd_{t}}$ and $\sd_{s} \liftrel
\sd_{t}$.
\end{definition}

%%% Local Variables: 
%%% mode: latex
%%% TeX-master: "pa-weak-bisim-poly"
%%% End: 

\section{Weak Transition Construction as a Linear Programming Problem}
\label{sec:weakTransitionAsLPP}
We now discuss key elements of a decision algorithm for weak
probabilistic bisimilarity.  As we will see, the core ingredient - and
the source of the exponential complexity of the decision algorithm
of~\cite{CS02} - is the recurring need to verify the step condition,
that is, given a challenging transition $\strongTransition{s}{a}{\sd}$
and $(s,t) \in \rel$, to check whether there exists a weak combined
transition $\weakCombinedTransition{t}{a}{\sd_{t}}$ such that $\sd
\liftrel \sd_{t}$.

With some inspiration from network flow problems, we will be able to
see a transition $\weakCombinedTransition{t}{a}{\sd_{t}}$ of the \PA{}
$\aut$ as a \emph{flow} where the initial probability mass $\dirac{t}$
flows and splits along internal transitions (and exactly one
transition with label $a$ for each stream when $a \neq \hidden$)
accordingly to the transition target distributions and the resolution
of the nondeterminism performed by the scheduler.

This will allow us to arrive at a polynomial time algorithm to verify
or refute the existence of a weak combined transition
$\weakCombinedTransition{t}{a}{\sd_{t}}$ such that $\sd \liftrel
\sd_{t}$.  This is the core ingredient of an efficient algorithm for
deciding weak probabilistic bisimilarity, stated in
Section~\ref{sec:decision},

\subsection{Allowed Transitions}
\label{ssec:AllowedTransitions}

For the construction we are going to develop, we consider a more
general case where we parametrize the scheduler so as to choose only
specific, allowed, transitions when resolving the nondeterministic
choices in a weak combined transition. This generalization will later
be exploited by enabling us to generate tailored and thereby smaller
LP-problems.  

For the intuition of this generalization, consider, for example, an
automaton $\aut[C]$ that models a communication channel: it receives
the information to transmit from the sender through an external
action, then it performs an internal transition to represent the
sending of the message on the communication channel, and finally it
sends the transmitted information to the receiver.  The communication
channel is chosen nondeterministically between a reliable channel and
an acknowledged lossy channel.  If we want to check whether $\aut[C]$
always ensures the correct transmission of the received information,
we can restrict the scheduler to choose only the lossy channel, i.e.,
we \emph{allow} only the transitions relative to the lossy channel; if
we impose this restriction and $\aut[C]$ is able to send eventually
the transmitted information to the receiver with probability $1$, then
we can say that $\aut[C]$ always ensures the correct transmission of
the received information.

\begin{definition}[Allowed weak combined transition]
Given a \PA{} $\aut$ and a set of \emph{allowed} transitions $\allowedTransitions \subseteq \transitionRelation$, we say that there is an \emph{allowed weak combined transition} from $s$ to $\sd$ with label $a$ respecting $\allowedTransitions$, denoted by $\weakCombinedAllowedTransition{s}{a}{\sd}{\allowedTransitions}$, if there exists a scheduler $\sched$ that induces $\weakCombinedTransition{s}{a}{\sd}$ such that for each $\alpha \in \finiteFrags{\aut}$, $\Supp{\schedeval{\sched}{\alpha}} \subseteq \allowedTransitions$.
\end{definition}
It is immediate to see that, when we consider every transition as allowed, i.e., $\allowedTransitions = \transitionRelation$, the allowed weak combined transition $\weakCombinedAllowedTransition{s}{a}{\sd}{\transitionRelation}$ is just the usual weak combined transition $\weakCombinedTransition{s}{a}{\sd}$.
\begin{proposition}
\label{pro:weakAllowedTransitionEquivalentToWeakTransitionWhenAllTransitionsAllowed}
	Given a \PA{} $\aut$, a state $s$, and action $a$, and a probability distribution $\sd \in \Disc{\stateSet}$, there exists a scheduler $\sched_{\transitionRelation}$ for $\aut$ that induces $\weakCombinedAllowedTransition{s}{a}{\sd}{\transitionRelation}$ if and only if there exists a scheduler $\sched$ for $\aut$ that induces $\weakCombinedTransition{s}{a}{\sd}$.
\end{proposition}

Similarly, we say that there is an \emph{allowed hyper-transition} from a distribution over states $\gd$ to a distribution over states $\sd$ labeled by $a$ respecting $\allowedTransitions$, denoted by $\hyperWeakCombinedAllowedTransition{\gd}{a}{\sd}{\allowedTransitions}$, if there exists a family of allowed weak combined transitions $\family{\weakCombinedAllowedTransition{s}{a}{\sd_{s}}{\allowedTransitions}}{s \in \Supp{\gd}}$ such that $\sd = \sum \probeval{\gd}{s} \cdot \sd_{s}$.

An equivalent definition of allowed hyper-transition $\hyperWeakCombinedAllowedTransition{\gd}{a}{\sd}{\allowedTransitions}$ is the following: given a \PA{} $\aut$, we say that there is an \emph{allowed hyper-transition} from a distribution over states $\gd$ to a distribution over states $\sd$ labeled by $a$ respecting $\allowedTransitions$ if there exists an allowed weak combined transition $\weakCombinedAllowedTransition{h}{a}{\sd}{\allowedTransitions_{h}}$ for the \PA{} $\aut_{h} = (\stateSet \cup \{h\}, \startState, \actionSet, \transitionRelation \cup \{\strongTransition{h}{\hidden}{\gd}\})$ where $h \notin \stateSet$ and $\allowedTransitions_{h} = \allowedTransitions \cup \{\strongTransition{h}{\hidden}{\gd}\}$.
\begin{proposition}
\label{pro:hyperTransitionEquivalentToWeakTransition}
	Given a \PA{} $\aut$, $h \notin \stateSet$, $a \in \actionSet$, $\allowedTransitions \subseteq \transitionRelation$, and $\gd,\sd \in \Disc{\stateSet}$, let $\aut_{h}$ be the \PA{} $\aut_{h} = (\stateSet \cup \{h\}, \startState, \actionSet, \transitionRelation \cup \{\strongTransition{h}{\hidden}{\gd}\})$ and $\allowedTransitions_{h}$ be $\allowedTransitions \cup \{\strongTransition{h}{\hidden}{\gd}\}$.
	
	$\hyperWeakCombinedAllowedTransition{\gd}{a}{\sd}{\allowedTransitions}$ exists in $\aut$ if and only if $\weakCombinedAllowedTransition{h}{a}{\sd}{\allowedTransitions_{h}}$ exists in $\aut_{h}$.
\end{proposition}

% AT: same dirty trick as in preliminaries...
% \vskip 1\baselineskip
\noindent \textit{Example~1 (cont.).\hspace{1ex}}
If we consider again the automaton $\aut[E]$ in Figure~\ref{fig:weakTransExample} and the set of allowed transitions $\allowedTransitions = \transitionRelation \setminus \{\strongTransition{t}{\hidden}{\dirac{\startState}}\}$, it is immediate to see that the weak combined transition $\weakCombinedTransition{\startState}{a}{\sd}$ where $\sd = \{\frac{1}{16}\stategreen, \frac{5}{16}\stateblue, \frac{10}{16}\statered\}$ is not an allowed weak combined transition respecting $\allowedTransitions$ and that the only allowed weak combined transition with label $a$ enabled by $\startState$ is $\weakCombinedAllowedTransition{\startState}{a}{\gd}{\allowedTransitions}$ having $\gd = \{\frac{1}{4}\stategreen, \frac{1}{4}\stateblue, \frac{1}{2}\statered\}$ as target distribution.
\vskip 1\baselineskip

\subsection{A Linear Programming Problem}
\label{ssec:LPProblem}

We now assume we are given the \PA{} $\aut$, the set of allowed transitions $\allowedTransitions \subseteq \transitionRelation$, the state $t$, the action $a$, the probability distribution $\sd$, and the equivalence relation $\rel$ on $\stateSet$. We intend to verify or refute the existence of a  weak combined transition $\weakCombinedAllowedTransition{t}{a}{\sd_{t}}{\allowedTransitions}$ of $\aut$  satisfying $\sd
\liftrel \sd_{t}$ via the construction  of a flow through  the network graph $\NetworkTBetaMuTransRel{t}{a}{\sd}{\allowedTransitions}{\rel} = (V,E)$ defined as follows:

\begin{definition}
Given the \PA{} $\aut$, the set of allowed transitions $\allowedTransitions \subseteq \transitionRelation$, the state $t$, the action $a$, the probability distribution $\sd$, and the equivalence relation $\rel$ on $\stateSet$, we define the network graph $\NetworkTBetaMuTransRel{t}{a}{\sd}{\allowedTransitions}{\rel} = (V,E)$ relative to $\weakCombinedAllowedTransition{t}{a}{\sd_{t}}{\allowedTransitions}$ of $\aut$ as follows:
for $a \neq \hidden$, the set of vertices is 
\[
	V = \{\netsource, \netsink\} \cup \stateSet \cup \stateSet^{\tr} \cup \stateSet_{a} \cup \stateSet^{\tr}_{a} \cup (\partitionset{\stateSet})
\]
where 
\[
\begin{array}{lll}
\stateSet^{\tr} & = & \{v^{\tr} \mid \tr = \strongTransition{v}{b}{\rho} \in \allowedTransitions, b \in \{a, \hidden\}\}\text{,} \\
\stateSet_{a} & = & \{v_{a} \mid v \in \stateSet\}\text{, and} \\
\stateSet^{\tr}_{a} & = & \{v^{\tr}_{a} \mid v^{\tr} \in \stateSet^{\tr}\}
\end{array}
\]
and the set of arcs is
\[
\begin{array}{rl}
E = & \{(\netsource,t)\} \cup \{(v_{a},\equivclass), (\equivclass,\netsink) \mid \equivclass \in \partitionset{\stateSet}, v \in \equivclass\} \\
& \cup \{(v,v^{\tr}), (v^{\tr}, v'), (v_{a},v^{\tr}_{a}), (v^{\tr}_{a},v'_{a}) \mid \tr = \strongTransition{v}{\hidden}{\rho} \in \allowedTransitions, v' \in \Supp{\rho}\} \\
& \cup \{(v,v^{\tr}_{a}), (v^{\tr}_{a}, v'_{a}) \mid \tr = \strongTransition{v}{a}{\rho} \in \allowedTransitions, v' \in \Supp{\rho}\}\text{.}
\end{array}
\]
\end{definition}
For $a = \hidden$ the definition is similar: $V = \{\netsource,\netsink\} \cup \stateSet \cup \stateSet^{\tr}\cup (\partitionset{\stateSet})$ and $E = \{(\netsource,t)\} \cup \{(v,\equivclass),(\equivclass,\netsink) \mid \equivclass \in \partitionset{\stateSet}, v \in \equivclass\} \cup \{(v,v^{\tr}), (v^{\tr}, v') \mid \tr = \strongTransition{v}{\hidden}{\rho} \in \allowedTransitions, v' \in \Supp{\rho}\}$.

$\netsource$ and $\netsink$ are two vertices that represent the source and the sink of the network, respectively. The graph encodes possible sequences of internal transitions, keeping track of which transition has happened by means of the vertices superscripted with $tr$, for this the set $\stateSet^{\tr}$ contains vertices that model the transitions of the automaton.  The subsets of vertices subscripted by $a$ are used to record that action $a$ has happened already.  Notably, not every vertex is used for defining arcs: the vertices $v^{\tr}$ where $\tr = \strongTransition{v}{b}{\rho} \in \allowedTransitions$ and $b = a \neq \hidden$ are used only to define the corresponding vertices $v^{\tr}_{a}$ that are actually involved in the definition of the set $E$ of arcs.
We could have removed these vertices from $\stateSet^{\tr}$ but this reduces the readability of the definition of $\stateSet^{\tr}_{a}$ without giving us a valuable effect on the computational complexity of the proposed solution.

% AT: same dirty trick as in preliminaries...
\vskip 1\baselineskip
\noindent \textit{Example~1 (cont.).\hspace{1ex}}
Consider the automaton $\aut[E]$ in Figure~\ref{fig:weakTransExample} and suppose that we want to check whether there exists an allowed weak combined transition $\weakCombinedAllowedTransition{\startState}{a}{\gd}{\transitionRelation}$ such that $\gd \liftrel \sd$ where $\sd = \{\frac{1}{16}\stategreen, \frac{5}{16}\stateblue, \frac{10}{16}\statered\}$ and the classes induced by $\rel$ are $\{\{\startState, t, u, v\}, \{\stategreen\}, \{\stateblue\}, \{\statered\} \}$.
Let $\tr_{0} = \strongTransition{\startState}{\hidden}{\{\frac{1}{4} t, \frac{1}{4} u, \frac{1}{2} v\}}$, $\tr_{1} = \strongTransition{t}{a}{\dirac{\stategreenInSubscript}}$, $\tr_{2} = \strongTransition{u}{a}{\dirac{\stateblueInSubscript}}$, $\tr_{3} = \strongTransition{v}{a}{\dirac{\stateredInSubscript}}$, and $\tr_{4} = \strongTransition{t}{\hidden}{\dirac{\startState}}$.
The network $\NetworkTBetaMuTransRel{\startState}{a}{\sd}{\transitionRelation}{\rel}$ is as follows, where we omit vertices $\stategreen$, $\stateblue$, and $\statered$ since they are not involved in any arc. Numbers attached to arcs indicate probabilities, and are not part of the graph.

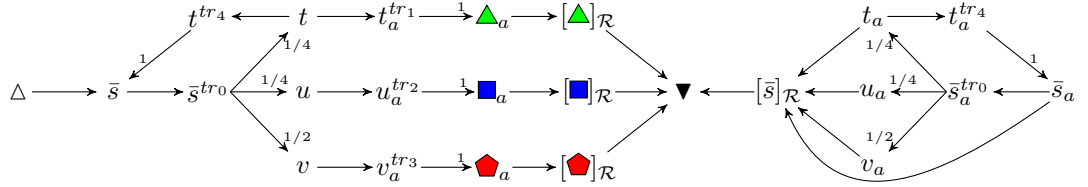
\begin{tikzpicture}[->,>=stealth',shorten >=1pt,auto]
\centering
\path[use as bounding box] (-0.85,-1.15) rectangle (6.5,1.35);
\tikzstyle{every state}=[fill=none,draw=black,text=black,shape=circle]

\node[vertex] (source) at (-1.25,0) {$\netsource$};
\node[vertex] (s) at (0,0) {$\startState$};
\node[vertex] (str0) at (1.25,0) {$\startState^{\tr_{0}}$};
\node[vertex] (t) at (2.5,1) {$t$};
\node[vertex] (u) at (2.5,0) {$u$};
\node[vertex] (v) at (2.5,-1) {$v$};
\node[vertex] (ttr4) at (1.25,1) {$t^{\tr_{4}}$};
\node[vertex] (tatr1) at (3.75,1) {$t^{\tr_{1}}_{a}$};
\node[vertex] (uatr2) at (3.75,0) {$u^{\tr_{2}}_{a}$};
\node[vertex] (vatr3) at (3.75,-1) {$v^{\tr_{3}}_{a}$};
\node[vertex] (greena) at (5,1) {$\stategreen_{a}$};
\node[vertex] (bluea) at (5,0) {$\stateblue_{a}$};
\node[vertex] (reda) at (5,-1) {$\statered_{a}$};
\node[vertex] (cgreena) at (6.25,1) {$\relclass{\stategreen}{\rel}$};
\node[vertex] (cbluea) at (6.25,0) {$\relclass{\stateblue}{\rel}$};
\node[vertex] (creda) at (6.25,-1) {$\relclass{\statered}{\rel}$};
\node[vertex] (sink) at (7.5,0) {$\netsink$};

\node[vertex] (cothersa) at (8.75,0) {$\relclass{\startState}{\rel}$};

\node[vertex] (sa) at (12.5,0) {$\startState_{a}$};
\node[vertex] (satr0) at (11.25,0) {$\startState^{\tr_{0}}_{a}$};
\node[vertex] (ta) at (10,1) {$t_{a}$};
\node[vertex] (ua) at (10,0) {$u_{a}$};
\node[vertex] (va) at (10,-1) {$v_{a}$};
\node[vertex] (tatr4) at (11.25,1) {$t^{\tr_{4}}_{a}$};

\tiny
\draw (source) to node {} (s);
\draw (s) to node {} (str0);
\draw (str0.east) to node [right, near end] {~$1/4$} (t);
\draw (str0.east) to node [above, near end] {$1/4$} (u);
\draw (str0.east) to node [right, near end] {~$1/2$} (v);
\draw (t) to node {} (ttr4);
\draw (ttr4) to node [above, near end] {$1$} (s);
\draw (t) to node {} (tatr1);
\draw (u) to node {} (uatr2);
\draw (v) to node {} (vatr3);
\draw (tatr1) to node [above, near end] {$1$} (greena);
\draw (uatr2) to node [above, near end] {$1$} (bluea);
\draw (vatr3) to node [above, near end] {$1$} (reda);
\draw (greena) to node {} (cgreena);
\draw (bluea) to node {} (cbluea);
\draw (reda) to node {} (creda);
\draw (cgreena) to node {} (sink);
\draw (cbluea) to node {} (sink);
\draw (creda) to node {} (sink);

\draw (cothersa) to node {} (sink);

\draw (sa) to node {} (satr0);
\draw (satr0.west) to node [left, near end] {$1/4$~~} (ta);
\draw (satr0.west) to node [above, near end] {$1/4$} (ua);
\draw (satr0.west) to node [left, near end] {$1/2$~~} (va);
\draw (ta) to node {} (tatr4);
\draw (tatr4) to node [above, near end] {$1$} (sa);
\draw (ta) to node {} (cothersa);
\draw (ua) to node {} (cothersa);
\draw (va) to node {} (cothersa);
\draw (sa) .. controls (10.75,-1.25) and (9.5,-1.75) .. (cothersa);
\end{tikzpicture}

Our intention is to use the network
$\NetworkTBetaMuTransRel{t}{a}{\sd}{\allowedTransitions}{\rel}$, in a
maximum flow problem, since solving the latter has polynomial
complexity.  Unfortunately, the resulting problem does not model an allowed
weak combined transition because probabilities are as such not
necessarily respected: In ordinary flow problems we can not enforce a
proportional balancing between the flows out of a given
vertex. Instead, the entire incoming flow might be sent over a single
outgoing arc, provided that the arc capacity is respected, while zero
flow is sent over other arcs. In particular, we have no way to force
the flow to split proportionally to the target probability
distribution of a transition when the flow is less than $1$.  Apart
from that, there is no obvious way to assign arc capacities since imposing capacity $1$ to arcs is not always correct even if this is the maximum value for a probability. This
problem is specifically caused by cycles of internal transitions. For
self loops like $\strongTransition{s}{\hidden}{\gd}$ with $\probeval{\gd}{s} > 0$, one might after
some reflection come up with a capacity $1/(1-p)$ where $p =
\probeval{\gd}{s}$, but this does not extend to arbitrary
$\hidden$-connected components.

For these reasons, we have to proceed differently: Since any maximum
flow problem can be expressed as a Linear Programming (LP) problem, we
follow this path, but then refine the LP problem further, in order to
eventually define a maximization problem whose solution is indeed
equivalent to an allowed weak combined transition, as we will show in
Section~\ref{ssec:LPequivalentToWeakTr}.  For this, we use the above
transformation of the automaton into a network graph as the starting
point for generating an LP problem, which is afterwards enriched with
additional constraints: We adopt the same notation of the max flow
problem so we use $f_{u,v}$ to denote the ``flow'' through the arc
from $u$ to $v$.  The \emph{balancing factor} is a new concept we
introduce to model a probabilistic choice and to ensure a balancing
between flows that leave a vertex representing a probabilistic choice,
i.e., leaving a vertex $v \in \stateSet^{\tr} \cup
\stateSet^{\tr}_{a}$.

\begin{definition}[The $\LPproblemTBetaMuTransRel{t}{a}{\sd}{\allowedTransitions}{\rel}$ LP problem]
For $a \neq \hidden$ we define the $\LPproblemTBetaMuTransRel{t}{a}{\sd}{\allowedTransitions}{\rel}$ LP problem associated to the network graph $(V,E) = \NetworkTBetaMuTransRel{t}{a}{\sd}{\allowedTransitions}{\rel}$ as follows:
\[
\begin{array}{lll}
		\max \sum_{(x,y) \in E} -f_{x,y} \\
	\text{under constraints} & \\ 	
		\multicolumn{2}{l}{f_{u,v} \geq 0 \qquad} & \text{for each $(u,v) \in E$}\\
		f_{\netsource,t} = 1 \qquad & \\
		\multicolumn{2}{l}{f_{\equivclass,\netsink} = \probeval{\sd}{\equivclass} \qquad} & \text{for each $\equivclass \in \partitionset{\stateSet}$}\\
		\multicolumn{2}{l}{\sum_{u \in \{x \mid (x,v) \in E\}} f_{u,v} - \sum_{u \in \{y \mid (v,y) \in E\}} f_{v,u} = 0 \qquad} & \text{for each $v \in V \setminus \{\netsource,\netsink\}$} \\
		f_{v^{\tr},v'} - \probeval{\rho}{v'} \cdot f_{v,v^{\tr}} = 0 \qquad & \multicolumn{2}{r}{\text{for each $\tr = \strongTransition{v}{\hidden}{\rho} \in \allowedTransitions$ and $v' \in \Supp{\rho}$}}\\
		f_{v^{\tr}_{a},v'_{a}} - \probeval{\rho}{v'} \cdot f_{v_{a},v^{\tr}_{a}} = 0 \qquad & \multicolumn{2}{r}{\text{for each $\tr = \strongTransition{v}{\hidden}{\rho} \in \allowedTransitions$ and $v' \in \Supp{\rho}$}}\\
		f_{v^{\tr}_{a},v'_{a}} - \probeval{\rho}{v'} \cdot f_{v,v^{\tr}_{a}} = 0 \qquad & \multicolumn{2}{r}{\text{for each $\tr = \strongTransition{v}{a}{\rho} \in \allowedTransitions$ and $v' \in \Supp{\rho}$}}
\end{array}
\]
\end{definition}

The constraints as $\sum_{u \in \{x \mid (x,v) \in E\}} f_{u,v} - \sum_{u \in \{y \mid (v,y) \in E\}} f_{v,u} = 0$ for $v \in V \setminus \{\netsource,\netsink\}$ are also known as \emph{conservation of the flow} constraints.
When $a$ is $\hidden$, the LP problem $\LPproblemTBetaMuTransRel{t}{\hidden}{\sd}{\allowedTransitions}{\rel}$ associated to $\NetworkTBetaMuTransRel{t}{\hidden}{\sd}{\allowedTransitions}{\rel}$ is defined as above without the last two groups of constraints.
Note that the constraints of $\LPproblemTBetaMuTransRel{t}{a}{\sd}{\allowedTransitions}{\rel}$ define a system of linear equations extended with the non-negativity of variables $f_{u,v}$ and this rules out solutions where some variable $f_{x,y}$ has an infinite value.
Moreover this may be used to improve the actual implementation of the solver.

We can define the objective function in several ways but this does not affect the equivalence of $\LPproblemTBetaMuTransRel{t}{a}{\sd}{\allowedTransitions}{\rel}$ and allowed weak combined transitions:
in fact, the equivalence is based on variables $f_{v_{a},\relclass{v}{\rel}}$ and $f_{\equivclass,\netsink}$ (where $v \in \stateSet$ and $\equivclass \in \partitionset{\stateSet}$) that represent the probability to reach each state $v$ (and then stopping) and each equivalence class $\equivclass$, respectively;
by definition of $\LPproblemTBetaMuTransRel{t}{a}{\sd}{\allowedTransitions}{\rel}$ we have that $\sum_{v \in \equivclass} f_{v_{a},\equivclass} = f_{\equivclass,\netsink}$ and $f_{\equivclass,\netsink} = \probeval{\sd}{\equivclass}$, thus their value does not strictly depend on the objective function.
When $a = \hidden$, we have the same result, just replacing $v_{a}$ with $v$.

The objective function we use allows us to rule out trivial self-loops: suppose that there exists a transition $\tr = \strongTransition{x}{\hidden}{\dirac{x}} \in \allowedTransitions$ that we model by arcs $(x, x^{\tr})$ and $(x^{\tr}, x)$.
The balancing constraint for such arcs is $f_{x^{\tr},x} - 1 \cdot f_{x,x^{\tr}} = 0$ that is satisfied for each value of $f_{x^{\tr},x} = f_{x,x^{\tr}}$;
however, the maximum for the objective function can be reached only when $f_{x,x^{\tr}} = 0$, that is, the self-loop is not used.
Similarly, we obtain that the value of the flow involving vertices that can not be reached from the vertex $t$ is null as well as when such vertices may be reached from $t$ but the solution of the problem requires that the flow from the vertex $t$ to them is null.

It is worthwhile to point out that the objective function $\max \sum_{(x,y) \in E} -f_{x,y}$ is actually equivalent to $\min \sum_{(x,y) \in E} f_{x,y}$, i.e., a weak transition can also be seen as a minimum cost flow problem plus balancing constraints.

% AT: same dirty trick as in preliminaries...
\vskip 1\baselineskip
\noindent \textit{Example~1 (cont.).\hspace{1ex}}
Consider again the automaton $\aut[E]$ in Figure~\ref{fig:weakTransExample} and suppose that we want to check whether there exists an allowed weak combined transition $\weakCombinedAllowedTransition{\startState}{a}{\gd}{\transitionRelation}$ such that $\gd \liftrel \sd$ where $\sd = \{\frac{1}{16}\stategreen, \frac{5}{16}\stateblue, \frac{10}{16}\statered\}$ and the classes induced by $\rel$ are $\{\{\startState, t, u, v\}, \{\stategreen\}, \{\stateblue\}, \{\statered\} \}$.
Let $\tr_{0} = \strongTransition{\startState}{\hidden}{\{\frac{1}{4} t, \frac{1}{4} u, \frac{1}{2} v\}}$, $\tr_{1} = \strongTransition{t}{a}{\dirac{\stategreenInSubscript}}$, $\tr_{2} = \strongTransition{u}{a}{\dirac{\stateblueInSubscript}}$, $\tr_{3} = \strongTransition{v}{a}{\dirac{\stateredInSubscript}}$, and $\tr_{4} = \strongTransition{t}{\hidden}{\dirac{\startState}}$.

Besides other constraints, the LP problem $\LPproblemTBetaMuTransRel{\startState}{a}{\sd}{\transitionRelation}{\rel}$ has the following constraints:
\[
\begin{array}{lclcl}
	f_{\netsource,\startState} = 1 & \quad & f_{\relclass{\stategreenInSubscript}{\rel},\netsink} = 1/16 & \quad & f_{\relclass{\stateblueInSubscript}{\rel},\netsink} = 5/16 \\
	f_{\relclass{\stateredInSubscript}{\rel},\netsink} = 10/16 & \multicolumn{2}{l}{\kern-20pt f_{\startState,\startState^{\tr_{0}}} - f_{\startState^{\tr_{0}},t} - f_{\startState^{\tr_{0}},u} - f_{\startState^{\tr_{0}},v} = 0} & \quad & f_{\netsource,\startState} + f_{t^{\tr_{4}},\startState} - f_{\startState,\startState^{\tr_{0}}} = 0 \\
	f_{\startState^{\tr_{0}}, t} - f_{t,t^{\tr_{1}}_{a}} - f_{t,{t^{\tr_{4}}}} = 0  & \quad & f_{\startState^{\tr_{0}}, u} - f_{u, u^{\tr_{2}}_{a}}= 0 & \quad &	f_{\startState^{\tr_{0}}, v} - f_{v, v^{\tr_{3}}_{a}}= 0  \\
	f_{t,t^{\tr_{1}}_{a}} - f_{t^{\tr_{1}}_{a}, \stategreenInSubscript_{a}} = 0 & \quad & f_{u,u^{\tr_{2}}_{a}} - f_{u^{\tr_{2}}_{a}, \stateblueInSubscript_{a}} = 0  & \quad & f_{v,v^{\tr_{3}}_{a}} - f_{v^{\tr_{3}}_{a}, \stateredInSubscript_{a}} = 0 \\
	f_{t,t^{\tr_{4}}} - f_{t^{\tr_{4}},\startState} = 0  & \quad & f_{t^{\tr_{1}}_{a}, \stategreenInSubscript_{a}} - f_{\stategreenInSubscript_{a},\relclass{\stategreenInSubscript}{\rel}} = 0 & \quad & f_{u^{\tr_{2}}_{a}, \stateblueInSubscript_{a}} - f_{\stateblueInSubscript_{a},\relclass{\stateblueInSubscript}{\rel}} = 0  \\ 
	f_{v^{\tr_{3}}_{a}, \stateredInSubscript_{a}} - f_{\stateredInSubscript_{a},\relclass{\stateredInSubscript}{\rel}} = 0 & \quad & f_{\stategreenInSubscript_{a},\relclass{\stategreenInSubscript}{\rel}} - f_{\relclass{\stategreenInSubscript}{\rel}, \netsink} = 0 & \quad & f_{\stateblueInSubscript_{a},\relclass{\stateblueInSubscript}{\rel}} - f_{\relclass{\stateblueInSubscript}{\rel}, \netsink} = 0  \\ 
	f_{\stateredInSubscript_{a},\relclass{\stateredInSubscript}{\rel}} - f_{\relclass{\stateredInSubscript}{\rel}, \netsink} = 0 & \quad & f_{\startState^{\tr_{0}},t} - 1/4 f_{\startState,\startState^{\tr_{0}}} = 0  & \quad & f_{\startState^{\tr_{0}},u} - 1/4 f_{\startState,\startState^{\tr_{0}}} = 0  \\
	f_{\startState^{\tr_{0}},v} - 1/2 f_{\startState,\startState^{\tr_{0}}} = 0 & \quad & f_{t^{\tr_{1}}_{a}, \stategreenInSubscript_{a}} - 1f_{t,t^{\tr_{1}}_{a}} = 0 & \quad & f_{u^{\tr_{2}}_{a}, \stateblueInSubscript_{a}} - 1f_{u,u^{\tr_{2}}_{a}} = 0 \\ 
	f_{v^{\tr_{3}}_{a}, \stateredInSubscript_{a}} - 1f_{v,v^{\tr_{3}}_{a}} = 0 & \quad & f_{t^{\tr_{4}},\startState} - 1f_{t,{t^{\tr_{4}}}} = 0
\end{array}
\]
A solution that maximizes the objective function sets all variables to value $0$ except for
\[
\begin{array}{lrclrclrclr}
	f_{\netsource,\startState} & = 16/16  & \quad & f_{\relclass{\stategreenInSubscript}{\rel},\netsink} & = 1/16  & \quad & f_{\relclass{\stateblueInSubscript}{\rel},\netsink} & = \phantom{1}5/16 & \quad & f_{\relclass{\stateredInSubscript}{\rel},\netsink} & = 10/16 \\
	f_{\startState,\startState^{\tr_{0}}} & = 20/16 & \quad & f_{\startState^{\tr_{0}},t} & = 5/16 & \quad & f_{\startState^{\tr_{0}},u} & = \phantom{1}5/16 & \quad & f_{\startState^{\tr_{0}},v} & = 10/16 \\
	f_{t, t^{\tr_{1}}_{a}} & = \phantom{2}1/16 & \quad & f_{t, t^{\tr_{4}}} & = 4/16 & \quad & f_{u, u^{\tr_{2}}_{a}} & = \phantom{1}5/16 & \quad & f_{v, v^{\tr_{3}}_{a}} & = 10/16 \\
	f_{t^{\tr_{1}}_{a},\stategreenInSubscript_{a}} & = \phantom{2}1/16 & \quad & f_{t^{\tr_{4}},\startState} & = 4/16 & \quad & f_{u^{\tr_{2}}_{a}, \stateblueInSubscript_{a}} & = \phantom{1}5/16 & \quad & f_{v^{\tr_{3}}_{a}, \stateredInSubscript_{a}} & = 10/16 \\
	f_{\stategreenInSubscript_{a},\relclass{\stategreenInSubscript}{\rel}} & = \phantom{2}1/16 & \quad & f_{\stateblueInSubscript_{a},\relclass{\stateblueInSubscript}{\rel}} & = 5/16 & \quad & f_{\stateredInSubscript_{a},\relclass{\stateredInSubscript}{\rel}} & = 10/16
\end{array}
\]
The variable $f_{\startState,\startState^{\tr_{0}}} = 20/16$ is part of a cycle and its value is greater than $1$, confirming that $1$, the maximum probability, in general is not a proper value for arc capacities.

\subsection{Complexity of the LP Problem}
\label{ssec:LPcomplexity}

We analyze the complexity of the $\LPproblemTBetaMuTransRel{t}{a}{\sd}{\allowedTransitions}{\rel}$ LP problem when $a \neq \hidden$ since $\LPproblemTBetaMuTransRel{t}{\hidden}{\sd}{\allowedTransitions}{\rel}$ is just a special case of $\LPproblemTBetaMuTransRel{t}{a}{\sd}{\allowedTransitions}{\rel}$.

Given the automaton $\aut$ and the set $\allowedTransitions \subseteq \transitionRelation$ of allowed transitions, let $N_{\stateSet} = |\stateSet|$, $N_{\allowedTransitions} = |\allowedTransitions|$, and $N = \max \{N_{\stateSet}, N_{\allowedTransitions}\}$.
Suppose that $a \neq \hidden$ and consider the network graph $\NetworkTBetaMuTransRel{t}{a}{\sd}{\allowedTransitions}{\rel} = (V,E)$.
The cardinality of $V$ is: $|V| \leq 2 + N_{\stateSet} + N_{\allowedTransitions} + N_{\stateSet} + N_{\allowedTransitions} + N_{\stateSet} \in \bigO{N}$ and the cardinality of $E$ is: $|E| \leq 1 + 2 N_{\stateSet} + 2 (N_{\stateSet} + 1) N_{\allowedTransitions} + (N_{\stateSet} + 1) N_{\allowedTransitions} \in \bigO{N^{2}}$.
Note that this is also the cost of generating the $\NetworkTBetaMuTransRel{t}{a}{\sd}{\allowedTransitions}{\rel}$ network graph from the automaton $\aut$.

Now, consider the $\LPproblemTBetaMuTransRel{t}{a}{\sd}{\allowedTransitions}{\rel}$ LP problem:
the number of variables is $|\{f_{u,v} \mid (u,v) \in E\}| = |E| \in \bigO{N^{2}}$ and the number of constraints is $|E| + 1 + N_{\stateSet} + N_{\stateSet} N_{\allowedTransitions} + N_{\stateSet} N_{\allowedTransitions} + N_{\stateSet} N_{\allowedTransitions} + |V| - 2 \in \bigO{N^{2}}$, so generating $\LPproblemTBetaMuTransRel{t}{a}{\sd}{\allowedTransitions}{\rel}$ is polynomial in $N$. 
Since there exist polynomial algorithms for solving LP problems \cite{Tod02}, solving the $\LPproblemTBetaMuTransRel{t}{a}{\sd}{\allowedTransitions}{\rel}$ problem is polynomial in $N$.

\begin{theorem}
\label{thm:LPProblemIsPolynomial}
	Given a \PA{} $\aut$, an equivalence relation $\rel$ on $\stateSet$, an action $a$, a probability distribution $\sd \in \Disc{\stateSet}$, a set of allowed transitions $\allowedTransitions \subseteq \transitionRelation$, and a state $t \in \stateSet$, consider the problem $\LPproblemTBetaMuTransRel{t}{a}{\sd}{\allowedTransitions}{\rel}$ as defined above.
	Let $N = \max \{|\stateSet|,|\allowedTransitions|\}$.
	
	Generating and checking the existence of a valid solution of the $\LPproblemTBetaMuTransRel{t}{a}{\sd}{\allowedTransitions}{\rel}$ LP problem is polynomial in $N$.
\end{theorem}

\subsection{Some Optimizations.}
The implementation of $\LPproblemTBetaMuTransRel{t}{a}{\sd}{\allowedTransitions}{\rel}$ can be optimized in several ways: 
we can safely remove each constraint $f_{u,v} \geq 0$ when $(u,v) \in \{(v^{\tr}, v') \mid \tr = \strongTransition{v}{\hidden}{\rho} \in \allowedTransitions, \probeval{\rho}{v'} > 0\}$ since it is implied by $f_{v,v^{\tr}} \geq 0$ and $f_{v^{\tr},v'} - \probeval{\rho}{v'} f_{v,v^{\tr}} = 0$ as well as when $(u,v) \in \{(v^{\tr}_{a},v'_{a}) \mid \tr = \strongTransition{v}{\hidden}{\rho} \in \allowedTransitions \text{ or } \tr = \strongTransition{v}{a}{\rho} \in \allowedTransitions, \probeval{\rho}{v'} > 0\}$;
as second optimization, we can avoid the constraint $f_{u,v} \geq 0$ when $u = \equivclass \in \partitionset{\stateSet}$ and $v = \netsink$ since this is implied by $f_{\equivclass,\netsink} = \probeval{\sd}{\equivclass}$.
These optimizations allow us to save up to $2|\stateSet|(1+|\allowedTransitions|)$ constraints but the advantage we gain from them depends on the actual implementation of the LP solver.

Constraints of the form $\sum_{u \in \{x \mid (x,v) \in E\}} f_{u,v} - \sum_{u \in \{y \mid (v,y) \in E\}} f_{v,u} = 0$ for $v \in \stateSet^{\tr}$ can be removed safely since they derive from $f_{v^{\tr},v'} - \probeval{\rho}{v'} f_{v,v^{\tr}} = 0$ and the fact that by construction there is only one arc that ends in $v^{\tr}$.
The same holds for $v^{\tr}_{a} \in \stateSet^{\tr}_{a}$ given $a \neq \hidden$, so we can skip the generation of up to $2|\allowedTransitions|$ constraints.

The last optimization does not involve the removal of a constraint but
only the generation of the LP problem itself. Given $a \neq
\hidden$, the subgraph whose arcs have both vertices in
$\stateSet_{a} \cup \stateSet^{\tr}_{a}$ is simply a copy of
the subgraph whose arcs have both vertices in $\stateSet \cup
\stateSet^{\tr}$, so we can speed up the LP problem generation by just
copying a previously generated encoding.  Similarly, the subgraph
obtained by encoding internal transitions like
$\strongTransition{s}{\hidden}{\gd}$ does not depend on neither the
state $t$, the action $a$, the probability distribution $\sd$, nor
the equivalence relation $\rel$, so it can be generated only once and
then is simply copied in the actual instance of the
$\LPproblemTBetaMuTransRel{t}{a}{\sd}{\allowedTransitions}{\rel}$ LP problem.
All these optimizations, however, do not change the complexity class
of generating and then finding a feasible solution of the
$\LPproblemTBetaMuTransRel{t}{a}{\sd}{\allowedTransitions}{\rel}$ LP problem, which remains
polynomial. In any case they can improve the actual computation time
of an implementation.

\subsection{Equivalence of LP Problems and Weak Transitions}
\label{ssec:LPequivalentToWeakTr}

In this section we present the main theorem that equates $\LPproblemTBetaMuTransRel{t}{a}{\sd}{\allowedTransitions}{\rel}$ with an allowed weak combined transition, that is, $\LPproblemTBetaMuTransRel{t}{a}{\sd}{\allowedTransitions}{\rel}$ has a solution if and only if there exists a scheduler $\sched$ for $\aut$ that induces an allowed weak combined transition $\weakCombinedAllowedTransition{t}{a}{\sd_{t}}{\allowedTransitions}$ such that $\sd \liftrel \sd_{t}$. 
This result easily extends to ordinary weak combined transitions and hyper-transitions.

\begin{theorem}
	\label{thm:LPequivalentToWeakAllowedTransitionLifting}
	Given a \PA{} $\aut$, an equivalence relation $\rel$ on $\stateSet$, an action $a$, a probability distribution $\sd \in \Disc{\stateSet}$, a set of allowed transitions $\allowedTransitions \subseteq \transitionRelation$, and a state $t \in \stateSet$, consider the problem $\LPproblemTBetaMuTransRel{t}{a}{\sd}{\allowedTransitions}{\rel}$ as defined above.
	
	$\LPproblemTBetaMuTransRel{t}{a}{\sd}{\allowedTransitions}{\rel}$ has a solution $f^{*}$ such that $f^{*}_{\equivclass,\netsink} = \probeval{\sd}{\equivclass}$ for each $\equivclass \in \partitionset{\stateSet}$ if and only if there exists a scheduler $\sched$ for $\aut$ that induces $\weakCombinedAllowedTransition{t}{a}{\sd_{t}}{\allowedTransitions}$ such that $\sd \liftrel \sd_{t}$.
\end{theorem}
\begin{myproof}[Proof outline]
	The scheduler $\sigma$ we define in the proof for the ``only if'' direction assigns to each execution fragment $\alpha$ with $\last{\alpha} = v$ the sub-probability distribution over transitions defined, for each transition $\tr \in \allowedTransitions$ such that $\source{\tr} = v$, as the ratio $f^{*}_{v_{\mathtt{t}},v^{\tr}_{\bar{\mathtt{t}}}}/\incomingflow^{*}_{v_{\mathtt{t}}}$, given that $\incomingflow^{*}_{v_{\mathtt{t}}} > 0$, where $\incomingflow^{*}_{v}$ is the total flow incoming $v$, $\mathtt{t} = \trace{\alpha}$, and $\bar{\mathtt{t}}$ is the concatenation of $\trace{\alpha}$ and of $\trace{\action{\tr}}$.
	The remaining probability of stopping in the state $v$ is exactly $f^{*}_{v_{\mathtt{t}},\relclass{v}{\rel}}/\incomingflow^{*}_{v_{\mathtt{t}}}$.
	The way we generate the network $\NetworkTBetaMuTransRel{t}{a}{\sd}{\allowedTransitions}{\rel}$ ensures that $f^{*}_{v_{\mathtt{t}},v^{\tr}_{\mathtt{t}}} = 0$ when $\mathtt{t} \notin \{\emptytrace,\trace{a}\}$ and that $f^{*}_{v_{\mathtt{t}},\relclass{v}{\rel}}/\incomingflow^{*}_{v_{\mathtt{t}}} = 0$ when $\mathtt{t} \neq \trace{a}$.
	The proof for the ``if'' direction is the dual, that is, we define a feasible solution $f^{*}$ as the sum of the probabilities of the cones of execution fragments, i.e., $\incomingflow^{*}_{v_{b}} = \sum_{\alpha \in \{\phi \in \finiteFrags{\aut} \mid \trace{\phi} = b \wedge \last{\phi} = v\}} \probeval{\sd_{\sched,t}}{\cone{\alpha}}$; then the existence of such feasible solution is enough to prove that there exists a (possibly different) solution $f^{o}$ that maximizes the objective function while preserving the property that for each $\equivclass \in \partitionset{\stateSet}$, $f^{o}_{\equivclass,\netsink} = \probeval{\sd}{\equivclass}$.

	For the detailed proof, see Appendix~\ref{app:thm:LPequivalentToWeakTransitionLifting}.
\end{myproof}

It is worth to observe that the resulting scheduler is a determinate scheduler and an immediate corollary of this theorem confirming and improving Proposition~3 of~\cite{CS02} is that each scheduler inducing $\weakCombinedAllowedTransition{t}{a}{\sd_{t}}{\allowedTransitions}$ can be replaced by a determinate scheduler inducing $\weakCombinedAllowedTransition{t}{a}{\sd_{t}}{\allowedTransitions}$ as well.

% AT: same dirty trick as usual...
\vskip 1\baselineskip
\noindent \textit{Example~1 (cont.).\hspace{1ex}}
It is interesting to observe that the same weak combined transition can be generated by different schedulers:
we already know from the first part of this example that there exists a scheduler $\sched$ inducing $\weakCombinedAllowedTransition{\startState}{a}{\sd}{\transitionRelation}$ where $\sd = \{\frac{1}{16}\stategreen, \frac{5}{16}\stateblue, \frac{10}{16}\statered\}$.

Let again $\tr_{0} = \strongTransition{\startState}{\hidden}{\{\frac{1}{4} t, \frac{1}{4} u, \frac{1}{2} v\}}$, $\tr_{1} = \strongTransition{t}{a}{\dirac{\stategreenInSubscript}}$, $\tr_{2} = \strongTransition{u}{a}{\dirac{\stateblueInSubscript}}$, $\tr_{3} = \strongTransition{v}{a}{\dirac{\stateredInSubscript}}$, and $\tr_{4} = \strongTransition{t}{\hidden}{\dirac{\startState}}$.
Theorem~\ref{thm:LPequivalentToWeakAllowedTransitionLifting} ensures that there exists a scheduler $\sched'$, possibly different from $\sched$, that induces $\weakCombinedAllowedTransition{\startState}{a}{\sd}{\transitionRelation}$; 
in particular, $\sched'$ is the determinate scheduler defined as follows: 
\[
\schedeval{\sched'}{\alpha} = 
\begin{cases}
	\dirac{\tr_{0}} & \text{if $\trace{\alpha} = \emptytrace$ and $\last{\alpha} = \startState$;} \\
	\{\frac{1}{5} \tr_{1}, \frac{4}{5} \tr_{4}\} & \text{if $\trace{\alpha} = \emptytrace$ and $\last{\alpha} = t$;}\\
	\dirac{\tr_{2}} & \text{if $\trace{\alpha} = \emptytrace$ and $\last{\alpha} = u$;} \\
	\dirac{\tr_{3}} & \text{if $\trace{\alpha} = \emptytrace$ and $\last{\alpha} = v$;} \\
	\dirac{\bot} & \text{otherwise.}
\end{cases}
\]

It is straightforward to check that $\sched'$ actually induces $\weakCombinedAllowedTransition{\startState}{a}{\sd}{\transitionRelation}$. 
For instance, state $\stategreen$ is reached with probability $\probeval{\sd_{\sched',\startState}}{\{ \alpha \in \finiteFrags{\aut[E]} \mid \last{\alpha} = \stategreen \}} = \probeval{\sd_{\sched',\startState}}{\{\startState \hidden t (\hidden \startState \hidden t)^{n} a \stategreen \mid n \in \nat\}} = 1 \cdot \frac{1}{4} \cdot\sum_{n \in \nat} (\frac{4}{5}\cdot 1 \cdot 1 \cdot \frac{1}{4})^{n} \cdot \frac{1}{5} \cdot 1 \cdot 1 = \frac{1}{4} \cdot \frac{1}{5} \cdot (1-\frac{1}{5})^{-1} = \frac{1}{4} \cdot \frac{1}{5} \cdot \frac{5}{4} = \frac{1}{16} = \probeval{\sd}{\stategreen}$, as required.
\vskip 1 \baselineskip

\begin{corollary}
	\label{cor:LPequivalentToTransitionXYZ}
	Given a \PA{} $\aut$, $t \in \stateSet$ and $h \notin \stateSet$, $a \in \actionSet$, $\gd, \sd, \sd_{t} \in \Disc{\stateSet}$, $\allowedTransitions \subseteq \transitionRelation$, an equivalence relation $\rel$ on $\stateSet$, %the identity relation $\idrel$ on $\stateSet \cup\{h\}$, 
	a transition $\strongTransition{h}{\hidden}{\gd}$, $\allowedTransitions_{h} = \allowedTransitions \cup \{\strongTransition{h}{\hidden}{\gd}\}$, $\transitionRelation_{h} = \transitionRelation \cup \{\strongTransition{h}{\hidden}{\gd}\}$, and the \PA{} $\aut_{h} = (\stateSet \cup \{h\}, \startState, \actionSet, \transitionRelation_{h})$, the following holds:
	\begin{enumerate}
	\item \label{pnt:LPequivalentToWeakTransition}
		$\LPproblemTBetaMuTransRel{t}{a}{\sd}{\transitionRelation}{\rel}$ has a solution $f^{*}$ such that $f^{*}_{\equivclass,\netsink} = \probeval{\sd}{\equivclass}$ for each $\equivclass \in \partitionset{\stateSet}$ if and only if there exists a scheduler $\sched$ for $\aut$ inducing $\weakCombinedTransition{t}{a}{\sd_{t}}$ such that $\sd \liftrel \sd_{t}$;
		
	\item \label{pnt:LPequivalentToHyperAllowedTransition}
		$\LPproblemTBetaMuTransRel{h}{a}{\sd}{\allowedTransitions_{h}}{\rel}$ ($\LPproblemTBetaMuTransRel{h}{a}{\sd}{\transitionRelation_{h}}{\rel}$) relative to $\aut_{h}$ has a solution $f^{*}$ such that $f^{*}_{\equivclass,\netsink} = \probeval{\sd}{\equivclass}$ for each $\equivclass \in \partitionset{\stateSet}$ if and only if there exists a scheduler $\sched$ for $\aut$ inducing $\hyperWeakCombinedAllowedTransition{\gd}{a}{\sd_{t}}{\allowedTransitions}$ ($\hyperWeakCombinedTransition{\gd}{a}{\sd_{t}}$, respectively) such that $\sd_{t} \liftrel \sd$.

	\end{enumerate}
	When $\rel$ is the identity relation $\idrel$, $\sd \liftrel[\idrel] \sd_{t}$ implies $\sd_{t} = \sd$.
\end{corollary}
\begin{myproof}[Proof outline]
	The corollary follows directly from a combination of Theorem~\ref{thm:LPequivalentToWeakAllowedTransitionLifting} for the equivalence between the LP problem and allowed weak combined transition, Proposition~\ref{pro:weakAllowedTransitionEquivalentToWeakTransitionWhenAllTransitionsAllowed} for weak combined transitions, and Proposition~\ref{pro:hyperTransitionEquivalentToWeakTransition} for hyper-transitions.
\end{myproof}

\subsection{Equivalence Matching}

Theorem~\ref{thm:LPequivalentToWeakAllowedTransitionLifting} and its corollary allow us to check in polynomial time whether it is possible to reach a given probability distribution $\sd$ from a state $t$ or a probability distribution $\gd$. 
We now consider a more general case where, given a \PA{} $\aut$, two distributions $\gd_{1}, \gd_{2} \in \Disc{\stateSet}$, two actions $a_{1}, a_{2} \in \actionSet$, two sets $\allowedTransitions_{1}, \allowedTransitions_{2} \subseteq \transitionRelation$ of allowed transitions, and an equivalence relation $\rel$ on $\stateSet$, we want to check in polynomial time whether there exist $\sd_{1}, \sd_{2} \in \Disc{\stateSet}$ such that $\hyperWeakCombinedAllowedTransition{\gd_{1}}{a_{1}}{\sd_{1}}{\allowedTransitions_{1}}$, $\hyperWeakCombinedAllowedTransition{\gd_{2}}{a_{2}}{\sd_{2}}{\allowedTransitions_{2}}$, and $\sd_{1} \liftrel \sd_{2}$.
In order to find $\sd_{1}$ and $\sd_{2}$, we can consider a family $\family{p_{\equivclass}}{\equivclass \in \partitionset{\stateSet}}$ of non-negative values such that $\sum_{\equivclass \in \partitionset{\stateSet}} p_{\equivclass} = 1$ and a probability distribution $\bar{\sd}$ satisfying $\probeval{\bar{\sd}}{\equivclass} = p_{\equivclass}$ for each $\equivclass \in \partitionset{\stateSet}$ and then solve $\LPproblemTBetaMuTransRel{\gd_{1}}{a_{1}}{\bar{\sd}}{\allowedTransitions_{1}}{\rel}$ and $\LPproblemTBetaMuTransRel{\gd_{2}}{a_{2}}{\bar{\sd}}{\allowedTransitions_{2}}{\rel}$ where $\LPproblemTBetaMuTransRel{\gd}{a}{\sd}{\allowedTransitions}{\rel}$ is the problem $\LPproblemTBetaMuTransRel{h}{a}{\sd}{\allowedTransitions_{h}}{\rel}$ relative to $\aut_{h} = (\stateSet \cup \{h\}, \startState, \actionSet, \transitionRelation \cup \{\strongTransition{h}{\hidden}{\gd}\})$ with $h \notin \stateSet$ and $\allowedTransitions_{h} = \allowedTransitions \cup \{\strongTransition{h}{\hidden}{\gd}\}$.
The main problem of this approach is to find a good family of values $p_{\equivclass}$;
since we do not care about actual values, we consider $p_{\equivclass}$ as variables satisfying $p_{\equivclass} \geq 0$ and $\sum_{\equivclass \in \partitionset{\stateSet}} p_{\equivclass} = 1$ and we define the LP problem 
% $P_{1,2} = \LPproblemMatchingRelGdaBetaaTranaGdbBetabTranb{\rel}{\gd_1}{a_1}{\allowedTransitions_1}{\gd_2}{a_2}{\allowedTransitions_2}$ 
$P_{1,2}$
derived from $P_{1} = \LPproblemTBetaMuTransRel{\gd_{1}}{a_{1}}{\bar{\sd}}{\allowedTransitions_{1}}{\rel}$ and $P_{2} = \LPproblemTBetaMuTransRel{\gd_{2}}{a_{2}}{\bar{\sd}}{\allowedTransitions_{2}}{\rel}$ as follows (after renaming of $P_{2}$ variables to avoid collisions): 
the objective function of $P_{1,2}$ is the sum of the objective functions of $P_{1}$ and $P_{2}$;
the set of constraints of $P_{1,2}$ is $\sum_{\equivclass \in \partitionset{\stateSet}} p_{\equivclass} = 1$ together with $p_{\equivclass} \geq 0$ for $\equivclass \in \partitionset{\stateSet}$ and the union of the sets of constraints of $P_{1}$ and $P_{2}$ where each occurrence of $\probeval{\bar{\sd}}{\equivclass}$ is replaced by $p_{\equivclass}$.

It is quite easy to verify that $P_{1,2}$ has a solution if and only if both $P_{1}$ and $P_{2}$ have a solution (with respect to the same $\bar{\sd}$) and thus, by Corollary~\ref{cor:LPequivalentToTransitionXYZ}(\ref{pnt:LPequivalentToHyperAllowedTransition}), if and only if $\gd_{1}$ and $\gd_{2}$ enable an allowed hyper-transition to $\sd_{1}$ and $\sd_{2}$, respectively, such that $\sd_{1} \liftrel \sd_{2}$, as required.
It is immediate to see that $P_{1,2}$ can still be solved in polynomial time, since it is just the union of $P_{1}$ and $P_{2}$ extended with at most $N$ variables and $2N$ constraints where $N = |\stateSet|$.
\begin{proposition}\label{prop:decideMatching}
	Given a \PA{} $\aut$, two distributions $\gd_{1}, \gd_{2} \in \Disc{\stateSet}$, two actions $a_{1}, a_{2} \in \actionSet$, two sets $\allowedTransitions_{1}, \allowedTransitions_{2} \subseteq \transitionRelation$ of allowed transitions, and an equivalence relation $\rel$ on $\stateSet$, the existence of $\sd_{1}, \sd_{2} \in \Disc{\stateSet}$ such that $\hyperWeakCombinedAllowedTransition{\gd_{1}}{a_{1}}{\sd_{1}}{\allowedTransitions_{1}}$, $\hyperWeakCombinedAllowedTransition{\gd_{2}}{a_{2}}{\sd_{2}}{\allowedTransitions_{2}}$, and $\sd_{1} \liftrel \sd_{2}$ can be checked in polynomial time.
\end{proposition}

The above proposition easily extends, by Corollary~\ref{cor:LPequivalentToTransitionXYZ}, to each combination of weak combined transitions, allowed hyper-transitions, and allowed weak combined transitions as well as to exact matching as induced by the identity relation $\idrel$.

%%% Local Variables: 
%%% mode: latex
%%% TeX-master: "pa-weak-bisim-poly"
%%% End: 

\section{Decision Procedure}
\label{sec:decision}

\begin{wrapfigure}[6]{r}{4.5cm}
  \centering \vspace{-14mm} \small
  \begin{tikzpicture}
    \path[use as bounding box] (-2,-0) rectangle (2,-2.5); \node[draw,
    text height=2ex, minimum width=4.5cm, inner sep = 0pt, minimum
    height=3ex] (heading)
    {\progHeader{$\proc{Quotient}(\aut_{1}, \aut_{2})$}};
    \node[below=0.5ex of heading] (text) {%
      \begin{minipage}[h]{1\linewidth}
        \begin{algorithmic}
          \STATE{$\partitioning = \{\stateSet_{1} \uplus \stateSet_{2}\}$;}
          \STATE{$(\equivclass,a,\sd) = \proc{FindSplit}(\partitioning)$;}
          \WHILE{$\equivclass \neq \emptyset$}
          \STATE{$\partitioning = \proc{Refine}(\partitioning,(\equivclass,a,\sd))$;}
          \STATE{$(\equivclass,a,\sd) = \proc{FindSplit}(\partitioning)$;}
          \ENDWHILE
          \RETURN{$\partitioning$}
        \end{algorithmic}
      \end{minipage}
    }; \coordinate (END) at ($(heading.north west)!21.5ex!(heading.south
    west)$); \draw (heading.south west) to (END) {}; \draw (END) to
    ($(END)+(2em,0)$) {};
  \end{tikzpicture}
%   \caption{Computes Bisimulation Quotient} \label{alg:mainLoop}
\end{wrapfigure}

In this section, we recast the decision procedure of \cite{CS02} that decides whether two probabilistic automata $\aut_{1}$ and $\aut_{2}$ are bisimilar according to $\weakBisim$, that is, whether $\aut_{1} \weakBisim \aut_{2}$, following the standard partition refinement approach~\cite{CS02,KS90,PT87,PLS00}.
More precisely, procedure \proc{Quotient} iteratively constructs the set $\quotienting{\stateSet}{\mathord{\weakBisim}}$, the set of equivalence classes of states $\stateSet = \stateSet_{1} \uplus \stateSet_{2}$ under $\weakBisim$, starting with the partitioning $\partitioning = \{\stateSet\}$ and refining it until $\partitioning$ satisfies the definition of weak probabilistic bisimulation and thus the resulting partitioning is the coarsest one, i.e., we compute the weak probabilistic bisimilarity.

\begin{wrapfigure}[7]{r}{6cm}
\vskip-5mm
\renewcommand{\algorithmicthen}{}
  \centering \vspace{-0.9cm} \small
  \begin{tikzpicture}
    \path[use as bounding box] (-2,3em) rectangle (2,-2.1); \node[draw,
    text height=2ex, minimum width=6cm, inner sep = 0pt,
    minimum height=3ex] (heading)
    {\progHeader{$\proc{FindSplit}(\partitioning)$}}; \node[below=0.5ex of heading]
    (text) {%
      \begin{minipage}[h]{1\linewidth}
        \begin{algorithmic}[1]
			\FORALL{$(s,a,\sd) \in \transitionRelation = \transitionRelation_{1} \uplus \transitionRelation_{2}$}
				\FORALL{$t \in \relclass{s}{\partitioning}$}
					\IF{$\LPproblemTBetaMuTransRel{t}{a}{\sd}{\transitionRelation}{\partitioning}$ has no solution} 
						\RETURN{$(\relclass{s}{\partitioning},a,\sd)$}
					\ENDIF
				\ENDFOR
			\ENDFOR
			\RETURN{$(\emptyset,\hidden,\dirac{\startState})$}
        \end{algorithmic}
      \end{minipage}
    }; \coordinate (END) at ($(heading.north west)!18.5ex!(heading.south
    west)$); \draw (heading.south west) to (END) {}; \draw (END) to
    ($(END)+(2em,0)$) {};
  \end{tikzpicture}
%  \caption{Find splitter} \label{alg:splitter}
\end{wrapfigure}

Deciding whether two automata are bisimilar then reduces to checking whether their start states belong to the same equivalence class.
In the following, we  treat $\partitioning$ both as a set of partitions and as an equivalence relation without further mentioning.

The partitioning is refined by procedure \proc{Refine} into a finer partitioning as long as there is a partition containing two states that violate the bisimulation condition, which is checked for in procedure~\proc{FindSplit}.
Procedure \proc{Refine}, that we do not provide explicitly as in~\cite{CS02}, splits partition $\equivclass$ into two new partitions according to the discriminating information $(\equivclass, a, \sd)$ identified by \proc{FindSplit} before. 
So far, the procedure is as the $\text{\sffamily\textsl{DecideBisim}}(\aut_{1}, \aut_{2})$ procedure proposed in~\cite{CS02}. 

The difference arises inside the procedure \proc{FindSplit}, where  we check directly the step condition by solving for each transition $\strongTransition{s}{a}{\sd}$ the LP problem $\LPproblemTBetaMuTransRel{t}{a}{\sd}{\transitionRelation}{\partitioning}$ that has a solution, according to Corollary~\ref{cor:LPequivalentToTransitionXYZ}(\ref{pnt:LPequivalentToWeakTransition}), if and only if there exists %a weak combined transition 
$\weakCombinedTransition{t}{a}{\sd_{t}}$ such that $\sd \liftrel[\partitioning] \sd_{t}$.

\subsection{Complexity Analysis of the Procedure}

Given two \PA{}s $\aut_{1}$ and $\aut_{2}$, let $\stateSet = \stateSet_{1} \uplus \stateSet_{2}$, $\transitionRelation = \transitionRelation_{1} \uplus \transitionRelation_{2}$, and $N = \max \{|\stateSet|,|\transitionRelation|\}$.

In the worst case (that occurs when the current $\partitioning$ satisfies the step condition), the \textbf{for} at line 1 of  procedure \proc{FindSplit} is performed at most $N$ times as well as the inner \textbf{for}, so $\LPproblemTBetaMuTransRel{t}{a}{\sd}{\transitionRelation}{\partitioning}$ is generated and solved at most $N^{2}$ times.
Since by Theorem~\ref{thm:LPProblemIsPolynomial} generating and checking the existence of a valid solution for $\LPproblemTBetaMuTransRel{t}{a}{\sd}{\transitionRelation}{\partitioning}$ is polynomial in $N$, this implies that also \proc{FindSplit} is polynomial in $N$; more precisely, denoted by $p(N)$ the complexity of $\LPproblemTBetaMuTransRel{t}{a}{\sd}{\transitionRelation}{\partitioning}$, $\proc{FindSplit} \in \bigO{N^{2}p(N)}$.
Note that we can improve the running time required to solve the $\LPproblemTBetaMuTransRel{t}{a}{\sd}{\transitionRelation}{\partitioning}$ LP problem by replacing $\transitionRelation$ with $\transitionRelation'$ at line 3 of \proc{FindSplit} where $\transitionRelation'$ contains only transitions with label $\hidden$ or $a$ enabled by states reachable from $t$.

The \textbf{while} loop in the procedure \proc{Quotient} can be performed at most $N$ times; this happens when in each loop the procedure~\proc{FindSplit} returns $(\equivclass, a, \sd)$ where $\equivclass \neq \emptyset$, that is, not every pair of states in $\equivclass$ satisfies the step condition.
Since in each loop the procedure \proc{Refine} splits such class $\equivclass$ in two classes  $\equivclass_{1}$ and $\equivclass_{2}$, after at most $N$ loops every class contains a single state and the procedure \proc{FindSplit} returns $(\emptyset, \hidden, \dirac{\startState})$ since each transition $\strongTransition{s}{a}{\sd_{s}}$ is obviously matched by $s$ itself.
Since \proc{Refine} and \proc{FindSplit} are polynomial in $N$, also \proc{Quotient} is polynomial in $N$, thus checking $\aut_{1} \weakBisim \aut_{2}$ is polynomial in $N$.
\begin{theorem}
	Given two \PA{}s $\aut_{1}$ and $\aut_{2}$, let $N = \max \{|\stateSet_{1} \uplus \stateSet_{2}|, |\transitionRelation_{1} \uplus \transitionRelation_{2}|\}$.
	
	Checking $\aut_{1} \weakBisim \aut_{2}$ is polynomial in $N$.
\end{theorem}

%%% Local Variables: 
%%% mode: latex
%%% TeX-master: "pa-weak-bisim-poly"
%%% End: 

\section{Concluding Remarks}
\label{sec:conclusion}
This paper has established a polynomial time decision algorithm for
\PA{} weak probabilistic bisimulation, closing the quest for an
effective decision algorithm coined in \cite{CS02}. The core
innovation is a novel characterization of weak combined transitions as
an LP problem, enabling us to check the existence of a weak combined transition
in polynomial time.  The algorithm can be exploited in an effective
compositional minimization strategy for \PA{} (or \MDP{}) and potentially
also for Markov automata.  Furthermore, the LP approach we developed is
readily extensible to related problems requiring to find a specific
weak transition. Another area of immediate applicability concerns
cost-related problems where transition costs may relate to power or
resource consumption in \PA{} or \MDP{}.

%%% Local Variables: 
%%% mode: latex
%%% TeX-master: "pa-weak-bisim-poly"
%%% End: 

\noindent{\bf Acknowledgments.}  The authors are grateful to
Christian Eisentraut (Saarland University) for insightful
discussions. This work has been supported by the DFG as part of the
SFB/TR~14~``Automatic Verification and Analysis of Complex Systems''
(AVACS), by the DFG/NWO Bilateral Research Programme ROCKS, and by the
European Union Seventh Framework Programme under grant agreement no.\@
295261 (MEALS). Andrea Turrini has received support from the Cluster
of Excellence ``Multimodal Computing and Interaction'' (MMCI), part of
the German Excellence Initiative.

% \bibliographystyle{abbrv}	
% \bibliography{biblio,bib}

\clearpage
\appendix

\section{Equivalences between Allowed Transitions and Ordinary Transitions}
\label{app:equivalenceHyperTransitionDefinitions}

\begin{result}[Proposition \ref{pro:hyperTransitionEquivalentToWeakTransition}]
	Given a \PA{} $\aut$, $h \notin \stateSet$, $a \in \actionSet$, $\allowedTransitions \subseteq \transitionRelation$, and $\gd,\sd \in \Disc{\stateSet}$, let $\aut_{h}$ be the \PA{} $\aut_{h} = (\stateSet \cup \{h\}, \startState, \actionSet, \transitionRelation \cup \{\strongTransition{h}{\hidden}{\gd}\})$ and $\allowedTransitions_{h}$ be $\allowedTransitions \cup \{\strongTransition{h}{\hidden}{\gd}\}$.
	
	$\hyperWeakCombinedAllowedTransition{\gd}{a}{\sd}{\allowedTransitions}$ exists in $\aut$ if and only if $\weakCombinedAllowedTransition{h}{a}{\sd}{\allowedTransitions_{h}}$ exists in $\aut_{h}$.
\end{result}
\begin{myproof}
	A common result we need is that for $\alpha = s_{0} a_{1} s_{1} \dots$ such that $\first{\alpha} = s_{0} \in \Supp{\gd}$, $\alpha \in \finiteFrags{\aut}$ if and only if $h\hidden\alpha \in \finiteFrags{\aut_{h}}$;
	denote by $s_{-1}$ the state $h$ and by $a_{0}$ the action $\hidden$ so $h\hidden\alpha$ is just $s_{-1} a_{0} s_{0} a_{1} s_{1} \dots$.
	Since $\alpha \in \finiteFrags{\aut}$ we have that for each $0 \leq i < |\alpha|$ there exists $(s_{i}, a_{i+1}, \sd_{i+1})$ such that $\probeval{\sd_{i+1}}{s_{i+1}} > 0$.
	Since $\probeval{\gd}{s_{0}} > 0$, then for each $-1 \leq i < |\alpha|$ there exists a transition $(s_{i}, a_{i+1}, \sd_{i+1})$ such that $\probeval{\sd_{i+1}}{s_{i+1}} > 0$, so $h\hidden\alpha \in \finiteFrags{\aut_{h}}$.
	
	Now, suppose that $h\hidden\alpha \in \finiteFrags{\aut_{h}}$.
	This implies that $-1 \leq i < |\alpha|$ there exists a transition $(s_{i}, a_{i+1}, \sd_{i+1})$ such that $\probeval{\sd_{i+1}}{s_{i+1}} > 0$; 
	in particular, it holds that $0 \leq i < |\alpha|$ there exists a transition $(s_{i}, a_{i+1}, \sd_{i+1})$ such that $\probeval{\sd_{i+1}}{s_{i+1}} > 0$ and this implies that $s_{0} \in \Supp{\gd}$ and $\alpha \in \finiteFrags{\aut}$.
	
	It is straightforward to check that given an automaton $\aut[B]$, a scheduler $\sched$, and a state $s$, for each $\alpha \in \finiteFrags{\aut[B]}$, $\probeval{\sd_{\sched,s}}{\cone{\alpha}} > 0$ implies $\first{\alpha} = s$ that is implied by $\probeval{\sd_{\sched,s}}{\alpha} > 0$ as well. 

\begin{description}
\item[($\Rightarrow$)]
	By definition of $\hyperWeakCombinedAllowedTransition{\gd}{a}{\sd}{\allowedTransitions}$ there exists a family $\family{\weakCombinedAllowedTransition{s}{a}{\sd_{s}}{\allowedTransitions}}{s \in \Supp{\gd}}$ of allowed weak transitions such that $\sd = \sum_{s \in \Supp{\gd}} \probeval{\gd}{s}\sd_{s}$.
	This implies that there exists a family of schedulers $\family{\sched_{s}}{s \in \Supp{\gd}}$ such that for each $s \in \Supp{\gd}$, $\sched_{s}$ induces the allowed weak transition $\weakCombinedAllowedTransition{s}{a}{\sd_{s}}{\allowedTransitions}$.
	
	Let $\sched$ be the scheduler for $\aut_{h}$ defined as follows:
	\[
		\schedeval{\sched}{\alpha} = 
		\begin{cases}
			\dirac{\strongTransition{h}{\hidden}{\gd}} & \text{if $\alpha = h$,} \\
			\schedeval{\sched_{s}}{\alpha'} & \text{if $\alpha = h\hidden\alpha' = h\hidden s a_{1} s_{1} \dots$,} \\
			\dirac{\bot} & \text{otherwise.}
		\end{cases}
	\]
	To prove that $\sched$ actually induces the allowed weak transition $\weakCombinedAllowedTransition{h}{a}{\sd}{\allowedTransitions_{h}}$, we need of some preliminary result:
	for each finite execution fragment $\alpha \in \finiteFrags{\aut_{h}}$, $\Supp{\schedeval{\sched}{\alpha}} \subseteq \allowedTransitions_{h}$.
	In fact, $\Supp{\schedeval{\sched}{h}} = \{\strongTransition{h}{\hidden}{\gd}\}\subseteq \allowedTransitions_{h}$; $\Supp{\schedeval{\sched}{h\hidden\alpha'}} = \Supp{\schedeval{\sched_{s}}{\alpha'}} \subseteq \allowedTransitions \subseteq \allowedTransitions_{h}$ where $s = \first{\alpha'}$; 
	for all other execution fragments, $\Supp{\schedeval{\sched}{\alpha}} = \Supp{\dirac{\bot}} = \emptyset \subseteq \allowedTransitions_{h}$.

	Another result we need is the following:
	for each $\alpha \in \finiteFrags{\aut}$, if $\first{\alpha} = s$, then $\probeval{\sd_{\sched,h}}{\cone{h\hidden\alpha}} = \probeval{\gd}{s}\probeval{\sd_{\sched_{s},s}}{\cone{\alpha}}$.
	We prove this result by induction on the length $n$ of $\alpha$: if $n = 0$, then $\probeval{\sd_{\sched,h}}{\cone{h\hidden s}} = \probeval{\sd_{\sched,h}}{\cone{h}} \sum_{\tr \in\transitionsWithLabel{\hidden}} \probeval{\schedeval{\sched}{h}}{\tr} \cdot \probeval{\sd_{\tr}}{s} = 1\sum_{\tr \in\transitionsWithLabel{\hidden}} \probeval{\schedeval{\sched}{h}}{\tr} \cdot \probeval{\sd_{\tr}}{s} = \probeval{\gd}{s} = \probeval{\gd}{s}\probeval{\sd_{\sched_{s},s}}{\cone{s}}$; 
	if $n > 0$, then there exists $\alpha'$ such that $\alpha = \alpha' a t$ for some action $a$ and state $t$, so $\probeval{\sd_{\sched,h}}{\cone{h\hidden\alpha}} = \probeval{\sd_{\sched,h}}{\cone{h\hidden\alpha'a t}} = \probeval{\sd_{\sched,h}}{\cone{h\hidden\alpha'}} \sum_{\tr \in\transitionsWithLabel{a}} \probeval{\schedeval{\sched}{h\hidden\alpha'}}{\tr} \cdot \probeval{\sd_{\tr}}{t} = \probeval{\gd}{s}\probeval{\sd_{\sched_{s},s}}{\cone{\alpha'}} \sum_{\tr \in\transitionsWithLabel{a}} \probeval{\schedeval{\sched_{s}}{\alpha'}}{\tr} \cdot \probeval{\sd_{\tr}}{t} = \probeval{\gd}{s}\probeval{\sd_{\sched_{s},s}}{\cone{\alpha'a t}} = \probeval{\gd}{s}\probeval{\sd_{\sched_{s},s}}{\cone{\alpha}}$.
	
	Now we are ready to show that the three conditions on the probabilistic execution fragment $\sd_{\sched,h}$ induced by $\sched$ are satisfied. 
	\begin{enumerate}
	\item 
		\begin{align*}
			& \probeval{\sd_{\sched,h}}{\finiteFrags{\aut_{h}}} \\
			& = \sum_{\alpha \in \finiteFrags{\aut_{h}}} \probeval{\sd_{\sched,h}}{\cone{\alpha}} \cdot \probeval{\schedeval{\sched}{\alpha}}{\bot} \\
			& = \probeval{\sd_{\sched,h}}{\cone{h}} \cdot \probeval{\schedeval{\sched}{h}}{\bot} + \sum_{h\hidden\alpha \in \finiteFrags{\aut_{h}}} \probeval{\sd_{\sched,h}}{\cone{h\hidden\alpha}} \cdot \probeval{\schedeval{\sched}{h\hidden\alpha}}{\bot} \\
			& = 0 + \sum_{h\hidden\alpha \in \finiteFrags{\aut_{h}}} \probeval{\gd}{\first{\alpha}}\probeval{\sd_{\sched_{\first{\alpha}},\first{\alpha}}}{\cone{\alpha}} \cdot \probeval{\schedeval{\sched}{h\hidden\alpha}}{\bot} \\
			& = \sum_{h\hidden\alpha \in \finiteFrags{\aut_{h}}} \probeval{\gd}{\first{\alpha}}\probeval{\sd_{\sched_{\first{\alpha}},\first{\alpha}}}{\cone{\alpha}} \cdot \probeval{\schedeval{\sched}{h\hidden\alpha}}{\bot} \\
			& = \sum_{s \in \stateSet} \sum_{\alpha \in \{\alpha' \in \finiteFrags{\aut} \mid \first{\alpha'} = s\}} \probeval{\gd}{s}\probeval{\sd_{\sched_{s},s}}{\cone{\alpha}} \cdot \probeval{\schedeval{\sched}{h\hidden\alpha}}{\bot} \\
			& = \sum_{s \in \stateSet} \probeval{\gd}{s}\sum_{\alpha \in \{\alpha' \in \finiteFrags{\aut} \mid \first{\alpha'} = s\}} \probeval{\sd_{\sched_{s},s}}{\cone{\alpha}} \cdot \probeval{\schedeval{\sched}{\alpha}}{\bot} \\
			& = \sum_{s \in \stateSet} \probeval{\gd}{s}\sum_{\alpha \in \{\alpha' \in \finiteFrags{\aut} \mid \first{\alpha'} = s\}} \probeval{\sd_{\sched_{s},s}}{\alpha} \\
			& = \sum_{s \in \stateSet} \probeval{\gd}{s}\sum_{\alpha \in \finiteFrags{\aut}} \probeval{\sd_{\sched_{s},s}}{\alpha} \\
			& = \sum_{s \in \stateSet} \probeval{\gd}{s} \probeval{\sd_{\sched_{s},s}}{\finiteFrags{\aut}} \\
			& = \sum_{s \in \stateSet} \probeval{\gd}{s} 1 \\
			& = 1\text{;}
		\end{align*}
	\item 
		let $\alpha' \in \finiteFrags{\aut_{h}}$ such that $\probeval{\sd_{\sched,h}}{\alpha'} > 0$;
		this implies that $\first{\alpha'} = h$ thus $\alpha' = h\hidden\alpha$ for some $\alpha \in \finiteFrags{\aut}$ since $\strongTransition{h}{\hidden}{\gd}$ is the only transition enabled by $h$.
		$\probeval{\sd_{\sched,h}}{\alpha'} > 0$ implies as well that $\first{\alpha} = s \in \Supp{\gd}$ and $\probeval{\sd_{\sched_{s},s}}{\alpha} > 0$ for some state $s$ hence, by definition of $\weakCombinedAllowedTransition{s}{a}{\sd_{s}}{\allowedTransitions}$, $\trace{a} = \trace{\alpha} = \trace{\alpha'}$, as required;
	\item 
		\begin{align*}
			& \probeval{\sd_{\sched,h}}{\{\alpha \in \finiteFrags{\aut_{h}} \mid \last{\alpha} = q\}} \\
			& = \sum_{\{\alpha \in \finiteFrags{\aut_{h}} \mid \last{\alpha} = q\}} \probeval{\sd_{\sched,h}}{\cone{\alpha}} \cdot \probeval{\schedeval{\sched}{\alpha}}{\bot} \\
			& = \sum_{\{h\hidden\alpha \in \finiteFrags{\aut_{h}} \mid \last{\alpha} = q\}} \probeval{\sd_{\sched,h}}{\cone{h\hidden\alpha}} \cdot \probeval{\schedeval{\sched}{h\hidden\alpha}}{\bot} \\
			& = \sum_{\{h\hidden\alpha \in \finiteFrags{\aut_{h}} \mid \last{\alpha} = q\}} \probeval{\gd}{\first{\alpha}}\probeval{\sd_{\sched_{\first{\alpha}},\first{\alpha}}}{\cone{\alpha}} \cdot \probeval{\schedeval{\sched}{h\hidden\alpha}}{\bot} \\
			& = \sum_{\{h\hidden\alpha \in \finiteFrags{\aut_{h}} \mid \last{\alpha} = q\}} \probeval{\gd}{\first{\alpha}}\probeval{\sd_{\sched_{\first{\alpha}},\first{\alpha}}}{\cone{\alpha}} \cdot \probeval{\schedeval{\sched}{h\hidden\alpha}}{\bot} \\
			& = \sum_{s \in \stateSet} \sum_{\alpha \in \{\alpha' \in \finiteFrags{\aut} \mid \first{\alpha'} = s \wedge \last{\alpha'} = q \}} \probeval{\gd}{s}\probeval{\sd_{\sched_{s},s}}{\cone{\alpha}} \cdot \probeval{\schedeval{\sched}{h\hidden\alpha}}{\bot} \\
			& = \sum_{s \in \stateSet} \probeval{\gd}{s}\sum_{\alpha \in \{\alpha' \in \finiteFrags{\aut} \mid \first{\alpha'} = s \wedge \last{\alpha'} = q \}} \probeval{\sd_{\sched_{s},s}}{\cone{\alpha}} \cdot \probeval{\schedeval{\sched}{\alpha}}{\bot} \\
			& = \sum_{s \in \stateSet} \probeval{\gd}{s}\sum_{\alpha \in \{\alpha' \in \finiteFrags{\aut} \mid \first{\alpha'} = s \wedge \last{\alpha'} = q \}} \probeval{\sd_{\sched_{s},s}}{\alpha} \\
			& = \sum_{s \in \stateSet} \probeval{\gd}{s}\sum_{\alpha \in \{\alpha' \in \finiteFrags{\aut} \mid \last{\alpha'} = q \}} \probeval{\sd_{\sched_{s},s}}{\alpha} \\
			& = \sum_{s \in \stateSet} \probeval{\gd}{s} \probeval{\sd_{s}}{q} \\
			& = \probeval{\sd}{q}\text{.}
		\end{align*}
	\end{enumerate}

\item[($\Leftarrow$)]
	For each $s \in \Supp{\gd}$, let $\sched_{s}$ be the scheduler for $\aut$ defined as follows:
	\[
		\schedeval{\sched_{s}}{\alpha} = 
		\begin{cases}
			\schedeval{\sched}{h\hidden\alpha} & \text{if $\first{\alpha} = s$,} \\
			\dirac{\bot} & \text{otherwise.}
		\end{cases}
	\]
	To prove that the family of schedulers $\sched_{s}$ induces the allowed hyper transition $\hyperWeakCombinedAllowedTransition{\gd}{a}{\sd}{\allowedTransitions}$, we need of some preliminary result:
	for each execution fragment $\alpha \in \finiteFrags{\aut}$, $\Supp{\schedeval{\sched}{\alpha}} \subseteq \allowedTransitions$.
	In fact, $\Supp{\schedeval{\sched_{s}}{\alpha}} = \Supp{\schedeval{\sched}{h\hidden\alpha}} \subseteq \allowedTransitions_{h}$ where $s = \first{\alpha}$; 
	by hypothesis, $h \notin \stateSet$ and this implies that for each $\strongTransition{s}{a_{s}}{\sd_{s}} \in \transitionRelation$, $h \notin \Supp{\sd_{s}}$, hence $\strongTransition{h}{\hidden}{\gd} \notin \SubDisc{\transitionsFromState{\last{\alpha}}}$, so $\strongTransition{h}{\hidden}{\gd} \notin \Supp{\schedeval{\sched}{h\hidden\alpha}}$ and thus $\Supp{\schedeval{\sched_{s}}{\alpha}} \subseteq \allowedTransitions$.
	For all other execution fragments, $\Supp{\schedeval{\sched}{\alpha}} = \Supp{\dirac{\bot}} = \emptyset \subseteq \allowedTransitions$.

	Another result we need is the following:
	for each $\alpha \in \finiteFrags{\aut}$, if $\first{\alpha} = s$, then $\probeval{\sd_{\sched_{s},s}}{\cone{\alpha}} = \dfrac{\probeval{\sd_{\sched,h}}{\cone{h\hidden\alpha}}}{\probeval{\gd}{s}}$.
	We prove this result by induction on the length $n$ of $\alpha$: if $n = 0$, then $\dfrac{\probeval{\sd_{\sched,h}}{\cone{h\hidden s}}}{\probeval{\gd}{s}} = \dfrac{\probeval{\sd_{\sched,h}}{\cone{h}} \sum_{\tr \in\transitionsWithLabel{\hidden}} \probeval{\schedeval{\sched}{h}}{\tr} \cdot \probeval{\sd_{\tr}}{s}}{\probeval{\gd}{s}} = \dfrac{1\sum_{\tr \in\transitionsWithLabel{\hidden}} \probeval{\schedeval{\sched}{h}}{\tr} \cdot \probeval{\sd_{\tr}}{s}}{\probeval{\gd}{s}} = \dfrac{\probeval{\gd}{s}}{\probeval{\gd}{s}} = 1 = \probeval{\sd_{\sched_{s},s}}{\cone{s}}$; 
	if $n > 0$, then we have that $\alpha = \alpha' a t$ for some action $a$ and state $t$, therefore $\dfrac{\probeval{\sd_{\sched,h}}{\cone{h\hidden\alpha}}}{\probeval{\gd}{s}} = \dfrac{\probeval{\sd_{\sched,h}}{\cone{h\hidden\alpha'a t}}}{\probeval{\gd}{s}} = \dfrac{\probeval{\sd_{\sched,h}}{\cone{h\hidden\alpha'}} \cdot \sum_{\tr \in\transitionsWithLabel{a}} \probeval{\schedeval{\sched}{h\hidden\alpha'}}{\tr} \cdot \probeval{\sd_{\tr}}{t}}{\probeval{\gd}{s}} = \dfrac{\probeval{\sd_{\sched,h}}{\cone{h\hidden\alpha'}}}{\probeval{\gd}{s}} \cdot \sum_{\tr \in\transitionsWithLabel{a}} \probeval{\schedeval{\sched_{s}}{\alpha'}}{\tr} \cdot \probeval{\sd_{\tr}}{t} = \probeval{\sd_{\sched_{s},s}}{\cone{\alpha'}} \cdot \sum_{\tr \in\transitionsWithLabel{a}} \probeval{\schedeval{\sched_{s}}{\alpha'}}{\tr} \cdot \probeval{\sd_{\tr}}{t} = \probeval{\sd_{\sched_{s},s}}{\cone{\alpha'a t}} = \probeval{\sd_{\sched_{s},s}}{\cone{\alpha}}$.
	
	Now we are ready to show that the three conditions on the probabilistic execution fragment $\sd_{\sched_{s},s}$ induced by $\sched_{s}$ are satisfied, where $\sd_{s}$ is defined for each $t \in \stateSet$, as follows: 
	\[
		\probeval{\sd_{s}}{t} = \dfrac{\probeval{\sd_{\sched,h}}{\{h\hidden\alpha' \in \finiteFrags{\aut_{h}} \mid \last{\alpha'} = t \wedge \first{\alpha'} = s\}}}{\probeval{\gd}{s}}
	\] 
	\begin{enumerate}
	\item 
		\begin{align*}
			& \probeval{\sd_{\sched_{s},s}}{\finiteFrags{\aut}} \\
			& = \sum_{\alpha \in \{\alpha' \in \finiteFrags{\aut} \mid \first{\alpha'} = s\}} \probeval{\sd_{\sched_{s},s}}{\cone{\alpha}} \cdot \probeval{\schedeval{\sched_{s}}{\alpha}}{\bot} \\
			& = \sum_{h\hidden\alpha \in \{h\hidden\alpha' \in \finiteFrags{\aut_{h}} \mid \first{\alpha'} = s\}} \dfrac{\probeval{\sd_{\sched,h}}{\cone{h\hidden\alpha}}}{\probeval{\gd}{s}} \cdot \probeval{\schedeval{\sched}{h\hidden\alpha}}{\bot} \\
			& = \sum_{h\hidden\alpha \in \{h\hidden\alpha' \in \finiteFrags{\aut_{h}} \mid \first{\alpha'} = s\}} \dfrac{\probeval{\sd_{\sched,h}}{\cone{h\hidden\alpha}} \cdot \probeval{\schedeval{\sched}{h\hidden\alpha}}{\bot}}{\probeval{\gd}{s}} \\
			& = \dfrac{\sum_{h\hidden\alpha \in \{h\hidden\alpha' \in \finiteFrags{\aut_{h}} \mid \first{\alpha'} = s\}} \probeval{\sd_{\sched,h}}{\cone{h\hidden\alpha}} \cdot \probeval{\schedeval{\sched}{h\hidden\alpha}}{\bot}}{\probeval{\gd}{s}} \\
			& = \dfrac{\probeval{\gd}{s}}{\probeval{\gd}{s}}\\
			& = 1\text{;}
		\end{align*}
	\item 
		let $\alpha \in \finiteFrags{\aut}$ such that $\probeval{\sd_{\sched_{s},s}}{\alpha} > 0$;
		this implies that $\first{\alpha} = s$ and $\probeval{\sd_{\sched,h}}{h\hidden\alpha} > 0$, hence $\trace{a} = \trace{h\hidden\alpha} = \trace{\alpha}$, as required;
	\item 
		\begin{align*}
			& \probeval{\sd_{\sched_{s},s}}{\{\alpha \in \finiteFrags{\aut} \mid \last{\alpha} = q \wedge \first{\alpha} = s\}} \\
			& = \sum_{\{\alpha \in \finiteFrags{\aut} \mid \last{\alpha} = q \wedge \first{\alpha} = s\}} \probeval{\sd_{\sched_{s},s}}{\cone{\alpha}} \cdot \probeval{\schedeval{\sched_{s}}{\alpha}}{\bot} \\
			& = \sum_{\{h\hidden\alpha \in \finiteFrags{\aut_{h}} \mid \last{\alpha} = q \wedge \first{\alpha} = s\}} \dfrac{\probeval{\sd_{\sched,h}}{\cone{h\hidden\alpha}}}{\probeval{\gd}{s}} \cdot \probeval{\schedeval{\sched}{h\hidden\alpha}}{\bot} \\
			& = \sum_{\{h\hidden\alpha \in \finiteFrags{\aut_{h}} \mid \last{\alpha} = q \wedge \first{\alpha} = s\}} \dfrac{\probeval{\sd_{\sched,h}}{\cone{h\hidden\alpha}} \cdot \probeval{\schedeval{\sched}{h\hidden\alpha}}{\bot}}{\probeval{\gd}{s}} \\
			& = \dfrac{\sum_{\{h\hidden\alpha \in \finiteFrags{\aut_{h}} \mid \last{\alpha} = q \wedge \first{\alpha} = s\}}\probeval{\sd_{\sched,h}}{\cone{h\hidden\alpha}} \cdot \probeval{\schedeval{\sched}{h\hidden\alpha}}{\bot}}{\probeval{\gd}{s}} \\
			& = \dfrac{\probeval{\sd_{\sched,h}}{\{h\hidden\alpha \in \finiteFrags{\aut_{h}} \mid \last{\alpha} = q \wedge \first{\alpha} = s\}}}{\probeval{\gd}{s}}\\
			& = \probeval{\sd_{s}}{q}\text{.}
		\end{align*}
	\end{enumerate}
	The final step is to prove that $\sd = \sum_{s \in \Supp{\gd}}\probeval{\gd}{s}\sd_{s}$, that is, for each state $t \in \stateSet$, it holds that $\probeval{\sd}{t} = \sum_{s \in \Supp{\gd}}\probeval{\gd}{s}\probeval{\sd_{s}}{t}$: 
	\begin{align*}
		& \sum_{s \in \Supp{\gd}}\probeval{\gd}{s}\probeval{\sd_{s}}{t} \\
		& = \sum_{s \in \Supp{\gd}} \probeval{\gd}{s}\probeval{\sd_{\sched_{s},s}}{\{\alpha \in \finiteFrags{\aut} \mid \last{\alpha} = t \wedge \first{\alpha} = s\}} \\
		& = \sum_{s \in \Supp{\gd}} \probeval{\gd}{s} \sum_{\alpha \in \{\alpha' \in \finiteFrags{\aut} \mid \last{\alpha'} = t \wedge \first{\alpha'} = s\}}\probeval{\sd_{\sched_{s},s}}{\alpha} \\
		& = \sum_{s \in \Supp{\gd}} \probeval{\gd}{s} \sum_{\alpha \in \{h\hidden\alpha' \in \finiteFrags{\aut} \mid \last{\alpha'} = t \wedge \first{\alpha'} = s\}}\dfrac{\probeval{\sd_{\sched,h}}{\alpha}}{\probeval{\gd}{s}} \\
		& = \sum_{s \in \Supp{\gd}} \sum_{\alpha \in \{h\hidden\alpha' \in \finiteFrags{\aut} \mid \last{\alpha'} = t \wedge \first{\alpha'} = s\}}\dfrac{\probeval{\gd}{s} \probeval{\sd_{\sched,h}}{	\alpha}}{\probeval{\gd}{s}} \\
		& = \sum_{\alpha \in \{h\hidden\alpha' \in \finiteFrags{\aut} \mid \last{\alpha'} = t\}}\probeval{\sd_{\sched,h}}{\alpha} \\
		& = \probeval{\sd_{\sched,h}}{\{\alpha \in \finiteFrags{\aut} \mid \last{\alpha} = t\}} \\
		& = \probeval{\sd}{t}\text{,}
	\end{align*}
	as required.
\
\end{description}
\end{myproof}

\begin{result}[Proposition~\ref{pro:weakAllowedTransitionEquivalentToWeakTransitionWhenAllTransitionsAllowed}]
	Given a \PA{} $\aut$, a state $s$, and a probability distribution $\sd \in \Disc{\stateSet}$, there exists a scheduler $\sched_{\transitionRelation}$ for $\aut$ that induces $\weakCombinedAllowedTransition{s}{a}{\sd}{\transitionRelation}$ if and only if there exists a scheduler $\sched$ for $\aut$ that induces $\weakCombinedTransition{s}{a}{\sd}$.
\end{result}
\begin{myproof}
	The fact that the existence of $\weakCombinedAllowedTransition{s}{a}{\sd}{\transitionRelation}$ implies that there is $\weakCombinedTransition{s}{a}{\sd}$ is immediate, since by definition of allowed transition, $\weakCombinedAllowedTransition{s}{a}{\sd}{\transitionRelation}$ requires the existence of a scheduler $\sched$ that induces $\weakCombinedTransition{s}{a}{\sd}$.
	
	For the other implication, it is enough to verify that $\sched$ satisfies the condition: \emph{for each $\alpha \in \finiteFrags{\aut}$, $\Supp{\schedeval{\sched}{\alpha}} \subseteq \transitionRelation$}.
	This is obviously true since by definition of scheduler $\schedeval{\sched}{\alpha} \in \SubDisc{\transitionRelation}$ holds, so $\Supp{\schedeval{\sched}{\alpha}} \subseteq \transitionRelation$.
\end{myproof}

%%% Local Variables: 
%%% mode: latex
%%% TeX-master: "pa-weak-bisim-poly"
%%% End: 

\section{Proof of Results Enunciated in Section~\ref{sec:weakTransitionAsLPP}}
\label{app:thm:LPequivalentToWeakTransitionLifting}

\begin{result}[Theorem~\ref{thm:LPequivalentToWeakAllowedTransitionLifting}]
	Given a \PA{} $\aut$, an equivalence relation $\rel$ on $\stateSet$, an action $a$, a probability distribution $\sd \in \Disc{\stateSet}$, a set of allowed transitions $\allowedTransitions \subseteq \transitionRelation$, and a state $t \in \stateSet$, consider the problem $\LPproblemTBetaMuTransRel{t}{a}{\sd}{\allowedTransitions}{\rel}$ as defined in Section~\ref{sec:weakTransitionAsLPP}.
	
	$\LPproblemTBetaMuTransRel{t}{a}{\sd}{\allowedTransitions}{\rel}$ has a solution $f^{*}$ such that $f^{*}_{\equivclass,\netsink} = \probeval{\sd}{\equivclass}$ for each $\equivclass \in \partitionset{\stateSet}$ if and only if there exists a scheduler $\sched$ for $\aut$ that induces $\weakCombinedAllowedTransition{t}{a}{\sd_{t}}{\allowedTransitions}$ such that $\sd \liftrel \sd_{t}$.
\end{result}
\begin{myproof}
	Given a solution $f^{*}$ of $\LPproblemTBetaMuTransRel{t}{a}{\sd}{\allowedTransitions}{\rel}$, denote by $\incomingflow^{*}_{v}$ the value $\incomingflow^{*}_{v} = \sum_{u \in V} f^{*}_{u,v}$, i.e., the total incoming flow in the node $v$.
	\begin{description}
		\item[($\Leftarrow$)]
			Let $\sched$ be the scheduler that induces the weak transition $\weakCombinedAllowedTransition{t}{a}{\sd_{t}}{\allowedTransitions}$ and $\sd_{\sched,t}$ be the probabilistic execution fragment generated by $\sched$ from $t$.
			For each finite execution fragment $\phi$ such that $\probeval{\sd_{\sched,t}}{\cone{\phi}} > 0$, denote by $\bar{\phi}$ the last state $\last{\phi}$ of $\phi$ and define $f^{\phi}_{x,y}$ as follows:
			\[f^{\phi}_{x,y} = 
			\begin{cases}
				1 & \text{if $x = \netsource$, $y = \phi = t$;} \\
				
				\probeval{\sd_{\sched,t}}{\cone{\phi}}\probeval{\schedeval{\sched}{\phi}}{\bot} & \text{if $x = \bar{\phi}$, $y = \relclass{\bar{\phi}}{\rel}$, and $a = \hidden$;} \\
				
				\probeval{\sd_{\sched,t}}{\cone{\phi}}\probeval{\schedeval{\sched}{\phi}}{\bot} & \text{if $x = \bar{\phi}_{a}$, $y = \relclass{\bar{\phi}}{\rel}$, and $a \neq \hidden$;} \\
				
				\probeval{\sd_{\sched,t}}{\cone{\phi}}\probeval{\schedeval{\sched}{\phi}}{\tr} & \text{if $x = \bar{\phi}$, $y = \bar{\phi}^{\tr}$, $\trace{\phi} = \emptytrace$, and $\tr = \strongTransition{\bar{\phi}}{\hidden}{\rho}$;} \\
				\probeval{\sd_{\sched,t}}{\cone{\phi}}\probeval{\schedeval{\sched}{\phi}}{\tr}\probeval{\rho}{q} & \text{if $x = \bar{\phi}^{\tr}$, $y = q$, $\trace{\phi} = \emptytrace$, and $\tr = \strongTransition{\bar{\phi}}{\hidden}{\rho}$;} \\
				
				\probeval{\sd_{\sched,t}}{\cone{\phi}}\probeval{\schedeval{\sched}{\phi}}{\tr} & \text{if $x = \bar{\phi}$, $y = \bar{\phi}^{\tr}_{a}$, $\trace{\phi} = \emptytrace$, $\tr = \strongTransition{\bar{\phi}}{a}{\rho}$, and $a \neq \hidden$;} \\
				\probeval{\sd_{\sched,t}}{\cone{\phi}}\probeval{\schedeval{\sched}{\phi}}{\tr}\probeval{\rho}{q} & \text{if $x = \bar{\phi}^{\tr}_{a}$, $y = q_{a}$, $\trace{\phi} = \emptytrace$, $\tr = \strongTransition{\bar{\phi}}{a}{\rho}$, and $a \neq \hidden$;} \\
				
				\probeval{\sd_{\sched,t}}{\cone{\phi}}\probeval{\schedeval{\sched}{\phi}}{\tr} & \text{if $x = \bar{\phi}_{a}$, $y = \bar{\phi}^{\tr}_{a}$, $\trace{\phi} = a \neq \hidden$, and $\tr = \strongTransition{\bar{\phi}}{\hidden}{\rho}$;} \\
				\probeval{\sd_{\sched,t}}{\cone{\phi}}\probeval{\schedeval{\sched}{\phi}}{\tr}\probeval{\rho}{q} & \text{if $x = \bar{\phi}^{\tr}_{a}$, $y = q_{a}$, $\trace{\phi} = a \neq \hidden$, and $\tr = \strongTransition{\bar{\phi}}{\hidden}{\rho}$;} \\
				
				0 & \text{otherwise.}
			\end{cases}
			\]
			Finally, define $f_{x,y}$ as 
			\[
			f_{x,y} = 
				\begin{cases}
					\probeval{\sd_{t}}{\equivclass} & \text{if $x = \equivclass \in \partitionset{\stateSet}$ and $y = \netsink$;} \\
					\sum_{\phi \in \finiteFrags{\aut}} f^{\phi}_{x,y} & \text{otherwise}
				\end{cases}
			\]
			
			It is straightforward to verify that the definition of $f_{x,y}$ given above implies that $f_{x,y} \geq 0$ for each $(x,y) \in E$, that $f_{\netsource,t} = 1$, and that $f_{\equivclass, \netsink} = \probeval{\sd_{t}}{\equivclass}$ for each $\equivclass \in \partitionset{\stateSet}$.
			
			Now consider the constraint $f_{v^{\tr},v'} = \probeval{\rho}{v'} f_{v,v^{\tr}}$ for $\tr = \strongTransition{v}{\hidden}{\rho} \in \transitionRelation$ and $v' \in \Supp{\rho}$.
			There are two cases depending on whether an execution fragment $\phi$ satisfies $v = \last{\phi}$ and $\probeval{\sd_{\sched,t}}{\cone{\phi}} > 0$.
			If $\phi$ satisfies $v = \last{\phi}$ and $\probeval{\sd_{\sched,t}}{\cone{\phi}} > 0$, then by definition we have $f^{\phi}_{v,v^{\tr}} = \probeval{\sd_{\sched,t}}{\cone{\phi}}\probeval{\schedeval{\sched}{\phi}}{\tr}$ and $f^{\phi}_{v^{\tr},v'} = \probeval{\sd_{\sched,t}}{\cone{\phi}}\probeval{\schedeval{\sched}{\phi}}{\tr}\probeval{\rho}{v'}$, thus $f^{\phi}_{v^{\tr},v'} = \probeval{\rho}{v'} f^{\phi}_{v,v^{\tr}}$, as required.
			If $\phi$ does not satisfy the conditions, then $f^{\phi}_{v,v^{\tr}} = 0$ and $f^{\phi}_{v^{\tr},v'} = 0$, hence again $f^{\phi}_{v^{\tr},v'} = \probeval{\rho}{v'} f^{\phi}_{v,v^{\tr}}$.
			This implies, together with the definition of $f_{x,y}$, that $f_{v^{\tr},v'} = \sum_{\phi \in \finiteFrags{\aut}} f^{\phi}_{v^{\tr},v'} = \sum_{\phi \in \finiteFrags{\aut}} \probeval{\rho}{v'}f^{\phi}_{v,v^{\tr}} = \probeval{\rho}{v'} f_{v,v^{\tr}}$, as required.
			The cases $f_{v^{\tr}_{a},v'_{a}} = \probeval{\rho}{v'_{a}} f_{v_{a},v^{\tr}_{a}}$ and $f_{v^{\tr}_{a},v'} = \probeval{\rho}{v'} f_{v,v^{\tr}_{a}}$ are similar.
			
			The remaining part of this proof considers the so called \emph{conservation of the flow} constraints, i.e., constraints of the kind $\sum_{u \in \{x \mid (x,v) \in E\}} f_{u,v} = \sum_{u \in \{y \mid (v,y) \in E\}} f_{v,u}$ for each $v \in V \setminus \{\netsource,\netsink\}$.
			There are several cases (comments refer to the previous equality):
			\begin{description}
				\item[case $v = \equivclass \in \partitionset{\stateSet}$:] 
					\begin{align*}
						\sum_{u \in \{y \mid (\equivclass,y) \in E\}} f_{\equivclass,u} 
						& = f_{\equivclass,\netsink} \\
						\intertext{by definition of $E$}
						& = \probeval{\sd_{t}}{\equivclass} \\
						\intertext{by constraint on $f_{\equivclass,\netsink}$}
						& = \sum_{\{\phi \in \finiteFrags{\aut} \mid \last{\phi} \in \equivclass\}} \probeval{\sd_{\sched,t}}{\cone{\phi}}\probeval{\schedeval{\sched}{\phi}}{\bot} \\
						\intertext{by definition of $\sd_{t}$ and of $\probeval{\sd_{\sched,t}}{\phi}$}
						& = \sum_{\{\phi \in \finiteFrags{\aut} \mid \bar{\phi} \in \equivclass\}} f^{\phi}_{\bar{\phi},\equivclass} +  f^{\phi}_{\bar{\phi}_{a},\equivclass} \\
						\intertext{by definition of $f^{\phi}_{x,y}$}
						& = \sum_{z \in \equivclass} \sum_{\{\phi \in \finiteFrags{\aut} \mid \bar{\phi} = z\}} f^{\phi}_{z,\equivclass} +  f^{\phi}_{z_{a},\equivclass}\\
						& = \sum_{u \in \{x \mid (x,\equivclass)\in E \}} \sum_{\{\phi \in \finiteFrags{\aut} \mid \bar{\phi} = u\}} f^{\phi}_{u,\equivclass}\\
						\intertext{by definition of $E$}
						& = \sum_{u \in \{x \mid (x,\equivclass)\in E \}} \sum_{\phi \in \finiteFrags{\aut}} f^{\phi}_{u,\equivclass}\\
						\intertext{by definition of $f^{\phi}_{x,y}$}
						& = \sum_{u \in \{x \mid (x,\equivclass) \in E\}} f_{u,\equivclass}
					\end{align*}
				
				\item[case $v = x^{\tr}$ for $\tr = \strongTransition{x}{\hidden}{\rho}$:]
					\begin{align*}
						\sum_{u \in \{z \mid (z,x^{\tr}) \in E\}} f_{u,x^{\tr}} 
						& = f_{x,x^{\tr}} \\
						\intertext{by definition of $E$}
						& = \sum_{\phi \in \finiteFrags{\aut}} f^{\phi}_{x,x^{\tr}} \\
						\intertext{by definition of $f_{x,y}$}
						& = \sum_{\{\phi \in \finiteFrags{\aut} \mid \last{\phi} = x\}} f^{\phi}_{x,x^{\tr}} \\
						\intertext{since $f^{\phi}_{x,y} = 0$ when $\last{\phi} \neq x$}
						& = \sum_{\{\phi \in \finiteFrags{\aut} \mid \last{\phi} = x\}} \probeval{\sd_{\sched,t}}{\cone{\phi}}\probeval{\schedeval{\sched}{\phi}}{\tr} \\
						\intertext{by definition of $f^{\phi}_{x,y}$}
						& = \sum_{\{\phi \in \finiteFrags{\aut} \mid \last{\phi} = x\}} \probeval{\sd_{\sched,t}}{\cone{\phi}}\probeval{\schedeval{\sched}{\phi}}{\tr} \sum_{x' \in \Supp{\rho}} \probeval{\rho}{x'}\\
						\intertext{since $\sum_{x' \in \Supp{\rho}} \probeval{\rho}{x'} = 1$}
						& = \sum_{\{\phi \in \finiteFrags{\aut} \mid \last{\phi} = x\}} \sum_{x' \in \Supp{\rho}} \probeval{\sd_{\sched,t}}{\cone{\phi}}\probeval{\schedeval{\sched}{\phi}}{\tr}\probeval{\rho}{x'} \\
						& = \sum_{x' \in \Supp{\rho}} \sum_{\{\phi \in \finiteFrags{\aut} \mid \last{\phi} = x\}} \probeval{\sd_{\sched,t}}{\cone{\phi}}\probeval{\schedeval{\sched}{\phi}}{\tr}\probeval{\rho}{x'} \\
						& = \sum_{x' \in \Supp{\rho}} \sum_{\{\phi \in \finiteFrags{\aut} \mid \last{\phi} = x\}} f^{\phi}_{x^{\tr},x'} \\
						\intertext{by definition of $f^{\phi}_{x^{\tr},x'}$}
						& = \sum_{x' \in \Supp{\rho}} \sum_{\phi \in \finiteFrags{\aut}} f^{\phi}_{x^{\tr},x'} \\
						\intertext{since $f^{\phi}_{x^{\tr},x'} = 0$ when $\last{\phi} \neq x$}
						& = \sum_{x' \in \Supp{\rho}} f_{x^{\tr},x'} \\
						\intertext{by definition of $f_{x^{\tr},x'}$}
						& = \sum_{u \in \{z \mid (x^{\tr},z) \in E\}} f_{x^{\tr},u} \\
						\intertext{by definition of $E$}
					\end{align*}

				\item[case $v = x^{\tr}_{a}$:] 
					the proof is analogous;
				
				\item[case $v = t$:]
					\begin{align*}
						\sum_{u \in \{y \mid (t,y) \in E\}} f_{t,u}
						= & \sum_{u \in \{y \mid (t,y) \in E\}} \sum_{\phi \in \finiteFrags{\aut}} f^{\phi}_{t,u}\\
						\intertext{by definition of $f_{t,u}$}
						= & \sum_{\phi \in \finiteFrags{\aut}} \sum_{u \in \{y \mid (t,y) \in E\}} f^{\phi}_{t,u}\\
						= & \sum_{\phi \in \finiteFrags{\aut}} \left(f^{\phi}_{t,\relclass{t}{\rel}} + \sum_{\{t^{\tr} \mid \tr = \strongTransition{t}{\hidden}{\rho}, \last{\phi} = t\}} f^{\phi}_{t,t^{\tr}}\right) \\
						\intertext{by definition of by definition of $f^{\phi}_{x^{\tr},x'}$}
						= & \sum_{\phi \in \finiteFrags{\aut}} f^{\phi}_{t,\relclass{t}{\rel}} + \sum_{\phi \in \finiteFrags{\aut}} \sum_{\{t^{\tr} \mid \tr = \strongTransition{t}{\hidden}{\rho}, \last{\phi} = t\}} f^{\phi}_{t,t^{\tr}} \\
						= & \sum_{\{\phi \in \finiteFrags{\aut}\mid \last{\phi} = t\}} \probeval{\sd_{\sched,t}}{\cone{\phi}}\probeval{\schedeval{\sched}{\phi}}{\bot} \\
						  & + \sum_{\phi \in \finiteFrags{\aut}} \sum_{\{\tr = \strongTransition{t}{\hidden}{\rho} \mid \last{\phi} = t\}} \probeval{\sd_{\sched,t}}{\cone{\phi}}\probeval{\schedeval{\sched}{\phi}}{\tr}\\
						\intertext{by definition of $f^{\phi}_{t,\relclass{t}{\rel}}$ and of $f^{\phi}_{t,t^{\tr}}$}
						= & \sum_{\{\phi \in \finiteFrags{\aut}\mid \last{\phi} = t\}} \probeval{\sd_{\sched,t}}{\cone{\phi}} \left(\probeval{\schedeval{\sched}{\phi}}{\bot} + \sum_{\{\tr = \strongTransition{t}{\hidden}{\rho}\}} \probeval{\schedeval{\sched}{\phi}}{\tr}\right)\\
						= & \sum_{\{\phi \in \finiteFrags{\aut}\mid \last{\phi} = t\}} \probeval{\sd_{\sched,t}}{\cone{\phi}} \\
						\intertext{since $\probeval{\schedeval{\sched}{\phi}}{\bot} = 1 - \sum_{\tr} \probeval{\schedeval{\sched}{\phi}}{\tr}$}
						= & \probeval{\sd_{\sched,t}}{\cone{t}} + \sum_{\{\phi \in \finiteFrags{\aut}\mid \phi = \phi'\hidden t\}} \probeval{\sd_{\sched,t}}{\cone{\phi}} \\
						= & 1 + \sum_{\{\phi \in \finiteFrags{\aut}\mid \phi = \phi'\hidden t\}}\probeval{\sd_{\sched,t}}{\cone{\phi'}} \sum_{\{\tr = \strongTransition{x}{\hidden}{\rho} \mid x = \last{\phi'}\}}\probeval{\schedeval{\sched}{\phi'}}{\tr}\probeval{\rho}{t} \\
						\intertext{by definition of $\probeval{\sd_{\sched,t}}{\cone{\phi}}$}
						= & f_{\netsource,t} + \sum_{\{\phi \in \finiteFrags{\aut}\mid \phi = \phi'\hidden t\}} \sum_{\{\tr = \strongTransition{x}{\hidden}{\rho} \mid x = \last{\phi'}\}}\probeval{\sd_{\sched,t}}{\cone{\phi'}}\probeval{\schedeval{\sched}{\phi'}}{\tr}\probeval{\rho}{t} \\
						\intertext{by definition of $f_{\netsource,t}$}
						= & f_{\netsource,t} + \sum_{\{\phi \in \finiteFrags{\aut}\mid \phi = \phi'\hidden t\}} \sum_{\{\tr = \strongTransition{x}{\hidden}{\rho} \mid x = \last{\phi'}\}} f^{\phi'}_{x^{\tr},t} \\
						\intertext{by definition of $f^{\phi'}_{x^{\tr},t}$}
						= & f_{\netsource,t} + \sum_{\{\phi \in \finiteFrags{\aut}\mid \phi = \phi'\hidden t\}} \sum_{\tr = \strongTransition{x}{\hidden}{\rho}} f^{\phi'}_{x^{\tr},t} \\
						\intertext{since $f^{\phi'}_{x^{\tr},t} = 0$ when $\last{\phi'} \neq x$} 
						= & f_{\netsource,t} + \sum_{\tr = \strongTransition{x}{\hidden}{\rho}} \sum_{\{\phi \in \finiteFrags{\aut}\mid \phi = \phi'\hidden t\}} f^{\phi'}_{x^{\tr},t} \\
						= & f_{\netsource,t} + \sum_{\{\tr = \strongTransition{x}{\hidden}{\rho}\}} f_{x^{\tr},t} \\
						\intertext{by definition of $f_{x^{\tr},t}$}
						= & \sum_{u \in \{x\mid (x,t) \in E\}} f_{u,t} \\
						\intertext{by definition of $E$}
					\end{align*}

				\item[case $v \in \stateSet \setminus \{t\} = V \setminus \{\netsource,\netsink,t\}$:]
					\begin{align*}
						\sum_{u \in \{y \mid (v,y) \in E\}} f_{v,u}
						= & \sum_{u \in \{y \mid (v,y) \in E\}} \sum_{\phi \in \finiteFrags{\aut}} f^{\phi}_{v,u}\\
						\intertext{by definition of $f_{v,u}$}
						= & \sum_{\phi \in \finiteFrags{\aut}} \sum_{u \in \{y \mid (v,y) \in E\}} f^{\phi}_{v,u}\\
						= & \sum_{\phi \in \finiteFrags{\aut}} \left(f^{\phi}_{v,\relclass{v}{\rel}} + \sum_{\{v^{\tr} \mid \tr = \strongTransition{v}{\hidden}{\rho}, \last{\phi} = v\}} f^{\phi}_{v,v^{\tr}}\right) \\
						\intertext{by definition of by definition of $f^{\phi}_{x^{\tr},x'}$}
						= & \sum_{\phi \in \finiteFrags{\aut}} f^{\phi}_{v,\relclass{v}{\rel}} + \sum_{\phi \in \finiteFrags{\aut}} \sum_{\{v^{\tr} \mid \tr = \strongTransition{v}{\hidden}{\rho}, \last{\phi} = v\}} f^{\phi}_{v,v^{\tr}} \\
						= & \sum_{\{\phi \in \finiteFrags{\aut}\mid \last{\phi} = v\}} \probeval{\sd_{\sched,t}}{\cone{\phi}}\probeval{\schedeval{\sched}{\phi}}{\bot} \\
						  & + \sum_{\phi \in \finiteFrags{\aut}} \sum_{\{\tr = \strongTransition{v}{\hidden}{\rho} \mid \last{\phi} = v\}} \probeval{\sd_{\sched,t}}{\cone{\phi}}\probeval{\schedeval{\sched}{\phi}}{\tr}\\
						\intertext{by definition of $f^{\phi}_{v,\relclass{v}{\rel}}$ and of $f^{\phi}_{v,v^{\tr}}$}
						= & \sum_{\{\phi \in \finiteFrags{\aut}\mid \last{\phi} = v\}} \probeval{\sd_{\sched,t}}{\cone{\phi}} \left(\probeval{\schedeval{\sched}{\phi}}{\bot} + \sum_{\{\tr = \strongTransition{v}{\hidden}{\rho}\}} \probeval{\schedeval{\sched}{\phi}}{\tr}\right)\\
						= & \sum_{\{\phi \in \finiteFrags{\aut}\mid \last{\phi} = v\}} \probeval{\sd_{\sched,t}}{\cone{\phi}} \\
						\intertext{since $\probeval{\schedeval{\sched}{\phi}}{\bot} = 1 - \sum_{\tr} \probeval{\schedeval{\sched}{\phi}}{\tr}$}
						= & \probeval{\sd_{\sched,t}}{\cone{v}} + \sum_{\{\phi \in \finiteFrags{\aut}\mid \phi = \phi'\hidden v\}} \probeval{\sd_{\sched,t}}{\cone{\phi}} \\
						= & 0 + \sum_{\{\phi \in \finiteFrags{\aut}\mid \phi = \phi'\hidden v\}}\probeval{\sd_{\sched,t}}{\cone{\phi'}} \sum_{\{\tr = \strongTransition{x}{\hidden}{\rho} \mid x = \last{\phi'}\}}\probeval{\schedeval{\sched}{\phi'}}{\tr}\probeval{\rho}{v} \\
						\intertext{by definition of $\probeval{\sd_{\sched,t}}{\cone{\phi}}$}
						= & \sum_{\{\phi \in \finiteFrags{\aut}\mid \phi = \phi'\hidden v\}} \sum_{\{\tr = \strongTransition{x}{\hidden}{\rho} \mid x = \last{\phi'}\}}\probeval{\sd_{\sched,t}}{\cone{\phi'}}\probeval{\schedeval{\sched}{\phi'}}{\tr}\probeval{\rho}{v} \\
						= & \sum_{\{\phi \in \finiteFrags{\aut}\mid \phi = \phi'\hidden t\}} \sum_{\{\tr = \strongTransition{x}{\hidden}{\rho} \mid x = \last{\phi'}\}} f^{\phi'}_{x^{\tr},v} \\
						\intertext{by definition of $f^{\phi'}_{x^{\tr},v}$}
						= & \sum_{\{\phi \in \finiteFrags{\aut}\mid \phi = \phi'\hidden v\}} \sum_{\tr = \strongTransition{x}{\hidden}{\rho}} f^{\phi'}_{x^{\tr},v} \\
						\intertext{since $f^{\phi'}_{x^{\tr},v} = 0$ when $\last{\phi'} \neq x$} 
						= & \sum_{\tr = \strongTransition{x}{\hidden}{\rho}} \sum_{\{\phi \in \finiteFrags{\aut}\mid \phi = \phi'\hidden v\}} f^{\phi'}_{x^{\tr},v} \\
						= & \sum_{\tr = \strongTransition{x}{\hidden}{\rho}} \sum_{\phi' \in \finiteFrags{\aut}} f^{\phi'}_{x^{\tr},v} \\
						= & \sum_{\{\tr = \strongTransition{x}{\hidden}{\rho}\}} f_{x^{\tr},v} \\
						\intertext{by definition of $f_{x^{\tr},v}$}
						= & \sum_{u \in \{x\mid (x,v) \in E\}} f_{u,v} \\
						\intertext{by definition of $E$}
					\end{align*}

					\item[case $v \in \stateSet_{a}$:] 
						the proof is analogous.
			\end{description}
			This concludes the proof that if there exists a scheduler $\sched$ that induces an allowed weak transition $\weakCombinedAllowedTransition{t}{a}{\sd_{t}}{\allowedTransitions}$ such that $\sd \liftrel \sd_{t}$, then $\LPproblemTBetaMuTransRel{t}{a}{\sd}{\allowedTransitions}{\rel}$ has a solution $f^{*}$ (the flow $f$ defined above) such that $f^{*}_{\equivclass,\netsink} = \probeval{\sd}{\equivclass}$ for each $\equivclass \in \partitionset{\stateSet}$.

			It is worth to note that for each state $v$, $\incomingflow_{v_{b}} = \sum_{\alpha \in \{\phi \in \finiteFrags{\aut} \mid \trace{\phi} = b \wedge \last{\phi} = v\}} \probeval{\sd_{\sched,t}}{\cone{\alpha}}$.
			This property derives from the definition of $f$, the conservation of the flow constraints, and the definition of probability of cones.
			
			Since $\LPproblemTBetaMuTransRel{t}{a}{\sd}{\allowedTransitions}{\rel}$ has a solution $f^{*}$, it has also a solution $f^{o}$ that maximizes the objective function;
			since $f^{o}$ is a valid solution, it must satisfy the constraint $f^{o}_{\equivclass,\netsink} = \probeval{\sd}{\equivclass}$ for each $\equivclass \in \partitionset{\stateSet}$, hence the statement \emph{if there exists a scheduler $\sched$ for $\aut$ that induces a weak transition $\weakCombinedAllowedTransition{t}{a}{\sd_{t}}{\allowedTransitions}$ such that $\sd \liftrel \sd_{t}$ then $\LPproblemTBetaMuTransRel{t}{a}{\sd}{\allowedTransitions}{\rel}$ has a solution $f^{*}$ such that $f^{*}_{\equivclass,\netsink} = \probeval{\sd}{\equivclass}$ for each $\equivclass \in \partitionset{\stateSet}$} still holds.
			
		\item[($\Rightarrow$)]
			For a state $x \in \stateSet$, let $\hat{x}$ be $x$ if $a = \hidden$ and be $x_{a}$ if $a \neq \hidden$.
			
			Given a solution $f^{*}$ of $\LPproblemTBetaMuTransRel{t}{a}{\sd}{\allowedTransitions}{\rel}$ such that $f^{*}_{\equivclass,\netsink} = \probeval{\sd}{\equivclass}$ for each $\equivclass \in \partitionset{\stateSet}$, define $\sd_{t}$ as follows: for each state $x \in \stateSet$, $\probeval{\sd_{t}}{x} = f^{*}_{\hat{x},\relclass{x}{\rel}}$ and for each $X \subseteq S$, $\probeval{\sd_{t}}{X} = \sum_{x \in X} \probeval{\sd_{t}}{x}$.
			
			It is straightforward to see that $\sd_{t} \in \Disc{\stateSet}$: for each $x$, $\probeval{\sd_{t}}{x} = f^{*}_{\hat{x},\relclass{x}{\rel}} \geq 0$ and $\probeval{\sd_{t}}{S} = \sum_{x \in \stateSet} \probeval{\sd_{t}}{x} = \sum_{x \in \stateSet} f^{*}_{\hat{x},\relclass{x}{\rel}} = \sum_{\equivclass \in \partitionset{\stateSet}} \sum_{x \in \equivclass} f^{*}_{\hat{x},\equivclass} = \sum_{\equivclass \in \partitionset{\stateSet}} f^{*}_{\equivclass,\netsink} = 1$.
			The following property holds for $\sd_{t}$: $\sd \liftrel \sd_{t}$.
			In fact, given an equivalence class $\equivclass$, $\probeval{\sd_{t}}{\equivclass} = \sum_{x \in \equivclass} \probeval{\sd_{t}}{x} = \sum_{x \in \equivclass}  f^{*}_{\hat{x},\equivclass} = f^{*}_{\equivclass, \netsink} = \probeval{\sd}{\equivclass}$.
			The second equality follows from the definition of $\sd_{t}$ while the last two equalities come from the constraints of $\LPproblemTBetaMuTransRel{t}{a}{\sd}{\allowedTransitions}{\rel}$.

			Let $\sched$ be a scheduler defined as follows: for each execution fragment $\phi \in \finiteFrags{\aut}$,
			\[
				\probeval{\schedeval{\sched}{\phi}}{x} =
					\begin{cases}
						f^{*}_{v,v^{\tr}}/\incomingflow^{*}_{v} & \text{if $\incomingflow^{*}_{v} \neq 0$, $\trace{\phi} = \emptytrace$, and $x = \tr = \strongTransition{v}{\hidden}{\rho} \in \allowedTransitions$;} \\
						f^{*}_{v,v^{\tr}_{a}}/\incomingflow^{*}_{v} & \text{if $\incomingflow^{*}_{v} \neq 0$, $\trace{\phi} = \emptytrace$, $a \neq \hidden$, and $x = \tr = \strongTransition{v}{a}{\rho} \in \allowedTransitions$;} \\
						f^{*}_{v_{a},v^{\tr}_{a}}/\incomingflow^{*}_{v_{a}} & \text{if $\incomingflow^{*}_{v_{a}} \neq 0$, $\trace{\phi} = a \neq \hidden$, and $x = \tr = \strongTransition{v}{\hidden}{\rho} \in \allowedTransitions$;} \\
						f^{*}_{v,\relclass{v}{\rel}}/\incomingflow^{*}_{v} & \text{if $\incomingflow^{*}_{v} \neq 0$, $\trace{\phi} = \emptytrace$, $a = \hidden$, and $x = \bot$;} \\
						f^{*}_{v_{a},\relclass{v}{\rel}}/\incomingflow^{*}_{v_{a}} & \text{if $\incomingflow^{*}_{v_{a}} \neq 0$, $\trace{\phi} = a \neq \hidden$, and $x = \bot$;} \\
						1 & \text{if $\trace{\phi} \notin \{\emptytrace, \trace{a}\}$ and $x = \bot$;} \\
						1 & \text{if $\incomingflow^{*}_{v} = 0$, $\trace{\phi} = \emptytrace$ and $x = \bot$;} \\
						1 & \text{if $\incomingflow^{*}_{v_{a}} = 0$, $\trace{\phi} = a \neq \hidden$ and $x = \bot$;} \\
						0 & \text{otherwise}
					\end{cases}
			\]
			where $v = \last{\phi}$.
				
			It is interesting to observe that the above scheduler is a determinate scheduler \cite{CS02} since for each $\phi, \phi' \in \finiteFrags{\aut}$ such that $\last{\phi} = \last{\phi'}$ and $\trace{\phi} = \trace{\phi'}$, we have $\schedeval{\sched}{\phi} = \schedeval{\sched}{\phi'}$.
			In fact, given $\phi, \phi' \in \finiteFrags{\aut}$ such that $\last{\phi} = \last{\phi'} = v$ and $\trace{\phi} = \trace{\phi'}$, if $\trace{\phi} = \trace{\phi'} = \emptytrace$, then $\probeval{\schedeval{\sched}{\phi}}{\bot} = f^{*}_{v,\relclass{v}{\rel}}/\incomingflow^{*}_{v} = \probeval{\schedeval{\sched}{\phi'}}{\bot}$, for each transition $\tr = \strongTransition{v}{\hidden}{\rho}$, $\probeval{\schedeval{\sched}{\phi}}{\tr} = f^{*}_{v,v^{\tr}}/\incomingflow^{*}_{v} = \probeval{\schedeval{\sched}{\phi'}}{\tr}$, and for each transition $\tr = \strongTransition{v}{a}{\rho}$, $\probeval{\schedeval{\sched}{\phi}}{\tr} = f^{*}_{v,v^{tr}_{a}}/\incomingflow^{*}_{v} = \probeval{\schedeval{\sched}{\phi'}}{\tr}$, as required.
			If $\trace{\phi} = \trace{\phi'} = a \neq \hidden$, then $\probeval{\schedeval{\sched}{\phi}}{\bot} = f^{*}_{v_{a},\relclass{v}{\rel}}/\incomingflow^{a}_{v} = \probeval{\schedeval{\sched}{\phi'}}{\bot}$ and for each transition $\tr = \strongTransition{v}{\hidden}{\rho}$, $\probeval{\schedeval{\sched}{\phi}}{\tr} = f^{*}_{v_{a},v^{tr}_{a}}/\incomingflow^{*}_{v_{a}} = \probeval{\schedeval{\sched}{\phi'}}{\tr}$;
			for all other cases, either $\probeval{\schedeval{\sched}{\phi}}{\bot} = 1 = \probeval{\schedeval{\sched}{\phi'}}{\bot}$ or $\probeval{\schedeval{\sched}{\phi}}{x} = 0 = \probeval{\schedeval{\sched}{\phi'}}{x}$, thus for each $\phi, \phi' \in \finiteFrags{\aut}$ such that $\last{\phi} = \last{\phi'}$ and $\trace{\phi} = \trace{\phi'}$, we have $\schedeval{\sched}{\phi} = \schedeval{\sched}{\phi'}$.
			
			Let $\sd_{\sched,t}$ be the probabilistic execution fragment generated by $\sched$ from $t$.
			In order to induce an allowed weak transition $\weakCombinedAllowedTransition{t}{a}{\sd_{t}}{\allowedTransitions}$, following conditions must be satisfied:
			\begin{enumerate}
				\item for each $\phi \in \finiteFrags{\aut}$, $\Supp{\schedeval{\sched}{\phi}} \subseteq \allowedTransitions$,
				\item $\probeval{\sd_{\sched,t}}{\finiteFrags{\aut}} = 1$, 
				\item for each $\phi \in \finiteFrags{\aut}$, if $\probeval{\sd_{\sched,t}}{\phi} > 0$ then $\trace{\phi} = \trace{a}$, and 
				\item for each state $t' \in \stateSet$, $\probeval{\sd_{\sched,t}}{\{\phi \in \finiteFrags{\aut} \mid \last{\phi} = t'\}} = \probeval{\sd_{t}}{t'}$.
			\end{enumerate}
			We now prove that such conditions are actually satisfied:
			\begin{enumerate}
				\item 
					this follows immediately from the definition of $\sched$ since for each transition $\tr$ such that $\probeval{\schedeval{\sched}{\phi}}{\tr} > 0$, $\tr \in \allowedTransitions$, thus $\Supp{\schedeval{\sched}{\phi}} \subseteq \allowedTransitions$.
				\item
					Suppose that condition~4 holds.
					This implies that for each state $v \in \stateSet$, $\probeval{\sd_{\sched,t}}{\{\phi \in \finiteFrags{\aut} \mid \last{\phi} = v\}} = \probeval{\sd_{t}}{v}$, hence $\probeval{\sd_{\sched,t}}{\finiteFrags{\aut}} = \sum_{v \in \stateSet} \probeval{\sd_{\sched,t}}{\{\phi \in \finiteFrags{\aut} \mid \last{\phi} = v\}} = \sum_{v \in \stateSet} \probeval{\sd_{t}}{v} = \sum_{v \in \stateSet} f^{*}_{\hat{v},\relclass{v}{\rel}} = \sum_{\equivclass \in \partitionset{\stateSet}}\sum_{v \in \equivclass} f^{*}_{\hat{v},\equivclass} = \sum_{\equivclass \in \partitionset{\stateSet}} f^{*}_{\equivclass,\netsink} = 1$, as required.
					
				\item
					Let $\phi$ be an execution fragment such that $\probeval{\sd_{\sched,t}}{\phi} > 0$.
					Since $\probeval{\sd_{\sched,t}}{\phi} = \probeval{\sd_{\sched,t}}{\cone{\phi}}\probeval{\schedeval{\sched}{\phi}}{\bot}$, $\probeval{\sd_{\sched,t}}{\phi} > 0$ holds if and only if $\probeval{\sd_{\sched,t}}{\cone{\phi}}\probeval{\schedeval{\sched}{\phi}}{\bot} > 0$, that is, $\probeval{\sd_{\sched,t}}{\cone{\phi}} > 0$ and $\probeval{\schedeval{\sched}{\phi}}{\bot} > 0$.
					Now, assume that $\probeval{\sd_{\sched,t}}{\cone{\phi}} > 0$.
					According to the definition of the scheduler, $\probeval{\schedeval{\sched}{\phi}}{\bot} > 0$ holds if 
					\begin{itemize}
						\item 
							$f^{*}_{v,\relclass{v}{\rel}}/\incomingflow^{*}_{v} > 0$, $\trace{\phi} = \emptytrace$, $a = \hidden$, and $v = \last{\phi}$;
						\item 
							$f^{*}_{v_{a},\relclass{v}{\rel}}/\incomingflow^{*}_{v_{a}} > 0$, $\trace{\phi} = a \neq \hidden$ and $v = \last{\phi}$;
						\item
							$\trace{\phi} \notin \{\emptytrace, \trace{a}\}$;
						\item 
							$\incomingflow^{*}_{v} = 0$, $\trace{\phi} = \emptytrace$ and $x = \bot$; or
						\item 
							$\incomingflow^{*}_{v_{a}} = 0$, $\trace{\phi} = a \neq \hidden$ and $x = \bot$;
					\end{itemize}
					The first and last two cases imply that $\trace{\phi} = \trace{a}$, as required;
					for the third case, we show that it can not occur if $\probeval{\sd_{\sched,t}}{\cone{\phi}} > 0$: suppose that $\trace{\phi} \notin \{\emptytrace, \trace{a}\}$.
					This implies that $\trace{\phi} = b$ for some sequence $b$ of external actions with $b \neq a$.
					Denote by $b_{1}$ the first action of $b$ and suppose that $b_{1} \neq a$.
					Let $\phi_{1}$ and $\phi_{2}$ be two execution fragments such that $\phi = \phi_{1}b_{1}\phi_{2}$ and $\trace{\phi_{1}} = \emptytrace$ and denote by $v_{1}$ and $v_{2}$ the last state of $\phi_{1}$ and the first state of $\phi_{2}$, respectively.
					The definition of probabilistic execution fragments and the fact that $\probeval{\sd_{\sched,t}}{\cone{\phi}} > 0$ imply that $\probeval{\sd_{\sched,t}}{\cone{\phi_{1}}} > 0$, $\probeval{\schedeval{\sched}{\phi_{1}}}{\tr} > 0$ and $\probeval{\rho}{v_{2}} > 0$ for some transition $\tr = \strongTransition{v_{1}}{b_{1}}{\rho}$.
					Since $b_{1} \neq a$ and $b_{1} \neq \hidden$, then by definition of the scheduler follows that $\probeval{\schedeval{\sched}{\phi_{1}}}{\tr} = 0$ for each transition $\tr = \strongTransition{v_{1}}{b_{1}}{\rho}$, thus $\probeval{\sd_{\sched,t}}{\cone{\phi}} = 0$.
					This contradicts the hypothesis that $\probeval{\sd_{\sched,t}}{\cone{\phi}} > 0$ and hence $\trace{\phi} \notin \{\emptytrace, \trace{a}\}$ can not occur.
					If $b_{1} = a$, consider $b_{2}$ and let $\phi_{1}$ and $\phi_{2}$ be two execution fragments such that $\phi = \phi_{1}b_{2}\phi_{2}$ and $\trace{\phi_{1}} = a$  and denote by $v_{1}$ and $v_{2}$ the last state of $\phi_{1}$ and the first state of $\phi_{2}$, respectively.
					The definition of probabilistic execution fragments and the fact that $\probeval{\sd_{\sched,t}}{\cone{\phi}} > 0$ imply that $\probeval{\sd_{\sched,t}}{\cone{\phi_{1}}} > 0$, $\probeval{\schedeval{\sched}{\phi_{1}}}{\tr} > 0$ and $\probeval{\rho}{v_{2}} > 0$ for some transition $\tr = \strongTransition{v_{1}}{b_{2}}{\rho}$.
					Since $\trace{\phi_{1}} = a \neq \hidden$ and $b_{2} \neq \hidden$, then by definition of the scheduler follows that $\probeval{\schedeval{\sched}{\phi_{1}}}{\tr} = 0$ for each transition $\tr = \strongTransition{v_{1}}{b_{2}}{\rho}$, thus $\probeval{\sd_{\sched,t}}{\cone{\phi}} = 0$.
					This contradicts the hypothesis that $\probeval{\sd_{\sched,t}}{\cone{\phi}} > 0$ and hence $\trace{\phi} \notin \{\emptytrace, \trace{a}\}$ can not occur.
					
				\item
					We first show by induction that for each $x \in \stateSet$ and each $n \in \nat$, $\incomingflow^{*}_{x}$ is an upper bound for the sum of the probabilities of the cones of execution fragments with empty trace and last state $x$ within $n$ steps, that is, denoted by $F_{n}(x)$ the set $\{\phi \in \finiteFrags{\aut} \mid \trace{\phi} = \emptytrace, \last{\phi} = x, \length{\phi} \leq n\}$, $\sum_{\phi \in F_{n}(x)} \probeval{\sd_{\sched,t}}{\cone{\phi}} \leq \incomingflow^{*}_{x}$;
					similarly $\incomingflow^{*}_{x_{a}}$ is an upper bound for the sum of the probabilities of the cones of execution fragments with trace $a \neq \hidden$ and last state $x$ within $n$ steps, that is, denoted by $F^{a}_{n}(x)$ the set $\{\phi \in \finiteFrags{\aut} \mid \trace{\phi} = a, \last{\phi} = x, \length{\phi} \leq n\}$, $\sum_{\phi \in F^{a}_{n}(x)} \probeval{\sd_{\sched,t}}{\cone{\phi}} \leq \incomingflow^{*}_{x_{a}}$.
					Note that for each $v \in \stateSet$ and $n \in \nat$, it holds that $F_{n}(v) \subseteq F_{n+1}(v)$ and $F^{a}_{n}(v) \subseteq F^{a}_{n+1}(v)$.
					
					We start showing that for each $x \in \stateSet$ and each $n \in \nat$, $\sum_{\phi \in F_{n}(x)} \probeval{\sd_{\sched,t}}{\cone{\phi}} \leq \incomingflow^{*}_{x}$:
					\begin{description}
						\item[Case $n=0$ and $x=t$:] 
							the only finite execution fragment that has length $0$ is $\phi = t$ and this implies that $\sum_{\phi \in F_{0}(t)} \probeval{\sd_{\sched,t}}{\cone{\phi}} = \probeval{\sd_{\sched,t}}{\cone{t}} = 1 = f^{*}_{\netsource,t} \leq \incomingflow^{*}_{t}$;
						\item[Case $n=0$ and $x \neq t$:] 
							as in the previous case we have $\phi = x$, thus $\sum_{\phi \in F_{0}(x)} \probeval{\sd_{\sched,t}}{\phi} = \probeval{\sd_{\sched,t}}{x} = \probeval{\sd_{\sched,t}}{\cone{x}} = 0 \leq \incomingflow^{*}_{x}$;
						\item[Case $n > 0$ and $x = t$:]
							\begin{align*}
								\sum_{\phi \in F_{n}(t)} \probeval{\sd_{\sched,t}}{\cone{\phi}} 
									& = \probeval{\sd_{\sched,t}}{\cone{t}} + \sum_{\phi'\hidden t \in F_{n}(t)} \probeval{\sd_{\sched,t}}{\cone{\phi'\hidden t}} \\
									& = 1 + \sum_{\phi'\hidden t \in F_{n}(t)} \probeval{\sd_{\sched,t}}{\cone{\phi'}}\sum_{\{\tr = \strongTransition{y}{\hidden}{\rho} \mid \last{\phi'} = y\}} \probeval{\schedeval{\sched}{\phi'}}{\tr}\probeval{\rho}{t}\\
									& = f^{*}_{\netsource,t} + \sum_{y \in \stateSet} \sum_{\phi' \in F_{n-1}(y)} \probeval{\sd_{\sched,t}}{\cone{\phi'}} \sum_{\tr = \strongTransition{y}{\hidden}{\rho}} \probeval{\schedeval{\sched}{\phi'}}{\tr}\probeval{\rho}{t}\\
									& = f^{*}_{\netsource,t} + \sum_{y \in \stateSet} \sum_{\tr = \strongTransition{y}{\hidden}{\rho}} \probeval{\rho}{t} \sum_{\phi' \in F_{n-1}(y)} \probeval{\sd_{\sched,t}}{\cone{\phi'}} \probeval{\schedeval{\sched}{\phi'}}{\tr} \\
									& = f^{*}_{\netsource,t} + \sum_{y \in \stateSet} \sum_{\tr = \strongTransition{y}{\hidden}{\rho}} \probeval{\rho}{t} \sum_{\phi' \in F_{n-1}(y)} \probeval{\sd_{\sched,t}}{\cone{\phi'}} \dfrac{f^{*}_{y,y^{\tr}}}{\incomingflow^{*}_{y}}\\
									& = f^{*}_{\netsource,t} + \sum_{y \in \stateSet} \sum_{\tr = \strongTransition{y}{\hidden}{\rho}} \probeval{\rho}{t} \dfrac{f^{*}_{y,y^{\tr}}}{\incomingflow^{*}_{y}} \sum_{\phi' \in F_{n-1}(y)} \probeval{\sd_{\sched,t}}{\cone{\phi'}}\\
									& \leq  f^{*}_{\netsource,t} + \sum_{y \in \stateSet} \sum_{\tr = \strongTransition{y}{\hidden}{\rho}} \probeval{\rho}{t} \dfrac{f^{*}_{y,y^{\tr}}}{\incomingflow^{*}_{y}} \incomingflow^{*}_{y} \\
									& =  f^{*}_{\netsource,t} + \sum_{y \in \stateSet} \sum_{\tr = \strongTransition{y}{\hidden}{\rho}} \probeval{\rho}{t} f^{*}_{y,y^{\tr}} \\
									& = f^{*}_{\netsource,t} + \sum_{y \in \stateSet} \sum_{\tr = \strongTransition{y}{\hidden}{\rho}} f^{*}_{y^{\tr},t} \\
									& = f^{*}_{\netsource,t} + \sum_{\tr = \strongTransition{z}{\hidden}{\rho}} f^{*}_{z^{\tr},t} \\
									& = \incomingflow^{*}_{t}
							\end{align*}
						\item[Case $n > 0$ and $x \neq t$:]
							\begin{align*}
								\sum_{\phi \in F_{n}(x)} \probeval{\sd_{\sched,t}}{\cone{\phi}} 
									& = \probeval{\sd_{\sched,t}}{\cone{x}} + \sum_{\phi'\hidden x \in F_{n}(x)} \probeval{\sd_{\sched,t}}{\cone{\phi'\hidden x}} \\
									& = \sum_{\phi'\hidden x \in F_{n}(x)} \probeval{\sd_{\sched,t}}{\cone{\phi'}}\sum_{\{\tr = \strongTransition{y}{\hidden}{\rho} \mid \last{\phi'} = y\}} \probeval{\schedeval{\sched}{\phi'}}{\tr}\probeval{\rho}{x}\\
									& = \sum_{y \in \stateSet} \sum_{\phi' \in F_{n-1}(y)} \probeval{\sd_{\sched,t}}{\cone{\phi'}} \sum_{\tr = \strongTransition{y}{\hidden}{\rho}} \probeval{\schedeval{\sched}{\phi'}}{\tr}\probeval{\rho}{x}\\
									& = \sum_{y \in \stateSet} \sum_{\tr = \strongTransition{y}{\hidden}{\rho}} \probeval{\rho}{x} \sum_{\phi' \in F_{n-1}(y)} \probeval{\sd_{\sched,t}}{\cone{\phi'}} \probeval{\schedeval{\sched}{\phi'}}{\tr} \\
									& = \sum_{y \in \stateSet} \sum_{\tr = \strongTransition{y}{\hidden}{\rho}} \probeval{\rho}{x} \sum_{\phi' \in F_{n-1}(y)} \probeval{\sd_{\sched,t}}{\cone{\phi'}} \dfrac{f^{*}_{y,y^{\tr}}}{\incomingflow^{*}_{y}}\\
									& = \sum_{y \in \stateSet} \sum_{\tr = \strongTransition{y}{\hidden}{\rho}} \probeval{\rho}{x} \dfrac{f^{*}_{y,y^{\tr}}}{\incomingflow^{*}_{y}} \sum_{\phi' \in F_{n-1}(y)} \probeval{\sd_{\sched,t}}{\cone{\phi'}}\\
									& \leq  \sum_{y \in \stateSet} \sum_{\tr = \strongTransition{y}{\hidden}{\rho}} \probeval{\rho}{x} \dfrac{f^{*}_{y,y^{\tr}}}{\incomingflow^{*}_{y}} \incomingflow^{*}_{y} \\
									& = \sum_{y \in \stateSet} \sum_{\tr = \strongTransition{y}{\hidden}{\rho}} \probeval{\rho}{x} f^{*}_{y,y^{\tr}} \\
									& = \sum_{y \in \stateSet} \sum_{\tr = \strongTransition{y}{\hidden}{\rho}} f^{*}_{y^{\tr},x} \\
									& = \sum_{\tr = \strongTransition{z}{\hidden}{\rho}} f^{*}_{z^{\tr},x} \\
									& = \incomingflow^{*}_{x}
							\end{align*}
					\end{description}
					This completes the proof that for each $x \in \stateSet$ and each $n \in \nat$, $\sum_{\phi \in F_{n}(x)} \probeval{\sd_{\sched,t}}{\cone{\phi}} \leq \incomingflow^{*}_{x}$.
					Now we consider the second result relative to $a \neq \hidden$, that is, for each $x \in \stateSet$ and each $n \in \nat$, $\sum_{\phi \in F^{a}_{n}(x)} \probeval{\sd_{\sched,t}}{\cone{\phi}} \leq \incomingflow^{*}_{x_{a}}$:
					\begin{description}
						\item[Case $n=0$:] 
							by definition of the trace of an execution fragment, we have that $F^{a}_{0}(x) = \emptyset$ and thus $\sum_{\phi \in F^{a}_{0}(x)} \probeval{\sd_{\sched,t}}{\cone{\phi}} = \sum_{\phi \in \emptyset} \probeval{\sd_{\sched,t}}{\cone{\phi}} = 0 \leq \incomingflow^{*}_{x_{a}}$;
						\item[Case $n > 0$:]
							\begin{align*}
								\sum_{\phi \in F^{a}_{n}(x)} \probeval{\sd_{\sched,t}}{\cone{\phi}} 
									& = \sum_{\phi'\hidden x \in F^{a}_{n}(x)} \probeval{\sd_{\sched,t}}{\cone{\phi'\hidden x}} + \sum_{\phi'a x \in F^{a}_{n}(x)} \probeval{\sd_{\sched,t}}{\cone{\phi'a x}} \\
									& = \sum_{\phi'\hidden x \in F^{a}_{n}(x)} \probeval{\sd_{\sched,t}}{\cone{\phi'}}\sum_{\{\tr = \strongTransition{y}{\hidden}{\rho} \mid \last{\phi'} = y\}} \probeval{\schedeval{\sched}{\phi'}}{\tr}\probeval{\rho}{x}\\
									& \phantom{=} + \sum_{\phi'a x \in F^{a}_{n}(x)} \probeval{\sd_{\sched,t}}{\cone{\phi'}}\sum_{\{\tr = \strongTransition{y}{a}{\rho} \mid \last{\phi'} = y\}} \probeval{\schedeval{\sched}{\phi'}}{\tr}\probeval{\rho}{x}\\
									& = \sum_{y \in \stateSet} \sum_{\phi' \in F^{a}_{n-1}(y)} \probeval{\sd_{\sched,t}}{\cone{\phi'}} \sum_{\tr = \strongTransition{y}{\hidden}{\rho}} \probeval{\schedeval{\sched}{\phi'}}{\tr}\probeval{\rho}{x}\\
									& \phantom{=} + \sum_{y \in \stateSet} \sum_{\phi' \in F_{n-1}(y)} \probeval{\sd_{\sched,t}}{\cone{\phi'}} \sum_{\tr = \strongTransition{y}{a}{\rho}} \probeval{\schedeval{\sched}{\phi'}}{\tr}\probeval{\rho}{x}\\
									& = \sum_{y \in \stateSet} \sum_{\tr = \strongTransition{y}{\hidden}{\rho}} \probeval{\rho}{x} \sum_{\phi' \in F^{a}_{n-1}(y)} \probeval{\sd_{\sched,t}}{\cone{\phi'}} \probeval{\schedeval{\sched}{\phi'}}{\tr} \\
									& \phantom{=} + \sum_{y \in \stateSet} \sum_{\tr = \strongTransition{y}{a}{\rho}} \probeval{\rho}{x} \sum_{\phi' \in F_{n-1}(y)} \probeval{\sd_{\sched,t}}{\cone{\phi'}} \probeval{\schedeval{\sched}{\phi'}}{\tr} \\
									& = \sum_{y \in \stateSet} \sum_{\tr = \strongTransition{y}{\hidden}{\rho}} \probeval{\rho}{x} \sum_{\phi' \in F^{a}_{n-1}(y)} \probeval{\sd_{\sched,t}}{\cone{\phi'}} \dfrac{f^{*}_{y_{a},y^{\tr}_{a}}}{\incomingflow^{*}_{y_{a}}}\\
									& \phantom{=} + \sum_{y \in \stateSet} \sum_{\tr = \strongTransition{y}{a}{\rho}} \probeval{\rho}{x} \sum_{\phi' \in F_{n-1}(y)} \probeval{\sd_{\sched,t}}{\cone{\phi'}} \dfrac{f^{*}_{y,y^{\tr}_{a}}}{\incomingflow^{*}_{y}}\\
									& = \sum_{y \in \stateSet} \sum_{\tr = \strongTransition{y}{\hidden}{\rho}} \probeval{\rho}{x} \dfrac{f^{*}_{y_{a},y^{\tr}_{a}}}{\incomingflow^{*}_{y_{a}}} \sum_{\phi' \in F^{a}_{n-1}(y)} \probeval{\sd_{\sched,t}}{\cone{\phi'}}\\
									& \phantom{=} + \sum_{y \in \stateSet} \sum_{\tr = \strongTransition{y}{a}{\rho}} \probeval{\rho}{x} \dfrac{f^{*}_{y,y^{\tr}_{a}}}{\incomingflow^{*}_{y}} \sum_{\phi' \in F_{n-1}(y)} \probeval{\sd_{\sched,t}}{\cone{\phi'}}\\
									& \leq  \sum_{y \in \stateSet} \sum_{\tr = \strongTransition{y}{\hidden}{\rho}} \probeval{\rho}{x} \dfrac{f^{*}_{y_{a},y^{\tr}_{a}}}{\incomingflow^{*}_{y_{a}}} \incomingflow^{*}_{y_{a}} + \sum_{y \in \stateSet} \sum_{\tr = \strongTransition{y}{a}{\rho}} \probeval{\rho}{x} \dfrac{f^{*}_{y,y^{\tr}_{a}}}{\incomingflow^{*}_{y}} \incomingflow^{*}_{y} \\
									& = \sum_{y \in \stateSet} \sum_{\tr = \strongTransition{y}{\hidden}{\rho}} \probeval{\rho}{x} f^{*}_{y_{a},y^{\tr}_{a}} + \sum_{y \in \stateSet} \sum_{\tr = \strongTransition{y}{a	}{\rho}} \probeval{\rho}{x} f^{*}_{y,y^{\tr}_{a}} \\
									& = \sum_{y \in \stateSet} \sum_{\tr = \strongTransition{y}{\hidden}{\rho}} f^{*}_{y^{\tr}_{a},x_{a}} + \sum_{y \in \stateSet} \sum_{\tr = \strongTransition{y}{a}{\rho}} f^{*}_{y^{\tr}_{a},x_{a}}\\
									& = \sum_{\tr = \strongTransition{z}{\hidden}{\rho}} f^{*}_{z^{\tr}_{a},x_{a}} + \sum_{\tr = \strongTransition{z}{a}{\rho}} f^{*}_{z^{\tr}_{a},x_{a}}\\
									& = \incomingflow^{*}_{x_{a}}
							\end{align*}
					\end{description}
					This completes the proof that for each $x \in \stateSet$ and each $n \in \nat$, $\sum_{\phi \in F^{a}_{n}(x)} \probeval{\sd_{\sched,t}}{\cone{\phi}} \leq \incomingflow^{*}_{x_{a}}$.
					
					For each $v \in \stateSet$, denote by $F(v)$ the set $\bigcup_{n \in \nat} F_{n}(v)$ and by $F^{a}(v)$ the set $\bigcup_{n \in \nat} F^{a}_{n}(v)$: we have again that $\sum_{\phi \in F(x)} \probeval{\sd_{\sched,t}}{\cone{\phi}} \leq \incomingflow^{*}_{x}$ and $\sum_{\phi \in F^{a}(x)} \probeval{\sd_{\sched,t}}{\cone{\phi}} \leq \incomingflow^{*}_{x_{a}}$.
					Now it is immediate to show that for each state $v \in \stateSet$, 
					\begin{align*}
						& \probeval{\sd_{\sched,t}}{\{\phi \in \finiteFrags{\aut} \mid \last{\phi} = v\}} \\
						& = \sum_{\{\phi \in \finiteFrags{\aut} \mid \last{\phi} = v\}} \probeval{\sd_{\sched,t}}{\cone{\phi}}\probeval{\schedeval{\sched}{\phi}}{\bot} \\
						& = \sum_{\phi \in F(v) \cup F^{a}(v)} \probeval{\sd_{\sched,t}}{\cone{\phi}}\probeval{\schedeval{\sched}{\phi}}{\bot} \\ 
						& \phantom{=} + \sum_{\{\phi \in \finiteFrags{\aut} \setminus (F(v) \cup F^{a}(v)) \mid \last{\phi} = v\}} \probeval{\sd_{\sched,t}}{\cone{\phi}}\probeval{\schedeval{\sched}{\phi}}{\bot} \\
						& = \sum_{\phi \in F(v) \cup F^{a}(v)} \probeval{\sd_{\sched,t}}{\cone{\phi}}\probeval{\schedeval{\sched}{\phi}}{\bot} \\ 
						& = \sum_{\phi \in F(v)} \probeval{\sd_{\sched,t}}{\cone{\phi}}\probeval{\schedeval{\sched}{\phi}}{\bot} + \sum_{\phi \in F^{a}(v)} \probeval{\sd_{\sched,t}}{\cone{\phi}}\probeval{\schedeval{\sched}{\phi}}{\bot} \\ 
						& \stackrel{\dag}{=} 
							\begin{cases}
								\displaystyle
								\sum_{\phi \in F(v)} \probeval{\sd_{\sched,t}}{\cone{\phi}} \dfrac{f^{*}_{v,\relclass{v}{\rel}}}{\incomingflow^{*}_{v}} & \text{if $a = \hidden$}\\
								\displaystyle
								\sum_{\phi \in F^{a}(v)} \probeval{\sd_{\sched,t}}{\cone{\phi}} \dfrac{f^{*}_{v_{a},\relclass{v}{\rel}}}{\incomingflow^{*}_{v_{a}}} & \text{otherwise}
							\end{cases}\\	
						& \leq \incomingflow^{*}_{\hat{v}} \dfrac{f^{*}_{\hat{v},\relclass{v}{\rel}}}{\incomingflow^{*}_{\hat{v}}}  \\
						& = f^{*}_{\hat{v},\relclass{v}{\rel}}  \\
						& = \probeval{\sd_{t}}{v}
					\end{align*}
					where the inequality is justified by the results about probabilities of cones we proved above and the equality $\stackrel{\dag}{=}$ by the definition of the scheduler $\sched$ that ensures that at least one between $\probeval{\schedeval{\sched}{\phi}}{\bot}$ and $\probeval{\schedeval{\sched}{\phi'}}{\bot}$ is $0$ provided that $\phi \in F(v)$ and $\phi' \in F^{a}(v)$.
					So we have that for each $v \in \stateSet$, $\probeval{\sd_{\sched,t}}{\{\phi \in \finiteFrags{\aut} \mid \last{\phi} = v\}} \leq \probeval{\sd_{t}}{v}$.
					
					Now, suppose for the sake of contradiction, that there exists a state $v$ such that $\probeval{\sd_{\sched,t}}{\{\phi \in \finiteFrags{\aut} \mid \last{\phi} = v\}} < \probeval{\sd_{t}}{v}$ and hence $\probeval{\sd_{\sched,t}}{\finiteFrags{\aut}} < 1 = \probeval{\sd_{t}}{\stateSet}$.
					This implies that there exists a set of infinite execution fragments $E$ that occurs with non-zero probability.
					Since the set of states $\stateSet$ is finite, there exists a set $C \subseteq E$ and a state $c$ (that can also be different from $v$) such that $c$ occurs infinitely many times in each execution fragment $\phi \in C$ and there exists a finite execution fragment $\phi_{c}$ with the following properties:
					\begin{itemize}
						\item $\last{\phi_{c}} = c$;
						\item $C \subseteq \cone{\phi_{c}}$;
						\item $\probeval{\sd_{\sched,t}}{\cup_{\phi \in C} \cone{\phi}} = \probeval{\sd_{\sched,t}}{\cone{\phi_{c}}}$; and 
						\item there exists a set $L \subseteq \finiteFrags{\aut}$ such that $\phi_{c} \notin L$, $\probeval{\sd_{\sched,t}}{\cup_{\phi \in L} \cone{\phi}} = \probeval{\sd_{\sched,t}}{\cone{\phi_{c}}}$, and for each $\phi \in L$, $\phi = \phi_{c} b_{1} s_{1} \dots b_{n} s_{n}$ for a family of actions $b_{i}$ and a family of states $s_{i}$ such that for each $0 < i < n$, $s_{i} \neq c$ and $s_{n} = c$.
					\end{itemize}
					Denote by $G$ the set $\{c b_{1} s_{1} \dots b_{n} s_{n} \mid \exists \phi \in L. \phi = \phi_{c} b_{1} s_{1} \dots b_{n} s_{n}\}$.
					Intuitively, the set $C$ models the fact that from $\phi_{c}$ we enter in a cycle such that the probability to reach again $c$ is $1$ (and the probability to leave the cycle is $0$) while the set $L$ contains the finite execution fragments $\phi$ that extend $\phi_{c}$ by an execution fragment in $G$ that can be seen as the generator of $C$, that is, it represents one loop of the cycle starting in $c$.
					Note that for each $\phi \in G$, $\trace{\phi} = \emptytrace$.
					Given an execution fragment $\phi$ such that $\last{\phi} = c$, let $\phi G^{n}$ be the set of execution fragments defined as follows:
					\[
						\phi G^{n} =
						\begin{cases}
							\{\phi\} & \text{if $n=0$ and}\\
							\{\phi'\phi'' \mid \phi'\in \phi G^{n-1}, \phi'' \in G\} & \text{if $n>0$.}
						\end{cases}
					\]
					It is immediate to verify that $L = \phi_{c}G^{1}$ and that for each $i \in \nat$, $\probeval{\sd_{\sched,t}}{\cup_{\phi \in \phi_{c}G^{i}} \cone{\phi}} = \probeval{\sd_{\sched,t}}{\cone{\phi_{c}}}$.
					Denote by $\phi G_{n}$ the set $\cup_{0\leq i \leq n} \phi G^{i}$.
					
					Now, suppose that $a = \hidden$ (the case $a \neq \hidden$ is analogous).
					Let $k_{c}$ be the length of $\phi_{c}$, that is, $k_{c} = \length{\phi_{c}}$;
					$p_{c}$ be the probability of $\cone{\phi_{c}}$, that is, $p_{c} = \probeval{\sd_{\sched,t}}{\cone{\phi_{c}}}$;
					$P_{c}$ be the sum of the probabilities of the cones of length at most $k_{c}$, that is, $P_{c} = \sum_{\phi \in F_{k_{c}}(c)} \probeval{\sd_{\sched,t}}{\cone{\phi}}$;
					and $\Delta_{c}$ be $\incomingflow^{*}_{c} - P_{c}$.
					Since $\incomingflow^{*}_{c}$ is finite and $p_{c} > 0$, $l = \lceil\Delta_{c} / p_{c}\rceil + 1$ is finite too;
					consider the set $F(c) = \cup_{n \in \nat} F_{n}(c)$: by definition of the set $F_{n}(c)$ we have that for each $0 \leq i \leq l$, $\phi_{c}G^{i} \subseteq F(c)$, thus 
					\begin{align*}
						\sum_{\phi \in F(c)} \probeval{\sd_{\sched,t}}{\cone{\phi}} 
						& = \sum_{\phi \in F_{k_{c}}(c)} \probeval{\sd_{\sched,t}}{\cone{\phi}} + \sum_{\phi \in \phi_{c}G_{l} \setminus \{\phi_{c}\}} \probeval{\sd_{\sched,t}}{\cone{\phi}} \\
						& \phantom{=} + \sum_{\phi \in F(c) \setminus (F_{k_{c}}(c) \cup \phi_{c}G_{l})} \probeval{\sd_{\sched,t}}{\cone{\phi}} \\
						& \geq \sum_{\phi \in F_{k_{c}}(c)} \probeval{\sd_{\sched,t}}{\cone{\phi}} + \sum_{\phi \in \phi_{c}G_{l} \setminus \{\phi_{c}\}} \probeval{\sd_{\sched,t}}{\cone{\phi}} \\
						& = P_{c} + \sum_{0 < i \leq l} \sum_{\phi \in \phi_{c}G^{i}} \probeval{\sd_{\sched,t}}{\cone{\phi}} \\
						& \geq P_{c} + \sum_{0 < i \leq l} \probeval{\sd_{\sched,t}}{\cup_{\phi \in \phi_{c}G^{i}} \cone{\phi}} \\
						& = P_{c} + \sum_{0 < i \leq l} \probeval{\sd_{\sched,t}}{\cone{\phi_{c}}} \\
						& = P_{c} + \sum_{0 < i \leq l} p_{c} \\
						& = P_{c} + lp_{c} \\
						& = P_{c} + (\lceil\Delta_{c} / p_{c}\rceil + 1)p_{c} \\
						& = P_{c} + \lceil\Delta_{c} / p_{c}\rceil p_{c} + p_{c} \\
						& \geq P_{c} + \dfrac{\Delta_{c}}{p_{c}}p_{c} + p_{c} \\
						& = P_{c} + \Delta_{c} + p_{c} \\
						& = P_{c} + \incomingflow^{*}_{c} - P_{c} + p_{c} \\
						& = \incomingflow^{*}_{c} + p_{c} \\
						& > \incomingflow^{*}_{c}
					\end{align*}
					but this contradicts the fact that $\sum_{\phi \in F(c)} \probeval{\sd_{\sched,t}}{\cone{\phi}} \leq \incomingflow^{*}_{c}$;
					thus for each $c \in \stateSet$, $\probeval{\sd_{\sched,t}}{\{\phi \in \finiteFrags{\aut} \mid \last{\phi} = c\}} = \probeval{\sd_{t}}{c}$, as required.
				\end{enumerate}
	\end{description}
\end{myproof}

\begin{corollary}
\label{cor:LPIncomingFlowEqualSumProbCones}
	Given a \PA{} $\aut$, an equivalence relation $\rel$ on $\stateSet$, an action $a$, a probability distribution $\sd \in \Disc{\stateSet}$, a set of allowed transitions $\allowedTransitions \subseteq \transitionRelation$, and a state $t \in \stateSet$, consider the problem $\LPproblemTBetaMuTransRel{t}{a}{\sd}{\allowedTransitions}{\rel}$ as defined in Section~\ref{sec:weakTransitionAsLPP}.
	
	$\LPproblemTBetaMuTransRel{t}{a}{\sd}{\allowedTransitions}{\rel}$ has a solution $f^{*}$ such that $f^{*}_{\equivclass,\netsink} = \probeval{\sd}{\equivclass}$ for each $\equivclass \in \partitionset{\stateSet}$ if and only if there exists a scheduler $\sched$ for $\aut$ that induces $\weakCombinedAllowedTransition{t}{a}{\sd_{t}}{\allowedTransitions}$ such that $\sd \liftrel \sd_{t}$ such that for each state $v$, $\incomingflow^{*}_{v} = \sum_{\alpha \in \{\beta \in \finiteFrags{\aut} \mid \last{\beta} = v\}} \probeval{\sd_{\sched,t}}{\cone{\alpha}}$.
\end{corollary}
\begin{myproof}
	Given a scheduler $\sched$ for $\aut$ that induces $\weakCombinedAllowedTransition{t}{a}{\sd_{t}}{\allowedTransitions}$, by the proof of Theorem~\ref{thm:LPequivalentToWeakAllowedTransitionLifting}, we know that there exists a solution $f^{*}$ such that $f^{*}_{\equivclass,\netsink} = \probeval{\sd}{\equivclass}$ for each $\equivclass \in \partitionset{\stateSet}$ and such that for each state $v$, $\incomingflow^{*}_{v} = \sum_{\alpha \in \{\beta \in \finiteFrags{\aut} \mid \last{\beta} = v\}} \probeval{\sd_{\sched,t}}{\cone{\alpha}}$.
	
	By the proof of Theorem~\ref{thm:LPequivalentToWeakAllowedTransitionLifting}, we know that given the optimal solution $f^{o}$ of the LP problem $\LPproblemTBetaMuTransRel{t}{a}{\sd}{\allowedTransitions}{\rel}$, we can define a scheduler $\sched$ inducing $\weakCombinedAllowedTransition{t}{a}{\sd_{t}}{\allowedTransitions}$ such that $\sd \liftrel \sd_{t}$ such that for each state $q$, $\sum_{\phi \in \{\alpha \in \finiteFrags{\aut} \mid \last{\alpha} = q\}} \probeval{\sd_{\sched,t}}{\cone{\phi}} \leq \incomingflow^{o}_{q}$.
	We claim that for each state $q$, $\sum_{\phi \in \{\alpha \in \finiteFrags{\aut} \mid \last{\alpha} = q\}} \probeval{\sd_{\sched,t}}{\cone{\phi}} = \incomingflow^{o}_{q}$.
	Suppose, for the sake of contradiction, that there exists a state $z$ such that $\sum_{\phi \in \{\alpha \in \finiteFrags{\aut} \mid \last{\alpha} = z\}} \probeval{\sd_{\sched,t}}{\cone{\phi}} < \incomingflow^{o}_{z}$.
	Theorem~\ref{thm:LPequivalentToWeakAllowedTransitionLifting} implies that the LP problem $\LPproblemTBetaMuTransRel{t}{a}{\sd}{\allowedTransitions}{\rel}$ has a feasible solution $f^{*}$ such that for each state $q$, $\incomingflow^{*}_{q} = \sum_{\phi \in \{\alpha \in \finiteFrags{\aut} \mid \last{\alpha} = q\}} \probeval{\sd_{\sched,t}}{\cone{\phi}} \leq \incomingflow^{o}_{q}$.
	Since $\incomingflow^{*}_{z} < \incomingflow^{o}_{z}$, we have that $\max \sum_{(x,y) \in E} -f^{o}_{x,y} < \max \sum_{(x,y) \in E} -f^{*}_{x,y}$ but this contradicts the fact that $f^{o}$ is an optimal solution.
	Hence it holds that $\sum_{\phi \in \{\alpha \in \finiteFrags{\aut} \mid \last{\alpha} = q\}} \probeval{\sd_{\sched,t}}{\cone{\phi}} = \incomingflow^{o}_{q}$, as required.
\end{myproof}

\begin{result}[Corollary~\ref{cor:LPequivalentToTransitionXYZ}]
	Given a \PA{} $\aut$, $t \in \stateSet$ and $h \notin \stateSet$, $a \in \actionSet$, $\gd, \sd, \sd_{t} \in \Disc{\stateSet}$, $\allowedTransitions \subseteq \transitionRelation$, an equivalence relation $\rel$ on $\stateSet$, the identity relation $\idrel$ on $\stateSet \cup\{h\}$, a transition $\strongTransition{h}{\hidden}{\gd}$, $\allowedTransitions_{h} = \allowedTransitions \cup \{\strongTransition{h}{\hidden}{\gd}\}$, $\transitionRelation_{h} = \transitionRelation \cup \{\strongTransition{h}{\hidden}{\gd}\}$, and the \PA{} $\aut_{h} = (\stateSet \cup \{h\}, \startState, \actionSet, \transitionRelation_{h})$, the following equivalences hold:
	\begin{enumerate}
	\item %\label{pnt:LPequivalentToWeakTransition}
		$\LPproblemTBetaMuTransRel{t}{a}{\sd}{\transitionRelation}{\rel}$ has a solution $f^{*}$ such that $f^{*}_{\equivclass,\netsink} = \probeval{\sd}{\equivclass}$ for each $\equivclass \in \partitionset{\stateSet}$ if and only if there exists a scheduler $\sched$ for $\aut$ inducing $\weakCombinedTransition{t}{a}{\sd_{t}}$ such that $\sd \liftrel \sd_{t}$;
		
	\item %\label{pnt:LPequivalentToHyperAllowedTransition}
		$\LPproblemTBetaMuTransRel{h}{a}{\sd}{\allowedTransitions_{h}}{\rel}$ ($\LPproblemTBetaMuTransRel{h}{a}{\sd}{\transitionRelation_{h}}{\rel}$) relative to $\aut_{h}$ has a solution $f^{*}$ such that $f^{*}_{\equivclass,\netsink} = \probeval{\sd}{\equivclass}$ for each $\equivclass \in \partitionset{\stateSet}$ if and only if there exists a scheduler $\sched$ for $\aut$ inducing $\hyperWeakCombinedAllowedTransition{\gd}{a}{\sd_{t}}{\allowedTransitions}$ ($\hyperWeakCombinedTransition{\gd}{a}{\sd_{t}}$, respectively) such that $\sd \liftrel \sd_{t}$;

	\end{enumerate}
\end{result}
\begin{myproof}
	The proof of the statement of the corollary involves Theorem~\ref{thm:LPequivalentToWeakAllowedTransitionLifting} for the equivalence between the LP problem and allowed weak combined transition, Proposition~\ref{pro:weakAllowedTransitionEquivalentToWeakTransitionWhenAllTransitionsAllowed} for ordinary transitions, and Proposition~\ref{pro:hyperTransitionEquivalentToWeakTransition} for hyper-transitions.
	
	More precisely, 
	\begin{enumerate}
	\item
		the statement follows immediately from Theorem~\ref{thm:LPequivalentToWeakAllowedTransitionLifting} and Proposition~\ref{pro:weakAllowedTransitionEquivalentToWeakTransitionWhenAllTransitionsAllowed}.
	\item
		By Theorem~\ref{thm:LPequivalentToWeakAllowedTransitionLifting}, $\LPproblemTBetaMuTransRel{h}{a}{\sd}{\allowedTransitions_{h}}{\rel}$ has a solution $f^{*}$ such that $f^{*}_{\equivclass,\netsink} = \probeval{\sd}{\equivclass}$ for each $\equivclass \in \partitionset{\stateSet}$ if and only if there exists a scheduler $\sched_{h}$ for $\aut_{h}$ that induces $\weakCombinedAllowedTransition{h}{a}{\sd_{t}}{\allowedTransitions_{h}}$ such that $\sd \liftrel \sd_{t}$ and the scheduler $\sched_{h}$ exists, by Proposition~\ref{pro:hyperTransitionEquivalentToWeakTransition}, if and only if there exists a scheduler $\sched$ for $\aut$ that induces $\hyperWeakCombinedAllowedTransition{\gd}{a}{\sd_{t}}{\allowedTransitions}$. 
		Since $\sd_{t}$ is reached also by $\sched$, $\sd \liftrel \sd_{t}$ still holds, as required.
		The case for $\LPproblemTBetaMuTransRel{h}{a}{\sd}{\transitionRelation_{h}}{\rel}$ follows immediately by Proposition~\ref{pro:weakAllowedTransitionEquivalentToWeakTransitionWhenAllTransitionsAllowed}.
	\end{enumerate}
\end{myproof}

\begin{result}[Proposition~\ref{prop:decideMatching}]
	Given a \PA{} $\aut$, two distributions $\gd_{1}, \gd_{2} \in \Disc{\stateSet}$, two actions $a_{1}, a_{2} \in \actionSet$, two sets $\allowedTransitions_{1}, \allowedTransitions_{2} \subseteq \transitionRelation$ of allowed transitions, and an equivalence relation $\rel$ on $\stateSet$, the existence of $\sd_{1}, \sd_{2} \in \Disc{\stateSet}$ such that $\hyperWeakCombinedAllowedTransition{\gd_{1}}{a_{1}}{\sd_{1}}{\allowedTransitions_{1}}$, $\hyperWeakCombinedAllowedTransition{\gd_{2}}{a_{2}}{\sd_{2}}{\allowedTransitions_{2}}$, and $\sd_{1} \liftrel \sd_{2}$ can be checked in polynomial time.
\end{result}
\begin{myproof}
	We remark that we denote by $\LPproblemTBetaMuTransRel{\gd}{a}{\sd}{\allowedTransitions}{\rel}$ the problem $\LPproblemTBetaMuTransRel{h}{a}{\sd}{\allowedTransitions_{h}}{\rel}$ relative to $\aut_{h} = (\stateSet \cup \{h\}, \startState, \actionSet, \transitionRelation \cup \{\strongTransition{h}{\hidden}{\gd}\})$ where $h \notin \stateSet$ and $\allowedTransitions_{h} = \allowedTransitions \cup \{\strongTransition{h}{\hidden}{\gd}\}$.
	
	Define the LP problem 
	$P_{1,2}$
	derived from the problems $P_{1} = \LPproblemTBetaMuTransRel{\gd_{1}}{a_{1}}{\bar{\sd}}{\allowedTransitions_{1}}{\rel}$ and $P_{2} = \LPproblemTBetaMuTransRel{\gd_{2}}{a_{2}}{\bar{\sd}}{\allowedTransitions_{2}}{\rel}$ as follows (after renaming of $P_{2}$ variables to avoid collisions): 
	the objective function of $P_{1,2}$ is the sum of the objective functions of $P_{1}$ and $P_{2}$;
	the set of constraints of $P_{1,2}$ is $\sum_{\equivclass \in \partitionset{\stateSet}} p_{\equivclass} = 1$ together with $p_{\equivclass} \geq 0$ for $\equivclass \in \partitionset{\stateSet}$ and the union of the sets of constraints of $P_{1}$ and $P_{2}$ where constraints $f_{\equivclass,\netsink} = \probeval{\bar{\sd}}{\equivclass}$ are replaced by $f_{\equivclass,\netsink} = p_{\equivclass}$.

	The proposition follows from the fact that $P_{1,2}$ has a solution if and only if both $P_{1}$ and $P_{2}$ have a solution for some common probability distribution $\bar{\sd}$ and thus, by Corollary~\ref{cor:LPequivalentToTransitionXYZ}(\ref{pnt:LPequivalentToHyperAllowedTransition}), if and only if $\gd_{1}$ and $\gd_{2}$ enable an allowed hyper-transition to $\sd_{1}$ and $\sd_{2}$, respectively, such that $\sd_{1} \liftrel \sd_{2}$, as required, since $\sd_{1} \liftrel \bar{\sd}$ as well as $\sd_{2} \liftrel \bar{\sd}$ and $\liftrel$ is transitive.
	It is immediate to see that $P_{1,2}$ can still be generated and solved in polynomial time, since it is just the union of $P_{1}$ and $P_{2}$ extended with at most $N$ variables and $2N$ constraints where $N = |\stateSet|$.

	We now prove the above claim:
	\begin{claim}
		$P_{1,2}$ has a solution if and only if there exists a probability distribution $\bar{\sd}$ such that both $P_{1}$ and $P_{2}$ have a solution.
	\end{claim}
	\begin{description}
		\item[($\Rightarrow$)]
			Suppose that $P_{1,2}$ has a solution and define $\bar{\sd}$ as follows: 
			for each $s \in \stateSet$, $\probeval{\bar{\sd}}{s} = \dfrac{p_{\equivclass}}{|\equivclass|}$ where $\equivclass = \relclass{s}{\rel}$.
			By hypothesis, $P_{1,2}$ has a solution, that is, there exists $f^{*}$ that maximizes the objective function of $P_{1,2}$ while satisfying constraints.
			In particular, $f^{*}$ satisfies constraints: $f^{*}_{u,v} \geq 0$ for each $(u,v) \in E$; $\sum_{(s,\equivclass) \in E} f^{*}_{s,\equivclass} - f^{*}_{\equivclass, \netsink} = 0 $ for each $\equivclass \in \partitionset{\stateSet}$ and $s \in \equivclass$; and $f^{*}_{\equivclass, \netsink} = p_{\equivclass}$ for each $\equivclass \in \partitionset{\stateSet}$.
			Now, consider $f^{*}_{1}$ and $f^{*}_{2}$ obtained by splitting $f^{*}$ according to variables relative to $P_{1}$ and $P_{2}$, respectively.
			It is straightforward to check that $f^{*}_{i}$ is a valid solution for $P_{i}$ with $i = 1, 2$, so, by Corollary~\ref{cor:LPequivalentToTransitionXYZ}(\ref{pnt:LPequivalentToHyperAllowedTransition}), it holds that $\sd_{1} \liftrel \bar{\sd}$ as well as $\sd_{2} \liftrel \bar{\sd}$ 
			
		\item[($\Leftarrow$)]
			Suppose that there exists $\bar{\sd}$ such that both problems $P_{1} = \LPproblemTBetaMuTransRel{\gd_{1}}{a_{1}}{\bar{\sd}}{\allowedTransitions_{1}}{\rel}$ and $P_{2} = \LPproblemTBetaMuTransRel{\gd_{2}}{a_{2}}{\bar{\sd}}{\allowedTransitions_{2}}{\rel}$ have a solution.
			Suppose that the set of variables of $P_{2}$ is disjoint from the set of variables of $P_{1}$.
			Let $f^{*}_{1}$ and $f^{*}_{2}$ the two solutions of $P_{1}$ and $P_{2}$ and denote by $f^{*}$ the union of $f^{*}_{1}$ and $f^{*}_{2}$ extended with the assignments $p_{\equivclass} = \probeval{\bar{\sd}}{\equivclass}$ for $\equivclass \in \partitionset{\stateSet}$.
			It is straightforward to check that $f^{*}$ satisfies all $P_{1,2}$ constraints since they are just the union of constraints of $P_{1}$ and $P_{2}$ that are satisfied by $f^{*}_{1}$ and $f^{*}_{2}$, respectively, and that the maximum of the objective function is given by $f^{*}$ since by definition the objective function is the sum of the two independent objective functions of $P_{1}$ and $P_{2}$ that are maximized by $f^{*}_{1}$ and $f^{*}_{2}$, respectively.
	\end{description}
	This concludes the proof of the claim and of the Proposition~\ref{prop:decideMatching}.
\end{myproof}

%%% Local Variables: 
%%% mode: latex
%%% TeX-master: "pa-weak-bisim-poly"
%%% End: 

\end{document}